\documentclass[longauth]{aa} 
\usepackage{graphicx}
\usepackage{txfonts}
\usepackage{natbib}

\def\eg{{\it e.g.\ }}

\def\ie{{\it i.e.\ }}
\def\vs{{\it versus\ }}

\def\hpc{$h^{-1}$Mpc }

\def\z{$\,${\it z}$\,$}

\begin{document}
   \title{The VIMOS VLT Deep Survey: the group catalogue}


   \author{O. Cucciati\inst{1,2}\thanks{\email{olga.cucciati@oamp.fr}}
           \and C. Marinoni \inst{3}
	   \and A. Iovino \inst{1}
	   \and S. Bardelli  \inst{4}
	   \and C. Adami \inst{2}
	   \and A. Mazure \inst{2}
	   \and M. Scodeggio \inst{5}
	   \and D. Maccagni \inst{5}
	   \and S. Temporin \inst{6}
	   \and E. Zucca    \inst{4}
	   \and G. De Lucia    \inst{7}
	   \and J. Blaizot    \inst{8}
	   \and B. Garilli \inst{5}
	   \and B. Meneux \inst{9,10}
	   \and G. Zamorani \inst{4} 
	   \and O. Le F\`evre \inst{2}
	   \and A. Cappi    \inst{4}
	   \and L. Guzzo \inst{1}
	   \and D. Bottini \inst{5}
	   \and V. Le Brun \inst{2}
	   \and L. Tresse \inst{2}
	   \and G. Vettolani \inst{11}
	   \and A. Zanichelli \inst{11}
	   \and S. Arnouts \inst{12,2}
	   \and M. Bolzonella  \inst{4} 
	   \and S. Charlot \inst{13,14}
	   \and P. Ciliegi    \inst{4}  
	   \and T. Contini \inst{15}
	   \and S. Foucaud \inst{16}
	   \and P. Franzetti \inst{5}
	   \and I. Gavignaud \inst{17}
	   \and O. Ilbert \inst{2}
	   \and F. Lamareille \inst{15}
	   \and H.J. McCracken \inst{14,18}
	   \and B. Marano     \inst{19}  
	   \and R. Merighi   \inst{4} 
	   \and S. Paltani \inst{20,21}
	   \and R. Pell\`o \inst{15}
	   \and A. Pollo \inst{22,23}
	   \and L. Pozzetti    \inst{4} 
	   \and D. Vergani \inst{5}
	   \and E. P\'erez-Montero \inst{24}
        }

   \offprints{O. Cucciati}

   \institute{
  INAF-Osservatorio Astronomico di Brera, Via Brera 28, I-20021, Milan, Italy 
  \and
  Laboratoire d'Astrophysique de Marseille, Universit\'e de Provence, CNRS, 
  38 rue Frederic Joliot-Curie, 
  F-13388 Marseille Cedex 13, France
  \and 
  Centre de Physique Th\'eorique, UMR 6207 CNRS-Universit\'e de Provence, F-13288, Marseille, France 
  \and 
  INAF-Osservatorio Astronomico di Bologna, Via Ranzani 1, I-40127, Bologna, Italy 
  \and 
  IASF-INAF, Via Bassini 15, I-20133, Milano, Italy
  \and
  Institute of Astro- and Particle Physics, Leopold-Franzens-University 
  Innsbruck, Technikerstra{\ss}e 25, A-6020 Innsbruck, Austria
  \and
  INAF - Osservatorio Astronomico di Trieste, via Tiepolo 11, I-34143, Trieste, Italy
  \and
  Universit\'e de Lyon, Lyon, F-69003, France ; Universit\'e Lyon 1,
  Observatoire de Lyon, 9 avenue Charles Andr\'e, Saint-Genis Laval,  
  F-69230, France ; CNRS, UMR 5574, Centre de Recherche Astrophysique de Lyon ;  
  Ecole Normale Sup\'erieure de Lyon, Lyon, F-69007, France.
  \and
  Max Planck Institut f\"ur Extraterrestrische Physik (MPE), Giessenbachstrasse 1,
  D-85748 Garching bei M\"unchen,Germany
  \and
  Universit\"atssternwarte M\"unchen, Scheinerstrasse 1, D-81679 M\"unchen, Germany
  \and
  IRA-INAF, Via Gobetti 101, I-40129, Bologna, Italy 
  \and 
  Canada France Hawaii Telescope corporation, Mamalahoa Hwy,  
  Kamuela, HI-96743, USA
  \and
  Max Planck Institut f\"ur Astrophysik, D-85741, Garching, Germany 
  \and 
  Institut d'Astrophysique de Paris, UMR 7095, 98 bis Bvd Arago, F-75014, Paris, France 
  \and 
  Laboratoire d'Astrophysique de Toulouse-Tarbes, Universit\'e de Toulouse, 
  CNRS, 14 av. E. Belin, F-31400 France
  \and 
  School of Physics \& Astronomy, University of Nottingham,
  University Park, Nottingham, NG72RD, UK
  \and
  Astrophysical Institute Potsdam, An der Sternwarte 16, D-14482, Potsdam, Germany 
  \and 
  Observatoire de Paris, LERMA, 61 Avenue de l'Observatoire, F-75014, Paris, France 
  \and 
  Universit\`a di Bologna, Dipartimento di Astronomia, Via Ranzani 1, I-40127, Bologna, Italy 
  \and 
  Integral Science Data Centre, ch. d'\'Ecogia 16, CH-1290, Versoix, Switzerland 
  \and
  Geneva Observatory, ch. des Maillettes 51, CH-1290, Sauverny, Switzerland 
  \and
  The Andrzej Soltan Institute for Nuclear Studies, ul. Hoza 69, 00-681 
  Warszawa, Poland
  \and
  Astronomical Observatory of the Jagiellonian University, ul Orla 171, PL-30-244, Krak{\'o}w, Poland 
  \and
  Instituto de Astrof\'\i sica de Andaluc\'\i a - CSIC.
  Apdo. de correos 3004. 18080. Granada (Spain) 	  }

   \date{Received -; accepted -}

 
  \abstract
  {}
   {We present a homogeneous and complete catalogue of optical groups identified in the 
   purely flux limited ($17.5 \leq I_{AB} \leq 24.0$) VIMOS-VLT Deep redshift Survey
   (VVDS).}
   {We use mock catalogues extracted from the MILLENNIUM simulation, 
   to correct for potential systematics that might affect the overall distribution  
   as well as the individual properties of the identified systems.
   Simulated samples allow us to forecast the number and properties of groups that can 
   be potentially  found in a survey with VVDS-like selection functions. We use them 
   to correct for the expected incompleteness and also to asses how well galaxy redshifts 
   trace the line-of-sight  velocity dispersion of the underlying mass overdensity.
   In particular, we train on these mock 
   catalogues the adopted group-finding technique \ie the
   Voronoi-Delaunay Method (VDM). The goal is to fine-tune its    
   free parameters, recover in a robust and unbiased way
   the redshift and velocity dispersion distributions of groups ($n(z)$ and
   $n(\sigma)$ respectively) and maximize, at the same time,  
   the level of completeness and purity of the group catalogue.}
   {We identify 318 VVDS groups  with at least 2 members in the range 
   $0.2 \leq z \leq 1.0$, among which 144 (/30) with at least 3 (/5) members. 
    The sample has an overall completeness of $\sim60$\% and purity of $\sim50$\%.  
    Nearly $45$\% of the groups with at least 3 members are still recovered  if we run 
    the algorithm with the particular parameter set which maximizes the purity 
    ($\sim75$\%) of the resulting catalogue.   
   We exploit the group sample to explore the redshift evolution of the fraction $f_b$ of
   blue galaxies ($U-B \leq 1$) in the redshift range $0.2\leq z \leq1$. 
   We find that the fraction of blue galaxies is significantly 
   lower in groups than in the global population (\ie in the whole ensemble of
   galaxies irrespectively of their environment). Both of these quantities increase
   with redshift, with the fraction of blue galaxies in groups showing a marginally     
  significant steeper increase. We also investigate the dependence of    
   $f_b$ on group richness: not only we confirm that, at any redshift, the blue 
   fraction decreases
   in systems with increasing richness, but we extend towards fainter luminosities 
   the magnitude range over which  this  result holds.}
   {}


   \maketitle
   
%

%
%

\section{Introduction}\label{intro}

\indent Galaxy groups and clusters are the
largest and most massive gravitationally bound systems in the
universe.  Because of this, they are very useful cosmological
probes. For example, the evolution of their abundance or baryon
fraction give insights into the value of fundamental cosmological
parameters (\eg \citealp{borgani1999,newman2002, allen2002, ettori2003,
zhang2006_baryon,ettori2009}), their mass and luminosity functions fix
the amplitude of the power spectrum at cluster scales
(\eg \citealp{rosati2002,finoguenov2010}), while their optical mass-to-light
ratio allows to constrain the matter density parameter $\Omega_m$
(\eg \citealp{girardi2000, marinoni2002_ML, sheldon2009_ML}).  Groups
and clusters are also ideal laboratories for astrophysical
studies. Several interesting physical processes are indeed triggered
on scales characterized by such extreme density conditions. Their
analysis is crucial in particular to understand the effects of local
environment on galaxy formation and evolution
(e.g. \citealp{oemler1974, dressler1980, postman1984, dressler1997,
garilli1999, treu2003, poggianti2006}).

\subsection{The detection of galaxy groups and clusters} 

A whole arsenal of
algorithms allows to identify and reconstruct galaxy systems.  They
range from the very first pioneer methods based on visual
identification on photometric plates
\citep{abell1958,zwicky1968_clusters} to more recent techniques which
exploit various physical properties of the systems as a guide for
identification. For example, the thermal bremsstrahlung emission from
the hot intracluster gas trapped inside the cluster gravitational
potential allows to spot them by means of X-ray band observations. On
the opposite side of the spectrum, in the centimetre regime, cluster
detection is made possible thanks to the Sunyaev-Zeldovich effect (SZE,
\citealp{sunyaev_zeldovich1972,sunyaev_zeldovich1980}).  Indeed, the
hot intracluster gas, by inverse-compton scattering the photons of the
Cosmic Microwave Background (CMB), leaves a characteristic imprint in
the CMB spectrum which can be exploited as a useful signature for
identification.  A cluster potential well can also be detected through
strong gravitational lensing or the cosmic shear induced by weak
gravitational lensing
\citep{kneib2003,gavazzi2009,limousin2009,richard2010,
limousin2010,morandi2010}. Clusters identification can be based also
on the properties of the member galaxies. It has been observed that
cluster cores host typically red galaxies, among which there are the
brightest cluster galaxies (BCG). Thus, a cluster center can be
identified as a {\it R.A.-dec} concentration of galaxies with typical
red colours (see for example the Red-Sequence Cluster Survey,
\citealp{gladders2000}, the first cluster survey based on this
method), in some case adding also the constraint of a high luminosity
(\eg the {\it maxBCG} method, \citealp{hansen2005},
\citealp{koester2007}).

An orthogonal
approach, based on geometrical algorithms, consists in identifying
systems from the 3D spatial distribution properties of their
members. These algorithms vary from the earlier hierarchical method
(\citealp{materne1978}, \citealp{tully1980h}) and the widely used
`friend of friend' (FOF) method \citep{huchra1982h}, to the more
recent 3D adaptive matched filter method \citep{kepner1999}, the `C4'
method \citep{miller2005} and the Voronoi-Delaunay Method
\citep[VDM,][]{marinoni2002}.  Finally, group-finding algorithms have also been
developed which exploit information extracted from photometric
redshifts (\eg \citealp{adami2005_CDFS,mazure2007}).

The availability of several identification protocols is not only
useful in order to confirm clusters detection by an a-posteriori
cross-correlation of various independent catalogues, but it is also
crucial for spotting eventual systematics which might affect
individual detection techniques.  For example, it has been shown by
the first joint X-ray/optical survey \citep{donahue2002_Xrayopt} that
only $\sim$ 20\% of the optically selected clusters were also
identified in X-rays, while $\sim$60\% of the X-ray clusters were
detected in the optical sample. Understanding the possible selection
effects hidden behind the different survey strategies is crucial in
order to interpret this small overlap between the two different
cluster catalogues (see for example \citealp{ledlow2003_Xrayopt,
gilbank2004_Xrayopt}). Moreover, using the RASS-SDSS galaxy cluster
catalogue \citet{popesso2004_I} show that a distinct class of `X-ray
underluminous Abell clusters' does exist, with an X-ray luminosity
$L_X$ which is one order of magnitude fainter than the one expected for
their mass according to the typical $L_X$-mass relation
\citep{popesso2007_V}. This supports the concern of
\cite{donahue2002_Xrayopt} about the possible existence of biases in
catalogues selected in different wavebands.

A major challenge we face is to extend cluster
searches at high redshift.  Indeed, most of the methods described
above suffer from major drawbacks when applied in this regime.  Both
the X-ray apparent surface brightness and the gravitational lensing
cross section of clusters decrease very rapidly with redshift. As a
consequence, only very massive clusters can be detected at high
$z$. On the contrary, the SZE detection efficiency does not depend on
redshift, but large SZ survey are yet to be completed.  For what
concerns cluster detection using the spatial distribution of members,
we emphasize the difference between photometric and spectroscopic
galaxy data sets. Several methods have been proposed to detect
clusters with photometric data, mainly exploiting galaxy colours in
different bands. On the one side, this method has been successfully
used both for surveys (see for example the above-mentioned
Red-Sequence Cluster Survey, \citealp{gladders2000}) and single
detections (\eg the very recent work by \citealp{andreon2008_z19}),
but on the other hand the selection of red galaxies implies the
selection of only the older structures, where galaxies lived enough
time to be affected by the physical processes typical of the group
environment (see for example the discussion in
\citealp{gerke2007_groupsblue}). Moreover, the depth required in
photometric surveys to identify high-$z$ groups and clusters increases
the number of foreground and background galaxies, due to the fact that
objects surface number density is enhanced by the faint flux
limit. This essentially limits the effectiveness of 2D identifications
at high-$z$.  The third dimensions is thus imperative if we want to
disentangle in an efficient way projection effects. Nonetheless, the
uncertainty on the line-of-sight position of galaxies may be a concern
when it is bigger (or even much bigger) than the typical velocity
dispersion of group galaxies, as in the case of
photometric redshifts.

\subsection{This work and the existing groups and clusters samples}

To date, many local, optically selected group
catalogues are available in literature. A review can be found in
\cite{eke2004}, where one of the largest catalogue of galaxy groups
detected in redshift space from the Two Degree Field Galaxy Redshift
Survey (2dFGRS) is presented. Similarly, several group catalogues have
been extracted from the Sloan Digital Sky Survey data (\eg
\citealp{miller2005, berlind2006, weinmann06}). Systematic searches of
groups in redshift space have been undertaken also at intermediate
redshift (\eg within the CNOC2 survey, up to redshift $z=0.55$,
\citealp{carlberg2001}). The compilation of optically selected and
complete samples of groups up to $z\sim 1$ and beyond has been made
possible only recently thanks to the completion of large and deep
spectroscopic surveys, such as the DEEP2 Galaxy Redshift Survey
\citep{Davis2003}, the VIMOS-VLT Deep Survey \citep{lefevre2005a}, and
the zCOSMOS survey \citep{lilly2007_zcosmos, lilly2009_zcosmos}.

\cite{gerke2005_groups} present the first DEEP2 group catalogue: it contains
899 groups with two or more members identified in the redshift range
$0.7\leq z \leq 1.4$ with the VDM method. The DEEP2 sample
reaches a limiting magnitude of $R_{AB}=24.1$, and its galaxies are
pre-selected in colour before being targeted for spectroscopic
observations, in order to reduce the number of galaxies at $z \lesssim
0.7$. The first zCOSMOS group catalogue \citep{knobel2009_groups} 
comprises   $\sim800$ groups with at least 2 members, covering the
redshift range $0.1\leq z \leq 1.0$. The parent galaxy sample is purely
flux limited ($15 \leq I_{AB} \leq 22.5$), and  groups are detected
with the FOF method, combined with the VDM.

In this work, we make use of the VIMOS-VLT Deep
Survey (VVDS, \citealp{lefevre2005a}) to compile an homogeneous
optically-selected group catalogue in the redshift range
($0.2<z<1.0$). We run the VDM code on a sample containing  more than 
6000 flux limited galaxies ($17.5 \leq I_{AB} \leq 24.0$)
for which reliable spectroscopic redshifts have been measured.  Particular
attention has been devoted to optimally tune the parameters of the 
group-finding algorithm using VVDS-like mock catalogues.
The selection function of the sample, essentially compensating only for
the flux limited nature of the survey, is simple and mostly insensitive
to possibly uncontrolled bias such as those which might affect
colour selected samples. Moreover, the magnitude depth of the VVDS 
allows us to sample a galaxy population which is fainter in luminosity 
than that currently probed by other flux-limited surveys of the deep universe.

The paper is organized as follows: in \S \ref{data_and_mocks} the
data sample and the mock catalogues are described.  The reliability of
the virial line of sight velocity dispersion as estimated using galaxies is discussed
in  \S \ref{preliminary_tests}. In \S \ref{the_algorithm} we
review the basics of the VDM group-finding algorithm, while the strategy
followed to fine tune  its parameters is presented in \S
\ref{VDM_optimization}.  In \S
\ref{real_VVDS_catalogue} we describe the properties of the VVDS group catalogue.
The redshift evolution of the $U-B$ colour of group galaxies is analyzed in 
\S \ref{gal_group_properties}. Conclusions are drawn in \S
\ref{Conclusions}.

We frame our analysis in the context of a $\Lambda$ Cold Dark Matter
model ($\Lambda$CDM) specified by the following parameters:
$\Omega_m=0.3$, $\Omega_{\Lambda}=0.7$, $H_0=70$ km s$^{-1}$
Mpc$^{-1}$. Magnitudes are expressed in the AB system.

%
%

\section{DATA SAMPLE and MOCK CATALOGUES}\label{data_and_mocks}
\subsection{The VVDS-02h sample}\label{real_data}

The VIMOS-VLT Deep Survey (VVDS) is a large spectroscopic survey whose
primary aim is to study galaxy evolution and large scale structure
formation.  The survey detailed strategy and goals are described in
\cite{lefevre2005a}.  VVDS is complemented by ancillary deep
photometric data that have been collected at the CFHT telescope
\citep[BVRI,][]{lefevre2004b,mccracken2003}, at the NTT telescope
\citep[JK,][]{iovino2005,temporin2008} and at the MPI telescope
\citep[U,][]{radovich2004}. Also {\it $u^{*}$, g', r', i', z'}-band
data are available as part of the CFHT Legacy Survey. The full suite
of spectroscopic and photometric data provides a superb database to
address in a wide redshift range many open questions of modern
observational cosmology.

In this paper we make use of the data collected in the VVDS-0226-04
Deep field (from now on ``VVDS-02h field''), where the spectroscopic
observations have targeted objects in the magnitude range $17.5 \leq
I_{AB} \leq 24.0$. In this range the parent photometric sample is
complete and free from surface brightness selection effects
\citep{mccracken2003}, resulting in a deep and purely flux-limited
spectroscopic sample. Spectroscopic observations (the so-called
``first epoch'' data) in the VVDS-02h field were carried out at the
ESO-VLT with the VIsible Multi-Object Spectrograph (VIMOS), a
4-channel imaging spectrograph, each channel (a {\it quadrant})
covering $\sim 7 \times 8$ arcmin${^2}$ for a total field of view (a
{\it pointing}) of $\sim 218$ arcmin${^2}$.  The observations used 1
arcsecond wide slits and the LRRed grism, covering the spectral range
5500\AA $< \lambda <$9400\AA. The resulting effective spectral
resolution is R $\sim$227, while the {\it rms} accuracy of the
redshift measurements is $\sim 275$ km/s \citep{lefevre2005a}.

\begin{figure}
\begin{center}
\includegraphics[width=8cm]{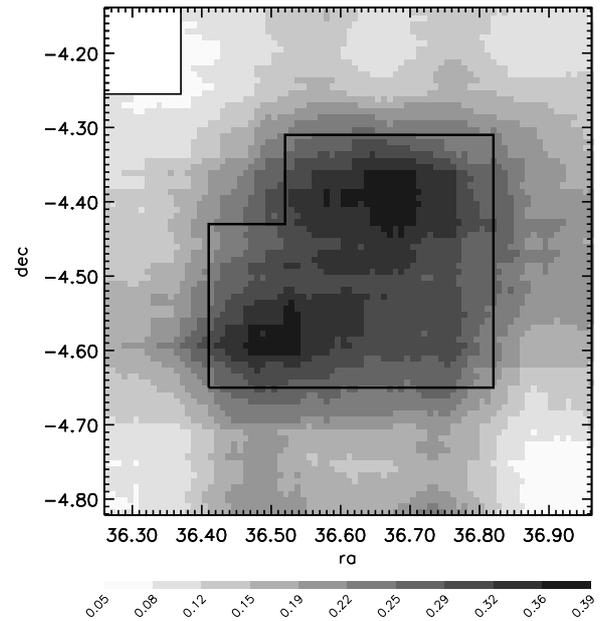}
\caption{Uneven spectrograph coverage in VVDS-02h field. 
The grey-scale from the lightest to the darkest grey indicates the 
sampling rate, with corresponding values shown in the label.
The grid used for the colour-code had steps 
of $30\arcsec$ in Right Ascension and Declination, and for each 
grid position we used squares of size $7\arcmin$  to estimate the sampling rate. 
The central area covered by 4 VIMOS passes 
is highlighted by a solid line.}
\label{VVDS02h_map_passes}
\end{center}
\end{figure}

The VVDS-02h field covers a total sky area of $0.7 \times 0.7$
deg$^2$, targeted by 1, 2 or 4 spectrograph passes. This strategy
produces an uneven target sampling rate as shown in Figure
\ref{VVDS02h_map_passes}. The multiple-pass
strategy assures that there is no serious undersampling of the 
denser regions, at least in the $\sim80$\% of the field covered by two
or more spectrograph passes. It should be noticed that some quadrants
had to be discarded due to their poor quality and not all the regions
of the field covered by the same number of passes have the same
sampling rate.  In average, spectra have
been obtained for a total of 22.8\% of the parent photometric catalogue.  Due
to low signal-to-noise ratio and/or to the absence of useful spectral
features, only $\sim80$\% of these targeted objects yield a redshift,
giving an overall sampling rate of $\sim18$\% ($\sim33$\% considering
only the area covered by 4 passes).

\vspace{1cm}

VVDS-02h field first epoch sample  probes a comoving volume (up to \z=1.5) of
nearly $1.5 \times 10^6$ $h^{-3}$Mpc$^{3}$ in a standard $\Lambda$CDM
cosmology. This volume has transversal dimensions $\sim$ 37$\times $37
\hpc at \z=1.5 and extends over 3060 \hpc in radial direction.

The collected sample contains 6615 galaxies and AGNs with secure redshifts, \ie
redshift determined with  a quality flag=2,3,4,9 (6058 with $0.2
\leq z \leq 1.5$). We refer the reader to \cite{lefevre2005a} for
further details about redshift quality flags. Here we only emphasize that,
comparing spectroscopic redshifts of objects observed
twice by means of independent observations, we conclude that redshifts with 
flag=2(/3/4) are correctly estimated with a likelihood of
81(/97/$>99$)\%. We assigned a flag=9 when in the spectrum there is only a single secure 
spectral feature in emission. Given the  spectral range covered by observations 
and the flux limits of the survey, this emission line is  typically 
[OII]3727\AA ~ or H$\alpha$ (in very rare cases
Ly$\alpha$). Thus flag=9 redshifts have a probability of
being correct of $\sim50$\%, being based on the choice between the two most probable 
emission lines.  
We  double-check
the robustness of the likelihood assigned to flag=2 and flag=9 objects, by contrasting
spectroscopic estimates against photometric
determinations. Photometric redshifts were computed as described in
\cite{ilbert2006zphot}, but now using the more recent T0005 release of CFHTLS data ({\it
$u^{*}$, g', r', i', z'} filters) and the latest data available from
WIRCAM (J, H and K filters, Bielby et al. in preparation). According
to this comparison, flag=2(/9) redshifts are correctly inferred with a 
likelihood of 78(/59)\%, a figure which is in good agreement with the 
independent determination discussed above. 

Note, also, that the conclusions of our work are unaffected by the fact of including or not in our 
analysis flag=9 low quality redshifts. 
As a matter of fact these objects constitute a small fraction ($<3$\%) of the whole sample.
Moreover, the effect of possible biases induced by wrong redshift  estimates are
weakened by the very existence of the galaxy correlation on small scales: if a
galaxy with flag=2 is located nearby (on the sky) to other galaxies
with similar (but more secure) redshift, the likelihood  that it shares
the same redshift actually increases with respect to the probability determined 
on the basis of our analysis.

\subsection{Mock catalogues}\label{mocks_description}

We made extensive use of mock catalogues, both to test the potential
for group searches of the VVDS-02h field data and to tune the
parameters of the group-finding algorithm for optimal detection.

Before introducing any particular group-finding algorithm, one needs
to test which limits in group reconstruction are imposed by the
specific characteristics of the VVDS survey design.  With mock
catalogues mimicking VVDS-02h field we were able to explore which
groups are lost irretrievably due to the survey sparse galaxy
sampling. Furthermore we were able to assess how our measurement of
the line of sight velocity dispersion of group galaxies is degraded by both the
sampling rate and the non negligible VVDS redshift measurement error.
After having explored these limits, we then moved to test and optimize
the group-finding algorithm, within the ranges in redshift and
velocity dispersion where we found that VVDS-02h data allow a reliable
group reconstruction.

Mock catalogues were obtained by applying the semi-analytic
prescriptions of \cite{delucia_blaizot2007} to the dark matter halo
merging trees extracted from the Millennium
Simulation\footnote{http://www.mpa-garching.mpg.de/galform/virgo/millennium/}
\citep{springel2005_MILL}.  The simulation contains $N = 2160^3$
particles of mass $8.6 \times 10^8 h^{-1} M_ {\odot}$ within a
comoving box of size 500 $h^{-1}$Mpc on a side. The cosmological model
is a $\Lambda CDM$ model with $\Omega_m = 0.25$, $\Omega_b = 0.045$,
$h = 0.73$, $\Omega_{\Lambda}= 0.75$, $ n = 1$ and $\sigma_8 = 0.9$.
The positions and velocities of all simulated particles were stored at
63 snapshots, spaced approximately logarithmically from $z=20$ to the
present day. Dark matter halos are identified using a standard
friends-of-friends (FOF) algorithm with a linking length of 0.2 in
units of the mean particle separation.

In this simulation, group galaxies are those in the same FOF halo,
identified with a unique ID.  For each simulated group a wealth of
physical information are available: galaxy membership, virial mass
(computed directly using the simulated particles), virial radius and
virial velocity dispersion (both computed from the virial mass,
through scaling laws and the virial theorem). The virial mass is
computed within the radius where the halo has an overdensity 200 times
the critical density of the simulation.

It is worth noticing that the model used to construct light-cones from
the MILLENNIUM simulation has been shown to be quite successful in
reproducing several basic properties of our real data set. The most
important are the average redshift distribution $n(z)$
\citep{meneux2008_sm} and the global Luminosity Function LF (Zucca et
al., in preparation), that are in good agreement with the real
VVDS-02h $n(z)$ and LF, with the only exception of a slight excess of
galaxies in the $n(z)$ mock samples for $z<0.5$.  It should be
noticed that such a small difference in $n(z)$ does not affect the
completeness and purity values (see Section \ref{success_criteria}) of
our group catalogue, as we specifically tested using separately the
mocks with the most similar and the most different $n(z)$. 
Moreover, in \cite{meneux2008_sm} it is shown that the galaxy
clustering in the MILLENNIUM simulation light cones is consistent with
the one measured using the VVDS-02h sample.

Through the Database built for the Millennium Simulation
\citep{lemson2006_database}, we selected 10 $(1\times 1)$ deg$^2$
independent MILLENNIUM light cones (generated with the code MoMaF,
\citealp{Blaizot2005}), from which we extracted several kinds of
mocks, according to our purposes. First of all, we extracted $(1\times
1)$ deg$^2$ flux limited samples, with the same flux limits as
VVDS-02h sample ($17.5 \leq I_{AB} \leq 24$). These catalogues have
100\% sampling rate, and no redshift measurement error added. We
called these catalogues $M(100,0)$, the first number in brackets
indicating the sampling rate and the second the redshift error. Then
we randomly depopulated these catalogues to obtain subsets with 33\%,
17\% and 10\% sampling rate, mimicking roughly the sampling rate of
the 4 passes, 2 passes and 1 pass areas of the VVDS-02h field. These
catalogues are called $M(33,0)$, $M(17,0)$ and $M(10,0)$ respectively.
With these mock catalogues and taking advantage of the known group
membership we were able to assess how much a group catalogue is
depopulated when sampling rate is lowered to values typical of those
of VVDS-02h field.

As a further step, we added redshift measurement errors to the 33\%
sampling rate mocks, randomly chosen from a Gaussian distribution centered on
0 with $\sigma=275$ km/s. This way we take into account the mean
redshift measurement error of our real data. We called these mock catalogues 
 $M(33,275)$.  With these mock catalogues, we were able to 
test how well we can determine group virial velocity dispersion when
the survey has flux limits, sampling rate and redshift measurement
errors mimicking those of the 4 passes areas of the VVDS-02h field.

As a last step we needed mock catalogues to test how effective is the
group-finding algorithm we adopted in identifying those groups
surviving in a sample like the VVDS-02h one. To test the efficiency of
our algorithm we used 20 ``VVDS-like'' mocks extracted from MILLENNIUM
simulation. These mocks have the same flux limits, geometry, uneven
sampling rate, redshift error measurement as the VVDS-02h sample (see
\citealp{pollo2005} and \citealp{meneux2008_sm} for the preparation of
these mocks).  Subtler effects, like those introduced by slit
positioning bias, have also been included, as the same slit
positioning tool used for VVDS-02h sample has been used, with the same
optimization criteria, to generate the VVDS-like mocks. Moreover,
the areas masked in the real photometric catalogue because of bright
stars and because of a beam of scattered light have also been masked
in the VVDS-like mocks.

For the sake of clarity, we emphasize that whenever we refer to the `FOF'
or `simulated' groups in all the above-mentioned mock catalogues, we
mean the sets of galaxies within the same original FOF halo provided
by the simulation itself, before any depopulating process: we never
ran any FOF algorithm on mocks after extracting $M(100,0)$, $M(33,0)$,
$M(17,0)$ and $M(10,0)$, $M(33,275)$ and ``VVDS-like'' mocks from
simulations.

%
%

\section{Preliminary tests}\label{preliminary_tests}

\subsection{Testing  the effects of VVDS survey strategy 
on groups}\label{group_survived_lowsampling}

In this section we explore how well the group catalogue extracted
from a VVDS-like survey trace the group population of 
an ideal survey which is  purely flux-limited.
In a real flux limited galaxy survey with a sampling rate lower than $100$\%, 
most groups will have a smaller number of members and some will even go undetected.
We want to assess the fraction of groups that
``survive'' as such (\ie with at least 2 members) in a survey with a
sampling rate like the one in VVDS-02h.  To identify groups, both in
the full flux limited and in the various `observed' catalogues, we
used at this phase the identification number of FOF groups in
the Millennium database.  In other words, we consider only the limitations
introduced by the survey strategy, neglecting for the moment further
complications introduced by the incompleteness/failures of the specific group
finding algorithm we used.

\begin{figure}
\centering
\includegraphics[width=8cm]{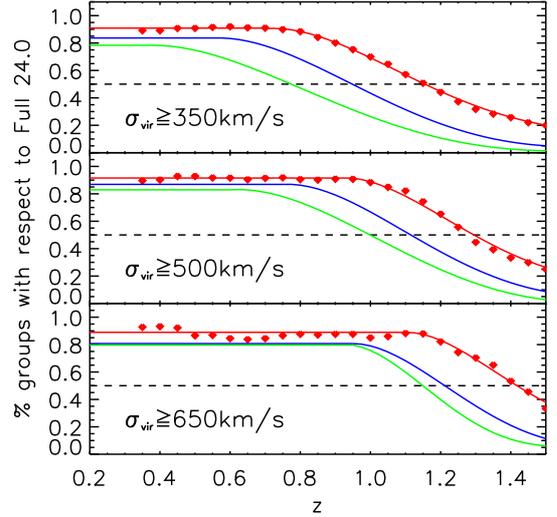}
\caption{The fraction, as a function of redshift, of ``surviving''
groups when the sampling rate is decreased from a purely flux limited
simulated sample $17.5 \leq I_{AB} \leq 24.0$ (\ie with 100\% sampling
rate) to $\sim$33\% (4 passes area, red line), $\sim$17\% (2
passes, blue line) and $\sim$10\% (1 pass area, green 
line).  The fraction has been computed in not-independent running redshift bins
of $\Delta z=0.3$: continuous lines are fits along all the bins, while for 
reference the fractions corresponding to the $M(33,0)$ catalogues are also shown 
for each redshift bin as red diamonds. Different panels show different 
cuts in $\sigma_{vir}$, as quoted in each panel. The horizontal dashed 
line at a fraction value equal to 50\% is for reference.}
\label{slos_potentaility}
\end{figure}

In Figure \ref{slos_potentaility} we plot the fraction of groups in
mock catalogues flux limited at $17.5 \leq I_{AB} \leq 24.0$ that
survived after applying a sampling rate corresponding to 1/2/4 passes
regions (\ie 10\%, 17\% and 33\% respectively) as indicated by
different lines. Practically, we plot the ratio between the number of
groups in $M(10,0)$, $M(17,0)$ and $M(33,0)$ catalogues and the number
of groups in $M(100,0)$ catalogues. This ratio has been computed in
not-independent running redshift bins of $\Delta z=0.3$: continuous
lines are fits along all the bins, while for reference the ratios
corresponding to the $M(33,0)$ catalogues are also shown for each
redshift bin as red diamonds. Note that the number of groups with
$\sigma_{vir} \geq 650$ km/s is quite low, mainly because of the small
field of view, thus the fraction of survived groups at $z \leq 0.8$
fluctuates around a mean value that we use to fit a straight
line. These fluctuations, however, only in the worst cases are as high
as 10\%. This is also true for $M(10,0)$ and $M(17,0)$ catalogues, for
which we do not plot single points not to crowd the figure. The
horizontal dashed line at a fraction value equal to 50\% is shown for
reference. The three panels correspond to different cuts in the virial
line of sight velocity dispersion ($\sigma_{vir}$) quoted in the
mocks, as indicated by the label (from now onwards all velocity
dispersions quoted will be line of sight velocity dispersions).

Figure \ref{slos_potentaility} shows that in 2 and 4 passes areas we
can recover the majority ($\geq 50\%$) of groups down to $\sigma_{vir}
\sim 350$ km/s in the full redshift range below $z=1.0$.  Obviously
going to higher values for $\sigma_{vir}$ allows to extend further the
redshift range. Such lower limit for $\sigma_{vir}$ is in agreement
with the one imposed by the non negligible redshift measurement error
of VVDS survey. As we will see in the next paragraph, for groups with
$\sigma_{vir} \leq 350$ km/s our measurements of velocity dispersion
are quite unreliable.

\subsection{Estimating group virial l.o.s. velocity dispersion}\label{sigma_tex}

A robust determination of the line-of-sight velocity dispersion of galaxies 
in group is essential if we are to infer the group mass in a reliable way.
When group members are sparsely sampled, as it is the case
for VVDS-02h data, the ``gapper method'', originally suggested by
\citet{beers1990}, has proved to be the most robust velocity
dispersion estimator \citep[see also][]{girardi1993_gapper}. This
method measures velocity dispersion exploiting the velocity gaps in
the given velocity distribution of galaxies, using the following
formula:

\begin{equation} \displaystyle
\sigma_G  =
\frac{\sqrt(\pi)}{N(N-1)} \sum_{i=1}^{N-1}i(N-i)(v_{i+1}-v_{i})
\label{sigma_gapper} 
\end{equation}

where the line of sight velocities $v_{i}$ are sorted into ascending
order. \citet{beers1990} show in their Table II that this method
reliably estimates the velocity dispersion with an efficiency $> 90$\%
for groups with $\sim 5-10$ elements, thanks to its robustness in
recovering the dispersion of a distribution also in the more general
case of a contaminated Gaussian distribution. It is important to
emphasize that this range of group members is well suited for the study
we present in this work. On the one hand, we consider the velocity
dispersion reliably measurable only for groups with at least 5
members, and on the other hand the large majority of groups surviving
in ``VVDS-like'' mocks have $\leq 10$ members.

\begin{figure*} 
\centering
\includegraphics[width=5.3cm]{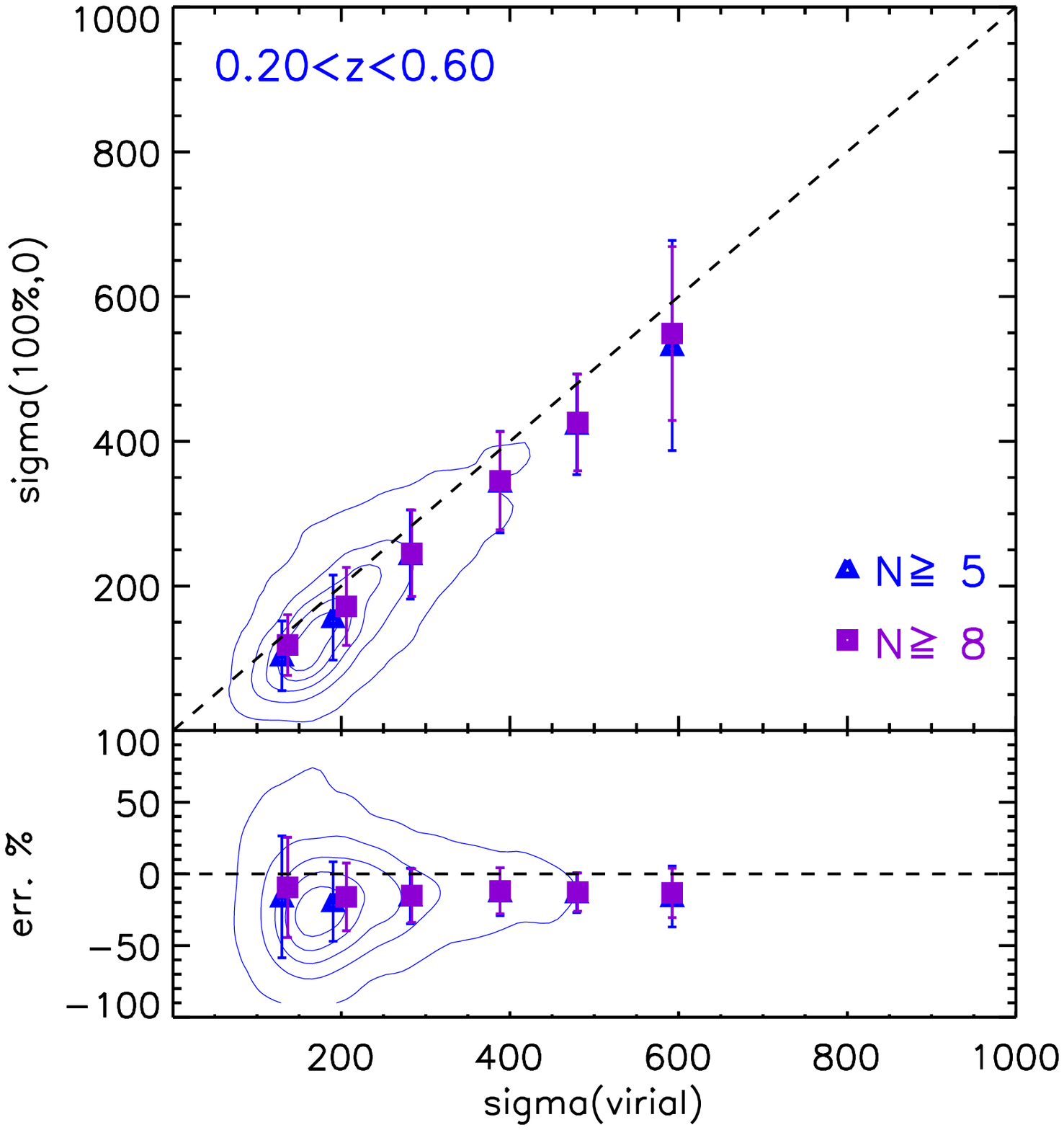}
\includegraphics[width=5.3cm]{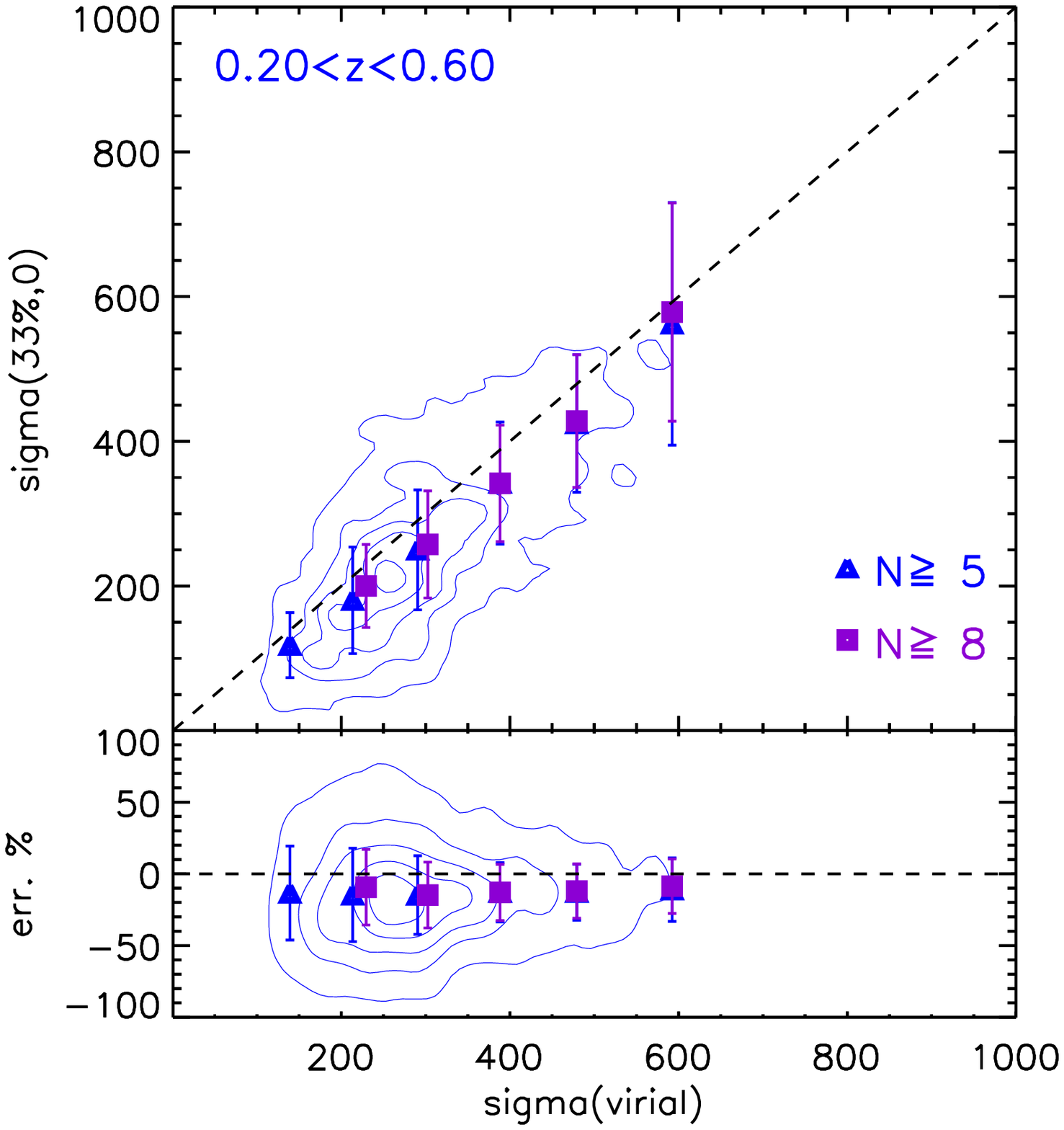}
\includegraphics[width=5.3cm]{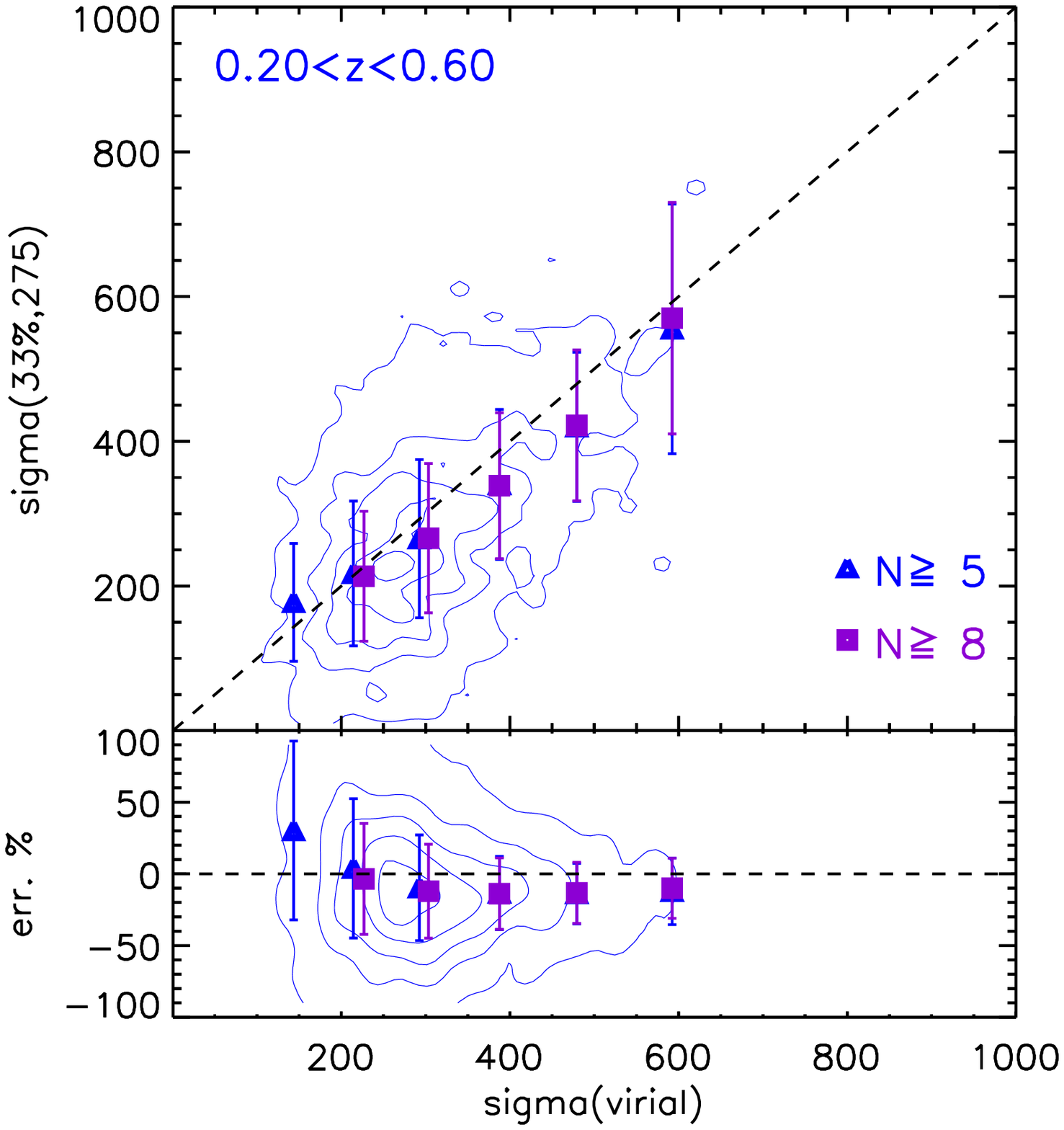}
\includegraphics[width=5.3cm]{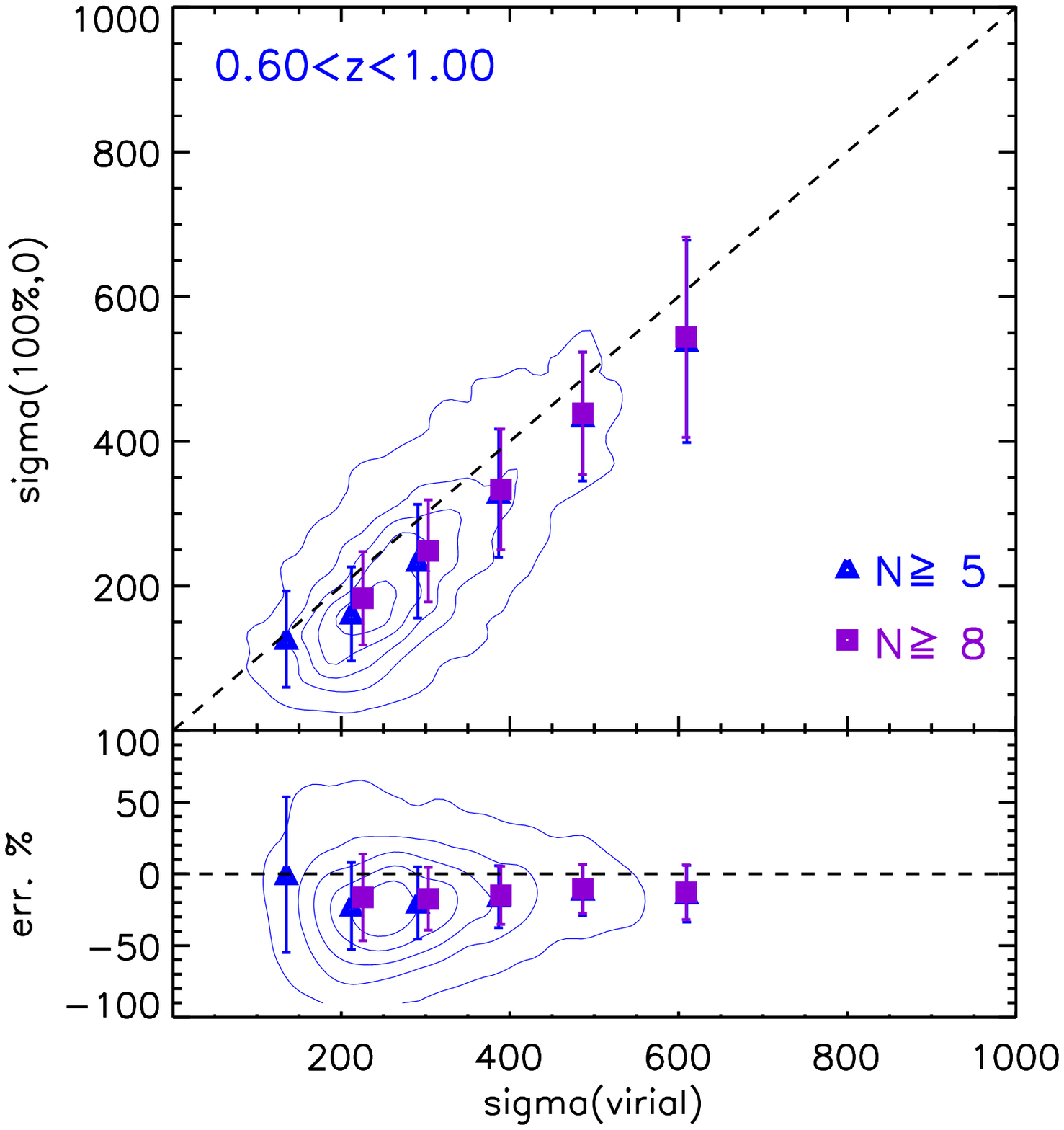}
\includegraphics[width=5.3cm]{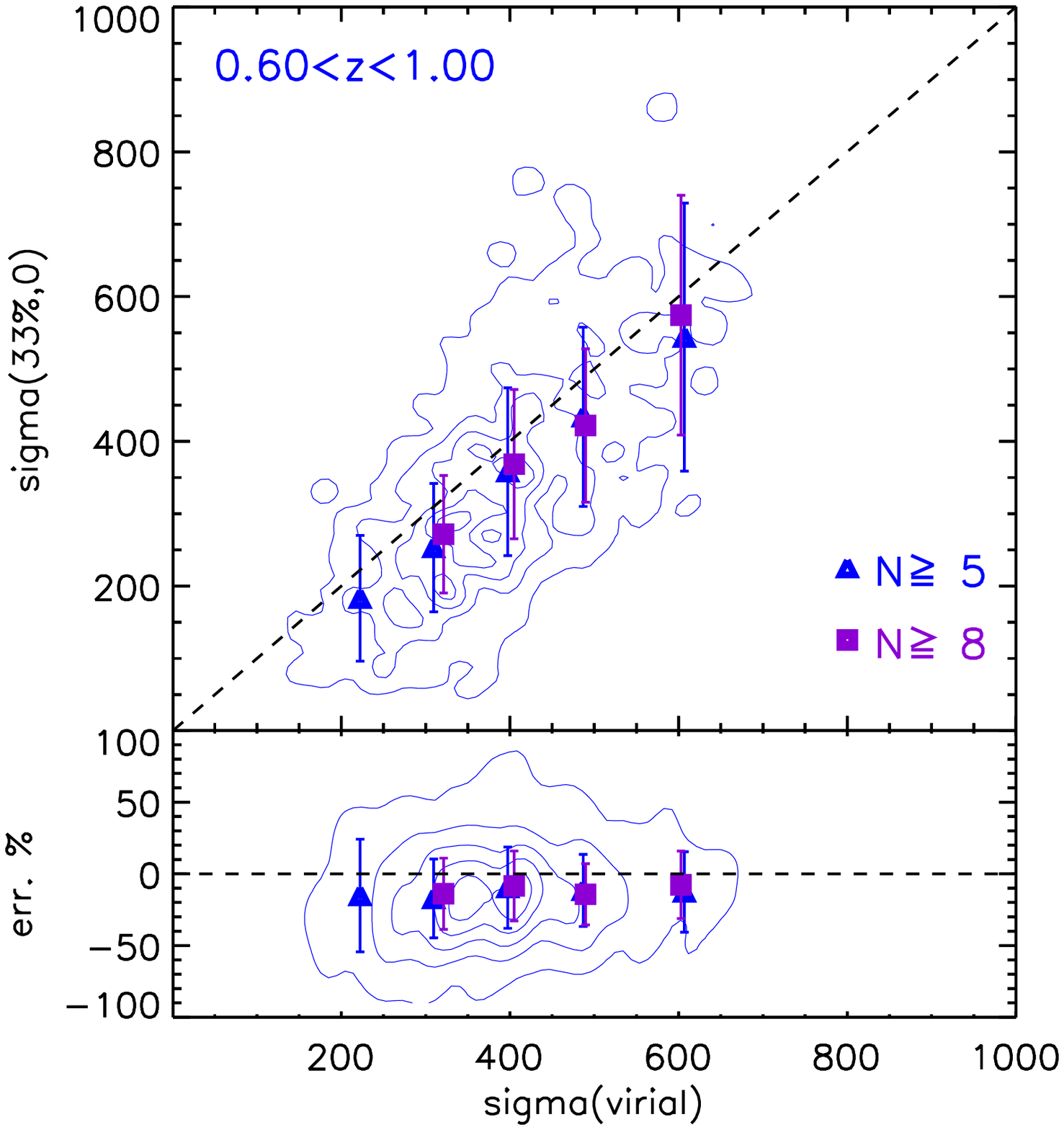}
\includegraphics[width=5.3cm]{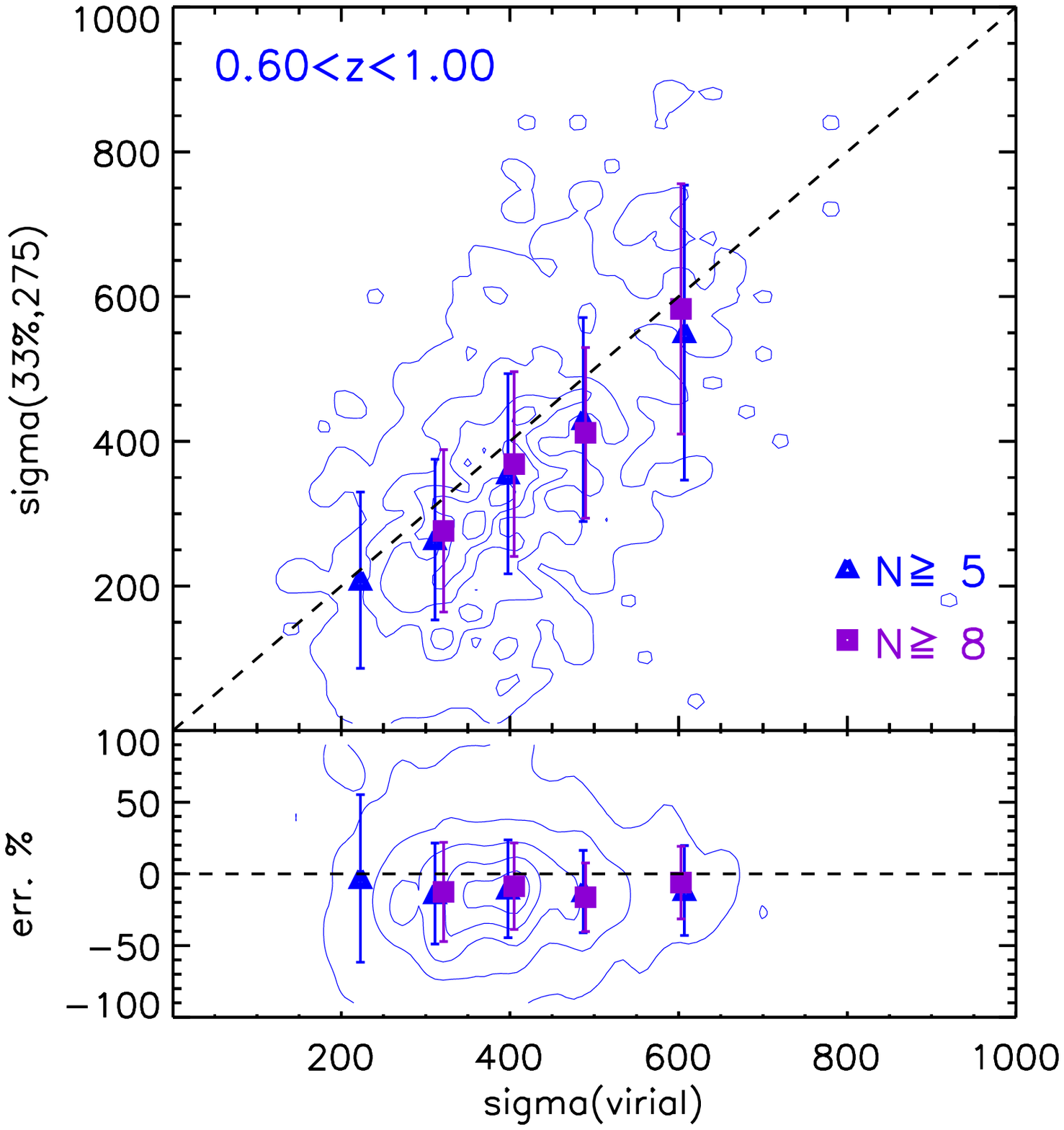}
\caption{Comparison of line of sight velocity dispersion
$\sigma_{meas}$ computed in $M(100,0)$, $M(33,0)$ and $M(33,275)$
mocks ($1^{st}$, $2^{nd}$ and $3^{rd}$ column respectively) with the
virial velocity dispersion $\sigma_{vir}$.  The first row is for the
redshift bin $0.2<z<0.6$, and the second for $0.6<z<1.0$.  In each
plot, the {\it upper panel} shows isodensity contours in the plane
$\sigma_{meas}$ \vs $\sigma_{vir}$ for groups with at least 5 members.
Blue triangles are the median (on $x$ axis) and mean (on $y$ axis)
values of single points grouped in bins of $\sigma_{vir}$, with
vertical error bars being the {\it rms} of mean values. The bins on the
$x$ axis have the following limits:
0.5 - 150 - 250 - 350 - 450 - 550 - 1100 km/s. Purple 
squares are the same as triangles but for groups with at least 8 members.
The {\it lower panel} in each plot shows the percentage
error (with its {\it rms}) when we compare $\sigma_{meas}$ and
$\sigma_{vir}$. Triangles and squares are plotted only when the corresponding 
$\sigma_{vir}$ bin contains at least
15 objects. See text for further details.}
\label{sigma_ratio_vsf_vir}
\end{figure*}

Hereafter, when discussing ``measured'' velocity dispersions
($\sigma_{meas}$) we will refer to velocity dispersions obtained
applying the gapper method to the members of the given group. Of
course, we corrected this velocity dispersion taking into account the
scaling between redshift and velocity. Thus we used:

\begin{equation} \displaystyle
\sigma_{meas}  =
\frac{\sigma_G}{1+z}
\label{sigma_meas_z} 
\end{equation}

where $z$ is the redshift of the group.

In this section we want to test whether our measurement of the line of
sight velocity dispersion $\sigma_{meas}$ is a reliable estimate of
the virial velocity dispersion $\sigma_{vir}$ (as listed in the mock
catalogues). For this comparison we used $M(100,0)$, $M(33,0)$ and
$M(33,275)$.  We called the $\sigma_{meas}$ of these three kinds of
catalogues $\sigma(100,0)$, $\sigma(33,0)$ and $\sigma(33,275)$
respectively. In the case of a non-zero redshift measurement error,
such as in $M(33,275)$ mock catalogues, we took the error itself into
account when computing $\sigma_{meas}$, so that the error ($v_{err}$)
was subtracted in quadrature as follows:

\begin{equation} \displaystyle
\sigma(33,275)^2  = \frac{max[0, \sigma(33,275)_G^2-v_{err}^2]}{(1+z)^2}
\label{sigma_gapper_err} 
\end{equation}

where $\sigma(33,275)_G$ is the velocity dispersion measured in
$M(33,275)$ mocks with Equation \ref{sigma_gapper}, $v_{err}=275$ km/s
and $z$ is the redshift of the group. When $\sigma(33,275)=0$ we
considered the velocity dispersion not measurable given the redshift
error.

Figure \ref{sigma_ratio_vsf_vir} shows the comparison of
$\sigma_{vir}$ with $\sigma(100,0)$, $\sigma(33,0)$ and
$\sigma(33,275)$, respectively in the first, second and third column.
The first row is for the redshift bin $0.2<z<0.6$, and the second for
$0.6<z<1.0$. In each plot, the {\it upper panel} shows isodensity
contours in the plane $\sigma_{meas}$ \vs $\sigma_{vir}$ for groups
with at least 5 members.  Blue triangles are the median (on $x$ axis)
and mean (on $y$ axis) values of single points grouped in bins of
$\sigma_{vir}$, with vertical error bars being the {\it rms} of mean
values. As a reference, purple squares are the same as triangles but
for groups with at least 8 members. The {\it lower panel} in each plot
shows the systematic offset of the relation in the upper panel; the
offset is expressed as a percentage error (with its {\it rms})
computed as follows:

\begin{equation} \displaystyle 
err.\% = \frac{[\sigma_{meas}-\sigma_{vir}]}{\sigma_{vir}} \times 100,
\label{err_perc_vsf} 
\end{equation}

where $\sigma_{meas}$ is $\sigma(100,0)$, $\sigma(33,0)$ and
$\sigma(33,275)$ in the three columns respectively.  Symbols have the
same meaning as in the upper panel.

Results graphically shown in Fig. \ref{sigma_ratio_vsf_vir} can be summarized as
follows:

\begin{itemize}

\item[1)] {\it Effects due to the VVDS-02h flux limit}. The plots in
the first column show that even in the ideal case of
purely flux limited mock catalogues with 100\% sampling rate and
without redshift measurement error, the measured velocity dispersion
$\sigma(100,0)$ systematically underestimates $\sigma_{vir}$. This
systematic offset, shown in the lower part of the plots, is always
below 20\%, with a smaller scatter for increasing $\sigma_{vir}$ and
for the lower redshift bin.  Such an offset can easily be understood
by noticing that in a flux limited survey, even with a 100\% sampling
rate, moving to higher redshifts groups will progressively lose the
fainter members that lie outside the selected flux range.  As a
consequence the measured velocity dispersion underestimates the real
virial velocity dispersion, as the ``surviving'' galaxies are the
brighter, usually found in group cores.

\item[2)] {\it Effects due to the lower sampling rate introduced by
VVDS-02h strategy}. The plots in the second column show that if we
decrease the sampling rate from 100\% to 33\%, our ability in
recovering $\sigma_{vir}$ decreases as well, as expected. The
systematic offset is not significantly worse than in mocks with 100\%
sampling rate, but the scatter around the systematic offset is larger,
especially for low $\sigma_{vir}$.

\item[3)] {\it Effects due to VVDS redshift measurement error}.
Finally, the plots of the third column illustrate the fact that when 
we add 275 km/s of redshift error, low $\sigma_{vir}$ are very difficult to recover,
while, for $\sigma_{vir}> 350$ km/s, the systematic offset and its {\it rms} 
remain below 25\% with a slightly higher scatter for the higher redshift bin.

\end{itemize}

\begin{figure}
\centering
\includegraphics[width=7cm]{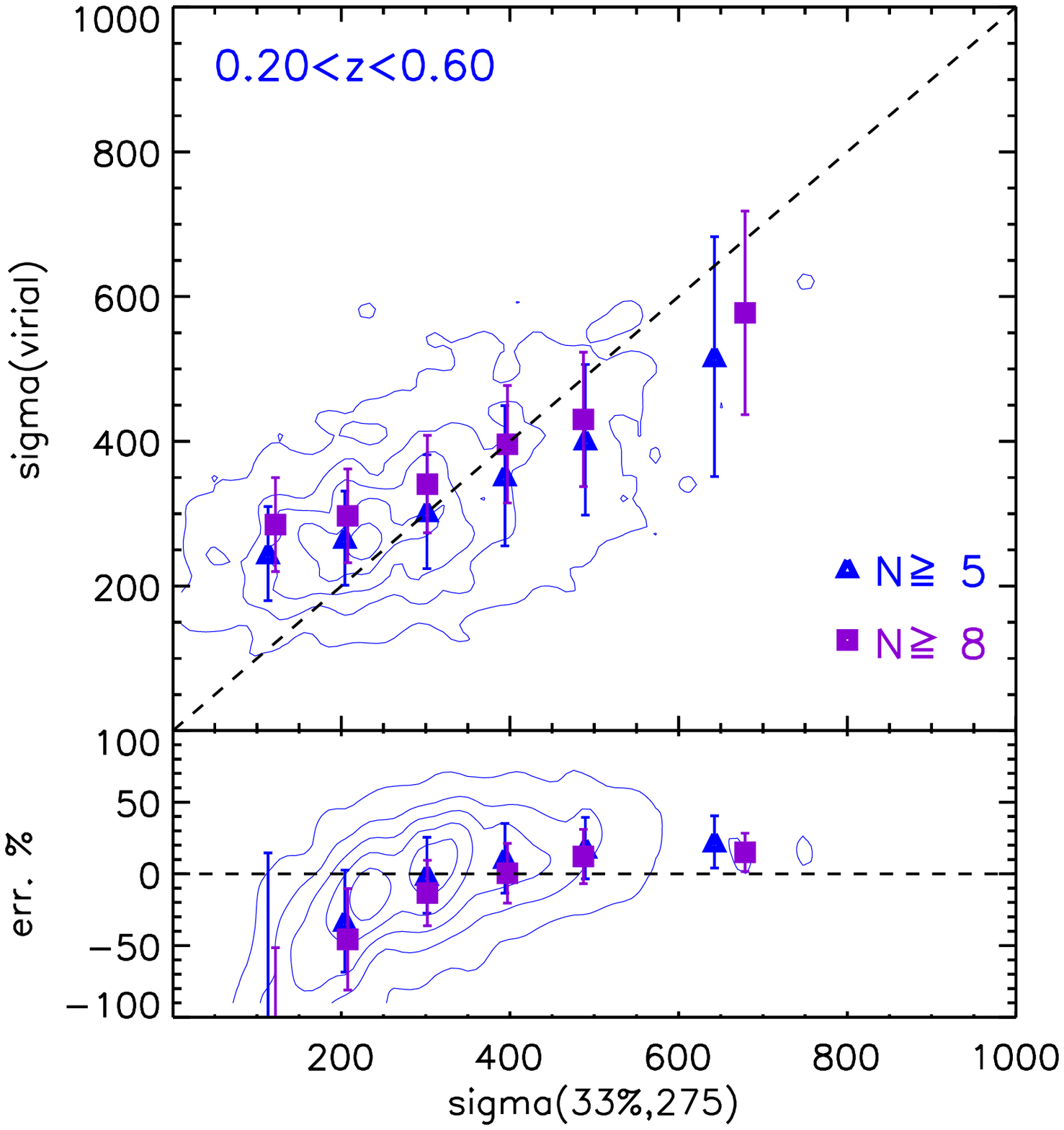}
\includegraphics[width=7cm]{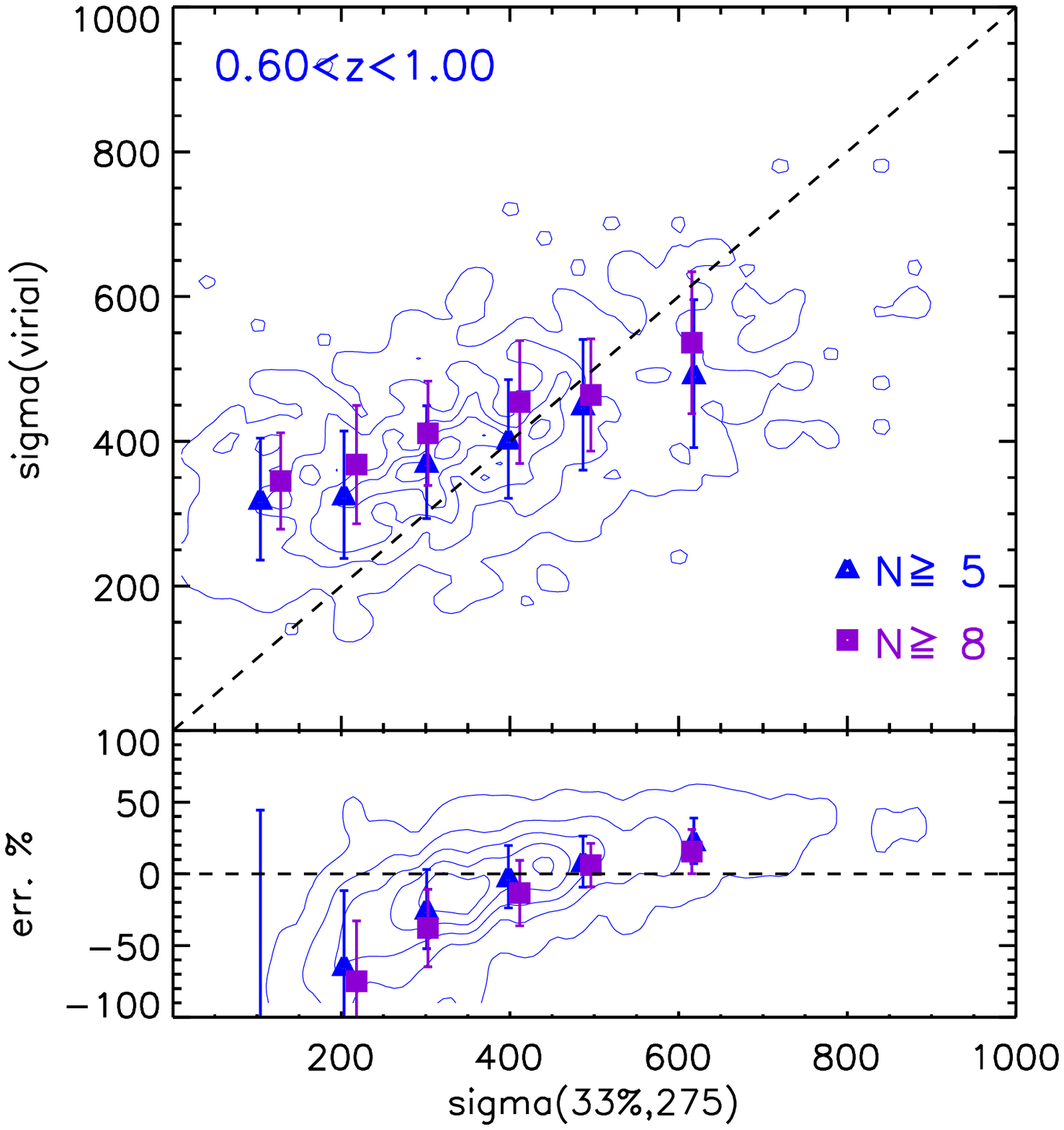}
\caption{As the plots in the last column of Figure 
\ref{sigma_ratio_vsf_vir}, but with   exchanged axes. In this case 
the binning is made according to $\sigma(33,275)$ and not $\sigma_{vir}$.}
\label{sigma_ratio_vsm_vir_33_275}
\end{figure}

Figure \ref{sigma_ratio_vsf_vir} convincingly demonstrates  that 
the estimate of the 
velocity dispersion is robust only in groups with $\sigma_{vir}> 350$ km/s. 
It also serves the following purpose: it forecasts the precision with which the 
measured velocity dispersion traces a specific $\sigma_{vir}$
of the matter particles in the halo once the VVDS sampling rate and 
spectroscopic uncertainties are taken into account.
As a matter of fact, when we analyze the
real VVDS-02h group catalogue,  only the estimate
$\sigma_{meas}$ are available, and nothing is known about $\sigma_{vir}$. We should
therefore ask also the reverse question of how far a given value of $\sigma_{vir}$ is from 
the observed $\sigma_{meas}$. This means that we have to take $\sigma_{meas}$ as
reference when we compute the percentage error. We show the results of this analysis
in Figure \ref{sigma_ratio_vsm_vir_33_275} in which, for
any given bin of $\sigma(33,275)$ we plotted the mean systematic
offset from the real underlying $\sigma_{vir}$, computed as a
percentage error on $\sigma(33,275)$ as follows:

\begin{equation} \displaystyle 
err.\% = \frac{[\sigma(33,275)-\sigma_{vir}]}{\sigma(33,275)} \times 100.
\label{err_perc_vsm_275} 
\end{equation}

Also in this case,  when considering $\sigma(33,275)$
greater than 350 km/s  we are able to recover $\sigma_{vir}$ with an
error of $<30$\% and $<20$\% for groups with $N\geq 5$ and $N\geq 8$
respectively. Therefore, for velocity dispersions above 350 km/s and
despite the relatively large error in redshift measurements 
causes a systematic increase in the estimated $\sigma(33,275)$, we can
still reconstruct a sensible value of $\sigma_{vir}$ in the VVDS-2h 4
passes region. We carried on a similar check for groups in 1 pass and
2 passes regions, and verified that results are qualitatively similar to 
those obtained in the 4 passes area.

Globally, our analysis suggests that we can use VVDS-02h data as a
suitable sample for extracting high-$z$ groups.

%
%

\section{The group-finding algorithm}\label{the_algorithm}

Several geometrical algorithms have been proposed to identify groups and
clusters from the 3-dimensional distribution of galaxies, that is by
optically identifying them within spectroscopic redshift survey (see Section \ref{intro}).

In this work we identified groups using the Voronoi-Delaunay Method \citep{marinoni2002}. 
In short, the VDM combines information from the three-dimensional
Voronoi diagram and its dual, the Delaunay triangulation. The Voronoi diagram
\citep{voronoi1908} is a polyhedral partition of 3D space,  each polyhedron 
surrounding a galaxy and defining 
the unique volume containing all the points
that are closer to that galaxy than to any other galaxy in the sample.
The Delaunay complex \citep{delaunay1934} also  contains proximity information. 
It is defined by the tetrahedra whose vertices are
sets of four galaxies, that have the property that the unique sphere
that circumscribe them does not contain any other galaxy. The center
of the sphere is a vertex of a Voronoi polyhedron, and each face of a
Voronoi polyhedron is the bisector plane of one of the segments that
link galaxies according to the Delaunay complex.

The basics of the VDM code is as follows.
The denser the environment in which a galaxy live, the smaller the Voronoi 
volume which is associated to it. Therefore the Voronoi
partition allows for an immediate identification of the central regions of structures. 
Complementarily, the Delaunay triangulation assigns galaxy members to the identified core. 
Note that a crucial difference between the VDM and other methods is that, 
since it preliminarily identifies group centers, group membership
reconstruction proceed radially outward, from the densest cores towards
the outskirts of the structures.

An advantage offered by the Voronoi-Delaunay method is that it
exploits the natural clustering of the galaxies in the sample. For example 
the dimension of the volume assigned to each galaxy locally depends on the
number density of the objects surrounding the galaxy itself. It is
thus adaptively and unparametrically  rescaled and not predefined on the basis 
of some fixed length parameter. Moreover, galaxies that are Delaunay connected 
to the central cores are processed with cylindrical windows whose dimensions 
are locally scaled  on the basis of physical  relations
observed in simulated (and real) samples of groups and clusters. 
The specific set of VDM parameters is thus designed in such a way to offer the maximum
of flexibility in selecting groups according to the {a-priori} physical information
we have on their structure. As a consequence,  a fine-tuned VDM
algorithm has been proven to be very efficient in reconstructing intrinsic characteristics of 
groups, such as for example the line of sight
velocity dispersion of their members \citep{marinoni2002}.

The Voronoi-Delaunay method was specifically designed to avoid some known drawbacks
characterizing standard group-finding algorithms such as for example the
FOF and the hierarchical methods. These methods are based on
user-specified parameters (the FOF linking length, the `affinity'
threshold in the hierarchical method) that do not depend on the real
distribution of galaxies. One of the negative  consequences is that 
spatially closed but unrelated
structures are often merged together in a single system.   
Moreover, some dynamical properties of
clusters are very sensitive to the adopted group-finding algorithm:
for example, the velocity dispersion of groups identified by the FOF
algorithm is found to be systematically higher (by nearly $30$\%) than that of groups 
found by the hierarchical algorithm, even when both algorithms are optimized
on the same galaxy sample \citep{giuricin2001}. 

We will now briefly summarize how the VDM works, although detailed 
descriptions can be found in \cite{marinoni2002} and
also in \cite{gerke2005_groups} (from which we adopted some technical
improvements).

At first, the algorithm computes the Voronoi-Delaunay mesh following
the prescriptions in \cite{barber1996quickhull} and
\cite{mirtich1996}. It then searches for groups with a 3-step procedure. 
At each step, new group members are identified by means of a
cylindrical window (of radius $R$ and half-length $L$) which is  used to 
scan Delaunay connected galaxies and to decide whether or not they are 
cluster members. Phase I concerns the 3-D identification of
group seeds. In Phase II the algorithm determines group central
richness, and finally in Phase III an adaptive scaling based on
N-$\sigma$ relation is used to rescale cylinder dimensions depending
on group richness found in Phase II. A detailed explanation of each of
these three steps follows in sections \ref{phaseI}, \ref{phaseII} and
\ref{phaseIII}.

The radius and the half-length of the cylinders of Phase I
($\mathcal{R}_{I}$ and $\mathcal{L}_{I}$) and of Phase II
($\mathcal{R}_{II}$ and $\mathcal{L}_{II}$), together with $r$ and
$l$, the scaling factor used to determine respectively the radius and
the half-length of the cylinder of Phase III ($\mathcal{R}_{III}$ and
$\mathcal{L}_{III}$), are free parameters of the algorithm. They
have to be optimized using physical information about clusters.

The choice of a cylindrical shape for the search window is physically motivated
by the fact that the gravitational field of galaxy overdensities  induces peculiar
velocities whose effect is to make the galaxy
distribution to appear elongated in the redshift direction. The only
way to take this into account is by using a search window with a radial
extension much longer than the transversal dimension, in order not to
miss group members. Note that we use a cylindrical window also in Phase I,
while in \cite{marinoni2002} Phase I search window had a spherical
shape. The original choice of a spherical window for the first phase
was physically motivated by the fact that galaxies 
residing in the highest density peaks, \ie the central cores of groups and 
clusters, are expected to have smaller
peculiar velocities.  However, we
verified that for less rich systems, \ie loose groups as those we expect to recover
in the VVDS sample, the best choice is a cylindrical
window. In particular, the survey 
quite large redshift measurement error together with
the sparse sampling rate motivate our new choice.

As we want the length of search cylinders to correspond roughly to the
peculiar velocity of the galaxies in the group, we have to consider
that the mapping between redshift interval and peculiar velocity
changes with redshift, and thus, following \cite{gerke2005_groups},
our algorithm automatically rescales cylinder lengths $\mathcal{L}(z)$
as a function of $z$, using the equation:

\begin{equation} 
\mathcal{L}(z) = [s(z)/s(z_0)]\mathcal{L}(z_0),  
\end{equation}

where $z_0$ is a reference redshift (see Section \ref{phaseII} for
details) and

\begin{equation}
s(z) = \frac{1+z}{\sqrt{\Omega_M (1+z)^3 + \Omega_{\Lambda} }}.
\end{equation}

This scaling as a function of redshift is applied to all
$\mathcal{L}_{I}$, $\mathcal{L}_{II}$ and $\mathcal{L}_{III}$.

\subsection{Phase I.}\label{phaseI}

In Phase I galaxies are at first ranked according to the increasing
size of their Voronoi volume. 
A cylinder with radius $\mathcal{R}_{I}$ and half length
$\mathcal{L}_{I}$ is then centered on the galaxy with smallest 
Voronoi volume. All galaxies
inside the cylinder and Delaunay-connected with the central galaxy are
considered group members and called {\it first-order Delaunay
neighbours}.  The central galaxy and its first-order Delaunay
neighbours are considered a group seed. In the case there are no other
galaxies in the cylinder, the central galaxy is 
rejected as potential seed. Thus, the choice of
$\mathcal{R}_{I}$ and $\mathcal{L}_{I}$ determines the final number of
identified groups.  At the end of this Phase the barycenter of the
seed is computed, using the positions of the central galaxy and its
first-order Delaunay neighbours. 

Then the algorithm processes the full 
sequence of Phases for the found seed. 
After Phase II and Phase III are completed, the whole procedure is reiterated 
by selecting from the sorted list the first galaxy not yet assigned to a group.

\subsection{Phase II.}\label{phaseII}

In the second phase a different cylindrical window with radius
$\mathcal{R}_{II}$ and half length $\mathcal{L}_{II}$ is centered on
the barycenter determined in Phase I, and it is used to determine the
central richness of the group.  All galaxies that fall in the Phase II
cylinder and are connected to the first-order Delaunay neighbours are
called {\it second-order Delaunay neighbours}, and are considered
further group members.  The total number of group members after this
phase (the central galaxy plus first- and second-order neighbours) is
considered as the central richness $N_{II}$ of the group.

A reliable estimate of $N_{II}$ is important as it controls the
adaptive search window used in Phase III (see below). From  one hand,
the fact of considering only Delaunay-connected galaxies minimizes the
inclusion of interlopers in $N_{II}$. On the other hand, in a flux
limited survey like VVDS, the $N_{II}$ distribution varies as a
function of redshift, because of the variation of the luminosity limit
with redshift. To ensure a uniform group population, $N_{II}$ has to
be corrected as a function of $z$:

\begin{equation}
N_{II}^{corr}(z) = N_{II} \frac{\langle \nu(z_0)\rangle}{\langle \nu(z)\rangle}
\end{equation}

Here $z_0$ is the redshift zero-point considered as reference, and
$\langle \nu(z)\rangle$ is the comoving number density, that we
calculated by smoothing the redshift distribution of the galaxy
sample, and then dividing it by the differential comoving volume
element at the considered redshift.  In \cite{gerke2005_groups}, $z_0$
is the lower limit of the DEEP2 galaxy redshift distribution $n(z)$,
\ie $z_0=0.7$. In the case of VVDS-02h sample, the lower limit in
$n(z)$ is $z=0.2$, but at this redshift the volume covered by the VVDS-02h 
is small. Because of this,  $\langle \nu(z=0.2)\rangle$ can  be poorly
constrained. Moreover, $\langle
\nu(z)\rangle$ decreases very rapidly from $z=0.2$ up to $z=1.0$.  
Thus we chose $z_0=0.7$ as a compromise between
high statistics (it is roughly the peak of our $n(z)$ distribution)
and not yet so large survey volume.

At the end of Phase II the barycenter position is readjusted using all $N_{II}$
members.

\subsection{Phase III.}\label{phaseIII}

Finally, in Phase III the algorithm reconstructs the full set of group
members, using a new search window which is centered on the group barycenter
determined at the end of Phase II and with dimensions determined
according to the following basic scaling relations.

Assuming that groups are 
singular isothermal spheres,  at any given distance
$r$ from the center the mass density distribution is
related to the velocity dispersion through the equation $\rho(r) =
\sigma^2/(2\pi G r^2)$ \citep{binney_tremaine1988}.
Since  $M(r)=4\pi r^3
\rho(r)/3$, and by defining $r_{vir}$ as the radius of a spherical
volume within which the mean density is $\Delta_c$ times the critical
density at the considered redshift, we find that $M_{vir} \propto
\sigma^3$, where $M_{vir}=M(r_{vir})$ is the virial mass. 
The virial theorem implies that $M_{vir} \propto
\sigma^3 \propto R^3$. therefore,  exploiting the correlation
between velocity dispersion and central richness, which has been
confirmed from loose groups up to massive clusters \citep[for example, see
][]{bahcall1981_Msigma}, we end up with the following chain of relations
$N_{II}^{corr} \propto M_{vir} \propto \sigma^3 \propto R^3$.

Accordingly, we let the
central richness $N_{II}^{corr}$ of each group to control both the
radius and the length of the cylindrical search window:

\begin{itemize}
\item[-] $\mathcal{R}_{III} = r(N_{II}^{corr})^{1/3}$; 
\item[-] $\mathcal{L}_{III} = l(N_{II}^{corr})^{1/3}$. 
\end{itemize}

Here, $r$ and $l$ are normalization parameters to be optimized
using simulations.  Note that  
the adaptive search window of Phase III  will differ from group to group and
that all galaxies enclosed within the cylinder are
considered as further group members, irrespectively of the order of their Delaunay
connections. From now on, we will call richness $N$ the final number
of members assigned to each group at the end of Phase III.

%
%

\section{Optimizing the group-finding algorithm}\label{VDM_optimization}

\subsection{Success criteria}\label{success_criteria}

In this section we detail the optimization strategy devised to
reconstruct groups in the most reliable and unbiased way. To this
purpose, we used VVDS-like mock catalogues. We applied the VDM
algorithm to them, and compared the groups found by the algorithm with
the groups present in the mocks identified by the same FOF
identification number (see Subsection \ref{mocks_description}). From
now on, we will refer to the FOF groups in the mocks as ``fiducial''
groups, while groups reconstructed by our algorithm will be called
``reconstructed'' groups, or simply ``VDM'' groups.

There are two levels of success we are interested in: 1) success in
finding groups, \ie to establish the level of contamination by
interlopers and fake groups, the percentage of missed galaxies and
missed groups and other statistics of this kind; 2) success in
reproducing group properties, \ie to accurately measure group
properties on a group-by-group basis, and also to reproduce their
statistical distribution as accurately as possible.

To test VDM algorithm success in finding the fiducial groups present
in the VVDS-like mocks, we used the following quality estimators (see
also \citealp{marinoni2002} and \citealp{gerke2005_groups} for more
details):

\begin{itemize}

\item[-] \textit{galaxy success rate $S_{gal}$}: fraction of galaxies
belonging to fiducial groups that are identified members of
reconstructed groups;

\item[-] \textit{interlopers fraction $f_{I}$}: fraction of galaxies
identified by the algorithm as members of reconstructed groups but
that are interlopers;

\item[-] \textit{completeness $C$}: fraction of fiducial groups that
are ``successfully'' identified in the reconstructed catalogue;

\item[-] \textit{purity $P$}: fraction of reconstructed groups that
``correspond'' to fiducial groups.

\end{itemize}

Hence, we now need a quantitative measure to determine whether a
fiducial group is detected ``successfully'' and whether a
reconstructed group ``corresponds'' to a fiducial one. To this
purpose, we consider a detection to be successful when more than half
of a fiducial group members are detected in the same VDM group.  On
the contrary, a VDM group corresponds to a fiducial one when more than
half of its members belongs to that fiducial group.  In general, these
two conditions can be verified independently. These general cases are
called {\it one-way matches} from one group catalogue to the other
(from fiducial to VDM or in the opposite direction). But when these
conditions are verified simultaneously involving the same fiducial and
VDM group in both directions, we have a {\it two-ways match}.  So we
can have a {\it one-way completeness ($C_1$)} and a {\it one-way
purity ($P_1$)} when we consider only one-way matches in the fiducial
and in the reconstructed group catalogue respectively, but also a {\it
two-way completeness ($C_2$)} and a {\it two-way purity ($P_2$)} can
be defined when considering two-ways matches.

On the one hand, knowing absolute value of completeness and purity
will help us in optimizing the algorithm, but on the other hand
comparing $C_1$ with $C_2$ and $P_1$ with $P_2$ we can establish the
kind of errors in the reconstructed group catalogue. In fact, when
$C_1 >> C_2$, it means that some fiducial groups are {\it one-way}
successes but not {\it two-ways} matches, and thus these fiducial
groups contain a low fraction of the members of their reconstructed
associated group. This is an indication that VDM algorithm tends to
overmerge separated groups in bigger reconstructed groups, or to
assign to reconstructed groups too many interlopers. On the other
hand, when $P_1 >> P_2$ we know that VDM algorithm is doing the
opposite error, \ie the reconstructed group catalogue is highly
fragmented with respect to the fiducial one.

We decided to use these indicators to search for the best set of
parameter for our algorithm following some guide lines. The basic idea
is to obtain $C_1$ and $C_2$ as high as possible, while keeping $P_1$
and $P_2$ at least above 50\%. Moreover, we would like not to produce
a highly overmerged ($C_1 >> C_2$) or a highly fragmented ($P_1 >>
P_2$) catalogue, and therefore we tried to obtain $C_1 \approx C_2$
and $P_1 \approx P_2$.

\subsection{Algorithm optimization}\label{algorithm_optimization}

We applied the VDM algorithm to 20 VVDS-like mocks, obtaining group
catalogues for the full redshift range $0.2 \leq z \leq 1.5$, but for
the reasons discussed in Section \ref{group_survived_lowsampling} we
implemented the optimization strategy only in the range $0.2 \leq z
\leq 1.0$.

With a trial and error approach we explored the flexibility of the 6
VDM parameters in recovering groups in a robust way.  We allowed each
parameter to vary in a wide range. In particular, 1) we let
$\mathcal{R}_{I}$ and $\mathcal{R}_{II}$ increase up to 1 $h^{-1}$Mpc,
with no lower limit: this because we wanted the radii to span
projected dimensions up to typical central radius of massive clusters
\citep{bahcall1981_Msigma}.  2) We let $r$ span the range $0.4 \leq r
\leq 1.5$, as we want the radius of the last search cylinder to be
equal or larger than small groups typical size ($\sim 0.5$
$h^{-1}$Mpc, see \citealp{borgani1997} and references therein) and
smaller than an Abell radius ($\sim 1.5$ $h^{-1}$Mpc, see
\citealp{borgani1997}). 3) We let $\mathcal{L}_{I}$,
$\mathcal{L}_{II}$ and $l$ vary from 4 to 20 $h^{-1}$Mpc, to include
clusters with velocity dispersion as high as 2000 km/s also at high
redshift ($z \sim 1$). In this case the lower limit is mainly
suggested by our redshift measurement error, that has to be added to
peculiar velocities.  We imposed to $\mathcal{R}_{III}$ and
$\mathcal{L}_{III}$ the same limits as $r$ and $l$. Nevertheless, we
also checked the performances of the algorithm when no limits are applied to
$\mathcal{R}_{III}$ and $\mathcal{L}_{III}$, and we verified that, 
with the exception of very few cases, $\mathcal{R}_{III}$
and $\mathcal{L}_{III}$ `behave well', as we expected as the whole
algorithm is based on physical scales and scaling laws.

Exploring the 6D parameter space, we found the parameter set that kept
$C_1$ and $C_2$ as high as possible and $P_1$ and $P_2$ at least above
50\%, while monitoring also the behavior of the group properties, both
on a group by group basis and on a statistical point of view.  Then we
moved slightly around these chosen values with smaller steps, to
search for a possible finer tuning.

\begin{figure*} 
\centering
\includegraphics[width=7cm]{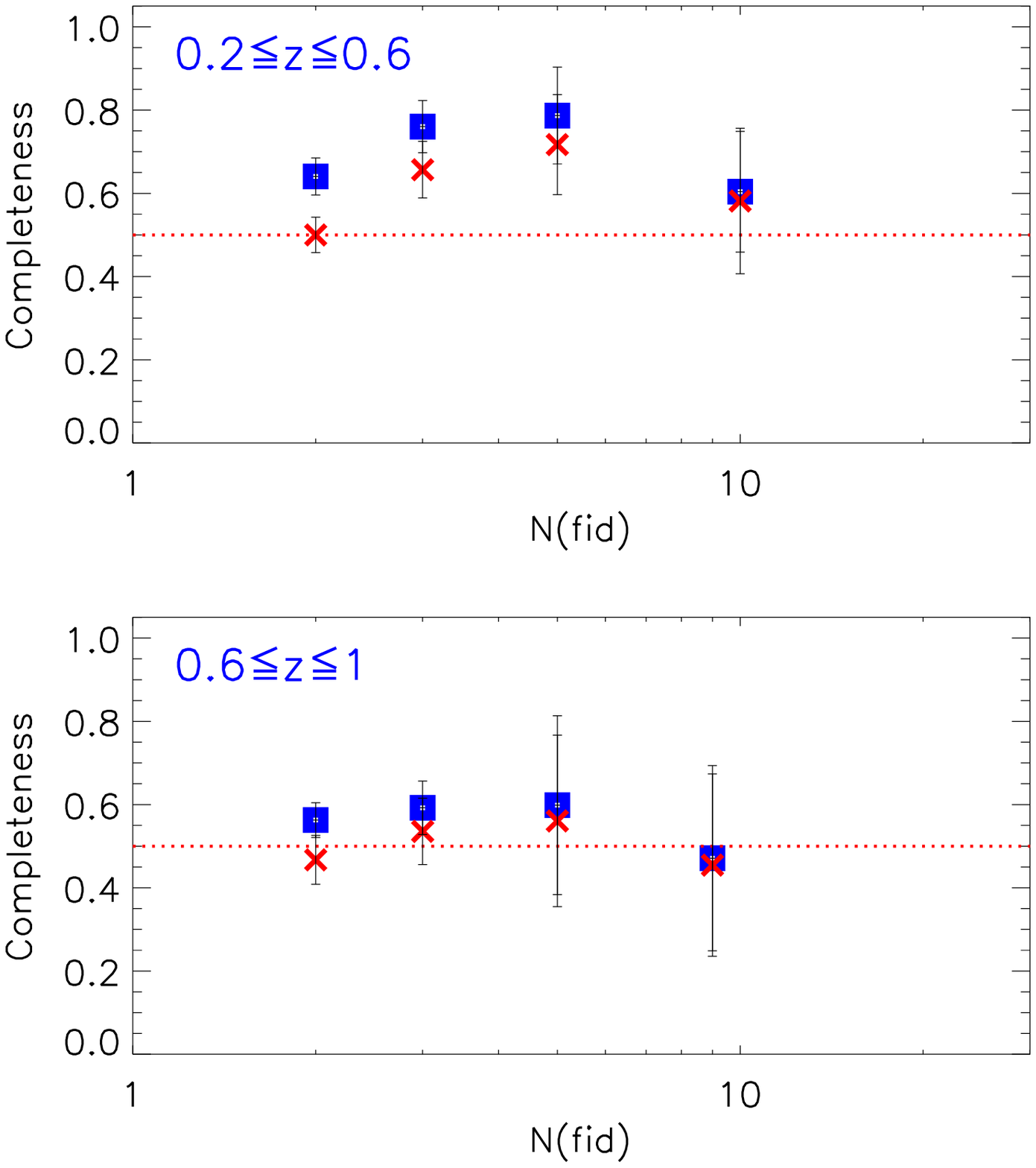}
\includegraphics[width=7cm]{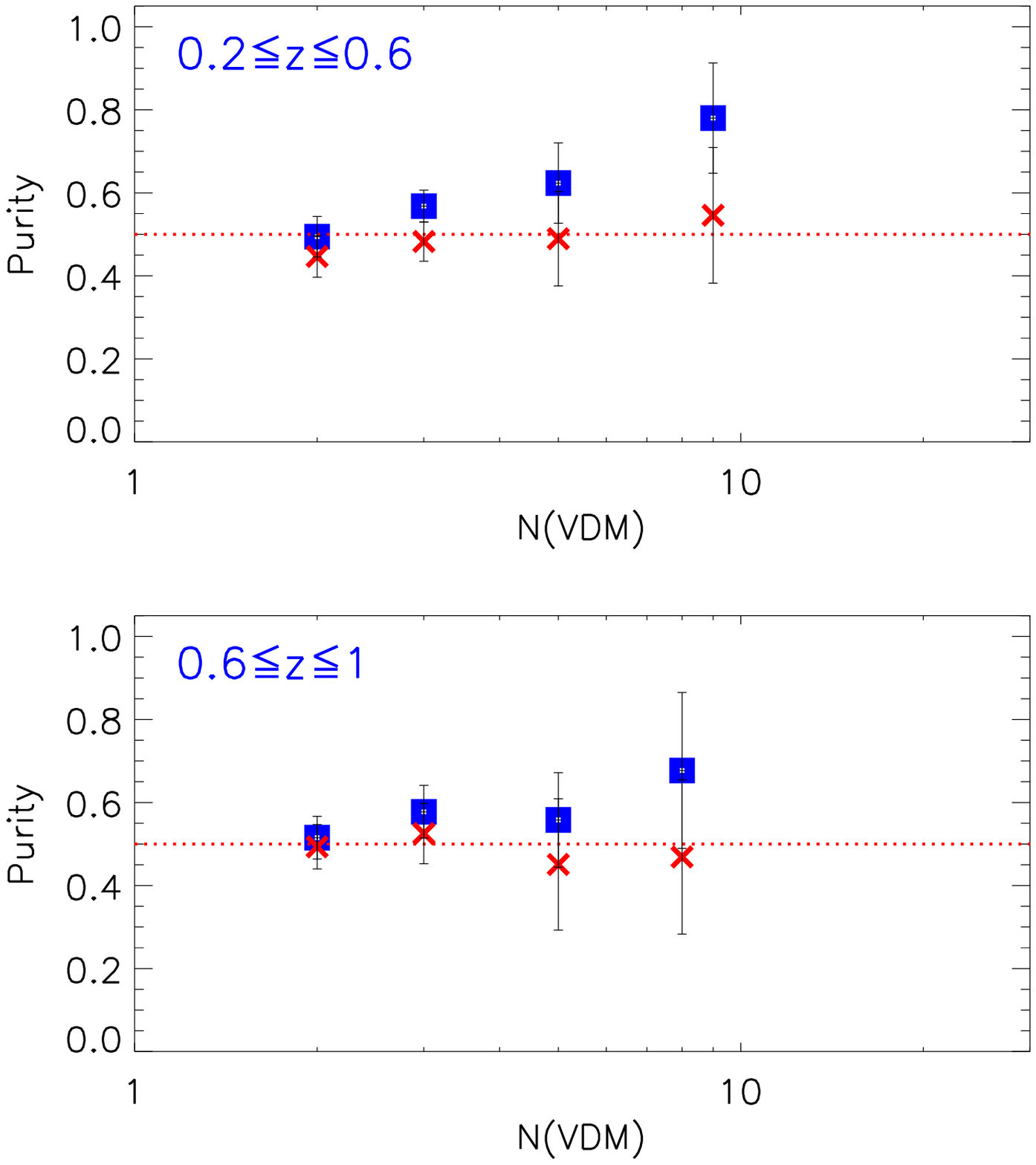}
\caption{$C_1$ and $C_2$ statistics as a function of ``fiducial''
group members (on the left) and $P_1$ and $P_2$ statistics as a
function of ``reconstructed'' group members (on the right). One-way
statistics are shown as blue squares, and two-way statistics are shown
as red crosses.  C and P have been computed separately in each mock:
in these plots points are C and P values averaged over all mocks,
while error bars are their $rms$. On the $x$ axis we grouped the
number of elements in the following way: [N=2],[N=3,4] ,[N=5,6]
 and [N$\geq$7]} 
\label{CP_vsN} 
\end{figure*}

At the end of this finer search, we found the following parameter set,
from now on called the {\it best set} of parameters:

- $\mathcal{R}_{I}=0.28$ $h^{-1}$Mpc 

- $\mathcal{L}_{I}=7.0$ $h^{-1}$Mpc 

- $\mathcal{R}_{II}=0.6$ $h^{-1}$Mpc 

- $\mathcal{L}_{II}=5.0$ $h^{-1}$Mpc 

- $r=0.55$ $h^{-1}$Mpc 

- $l=14.0$ $h^{-1}$Mpc 

We assigned to each group a redshift and a position in the {\it
R.A.-Dec.} plane, respectively defined as the median values of
redshift, Right Ascension and Declination of the group members.

Values of the quality parameters $C_1$, $C_2$, $P_1$, $P_2$, $S_{gal}$
and $f_{I}$ can be found in Table \ref{table_stat}. Note that, to test
the quality of the algorithm also as a function of redshift, we
considered separately two redshift bins ($0.2 \leq z \leq 0.6$ and
$0.6 \leq z \leq 1.0$).

\begin{figure} 
\centering
\includegraphics[width=7cm]{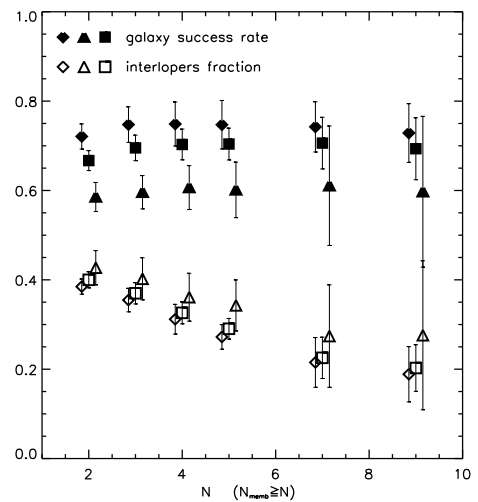}
\caption{Interlopers fraction $f_{I}$ (empty symbols) and galaxy
success rate $S_{gal}$ (full symbols) for different lower limits of
group richness ($x$ axis). Diamonds are for the redshift bin $0.2\leq
z \leq 0.6$, triangles for $0.6\leq z \leq 1.0$ and squares for the
entire range $0.2\leq z \leq 1.0$.  $S_{gal}$ and $f_{I}$ have been
computed separately in each mock. In this plot, points are $S_{gal}$
and $f_{I}$ values averaged over all mocks, while error bars are their
$rms$. } 
\label{int_frac_fig} 
\end{figure}

We analyzed completeness and purity also as a function of group
richness.  Figure \ref{CP_vsN} shows $C_1$ and $C_2$ as a function of
``fiducial'' group members and $P_1$ and $P_2$ as a function of
``reconstructed'' group members.  One-way statistics are shown as blue
squares, and two-ways statistics are shown as red crosses.  C and P
have been computed separately in each mock. In Figure \ref{CP_vsN} we
plot C and P values averaged over all mocks, while error bars are
their rms.  The differences between $C_1$ and $C_2$ and between $P_1$
and $P_2$ indicate that our group catalogue will not be completely
free neither from overmerging nor from fragmentation.  Moreover,
Figure \ref{int_frac_fig} shows that, while the galaxy success rate
$S_{gal}$ does not vary much as a function of $N$, the interloper
fraction $f_{I}$ decreases by a factor of $\sim 2$ from $N \geq 2$ to
$N \geq 9$.

\begin{table}
\caption{Quality statistics ($C_1$, $C_2$, $P_1$, $P_2$, $S_{gal}$ and
$f_{I}$, see text for details) of the reconstructed group catalogue,
for two different redshift bins and for the whole redshift range. The
first table shows $C_1$, $C_2$, $P_1$, $P_2$, $S_{gal}$ and $f_{I}$
for all groups, while the second table for groups with at least 3
members. Each parameter is computed as the mean over the 20 VVDS-like
mocks, and the associated error is its $rms$.}
\label{table_stat}      
\centering                          
\begin{tabular}{c c c c}        
 \multicolumn{4}{c}{}\\  
 \multicolumn{4}{c}{Quality statistics for N$\geq$2}\\  
\hline\hline 
Quality parameter & $0.2\leq z \leq 0.6$ & $0.6\leq z \leq 1.0$ &  $0.2\leq z \leq 1.0$\\    
\hline                        
   $C_1$          & $0.68 \pm 0.03$      & $0.57 \pm 0.04$      &  $0.63 \pm 0.03$     \\
   $C_2$          & $0.56 \pm 0.04$      & $0.49 \pm 0.05$      &  $0.53 \pm 0.03$     \\
   $P_1$          & $0.56 \pm 0.02$      & $0.55 \pm 0.04$      &  $0.56 \pm 0.02$     \\
   $P_2$          & $0.48 \pm 0.04$      & $0.50 \pm 0.04$      &  $0.49 \pm 0.03$      \\
   $S_{gal}$      & $0.72 \pm 0.03$      & $0.59 \pm 0.03$      &  $0.67 \pm 0.02$        \\
   $f_{I}$        & $0.38 \pm 0.02$      & $0.43 \pm 0.04$      &  $0.40 \pm 0.02$        \\
\hline                                   
\end{tabular}

\vspace{0.5cm}

\begin{tabular}{c c c c}        
 \multicolumn{4}{c}{Quality statistics for N$\geq$3}\\  
 \hline\hline 
Quality parameter & $0.2\leq z \leq 0.6$ & $0.6\leq z \leq 1.0$ &  $0.2\leq z \leq 1.0$\\    
\hline                        
   $C_1$          & $0.73 \pm 0.06$      & $0.57 \pm 0.05$      &  $0.67 \pm 0.04$     \\
   $C_2$          & $0.65 \pm 0.06$      & $0.52 \pm 0.06$      &  $0.60 \pm 0.03$     \\
   $P_1$          & $0.61 \pm 0.03$      & $0.58 \pm 0.05$      &  $0.60 \pm 0.03$     \\
   $P_2$          & $0.50 \pm 0.05$      & $0.51 \pm 0.07$      &  $0.50 \pm 0.05$      \\
   $S_{gal}$      & $0.75 \pm 0.04$      & $0.60 \pm 0.04$      &  $0.70 \pm 0.03$        \\
   $f_{I}$        & $0.35 \pm 0.03$      & $0.40 \pm 0.05$      &  $0.37 \pm 0.02$        \\
\hline                                   
\end{tabular}
\end{table}

\subsection{Tests on recovered group properties}\label{other_tests}

\begin{figure} 
\centering
\includegraphics[width=7.cm]{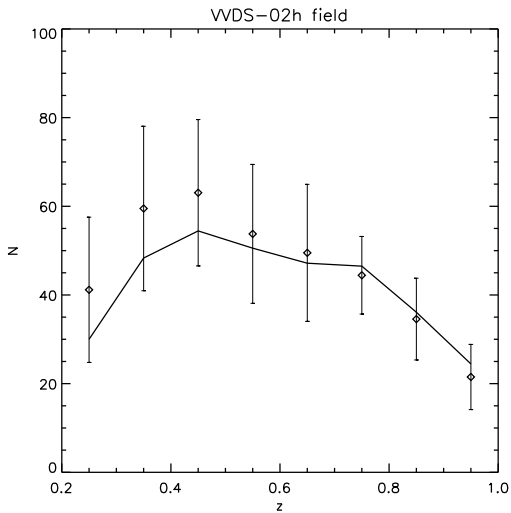}
\caption{Mean distribution of fiducial groups as a function of
redshift (continuous line), computed as the average over 20 VVDS-like
mocks. The mean distribution of VDM-reconstructed groups over the same
20 mocks is over-plotted as black points.}  
\label{nz_variance}
\end{figure}

{\it The n(z) distribution.}  We analyzed how well the ``fiducial''
group distribution $n_{fid}(z)$ as a function of redshift is recovered
by the distribution $n_{VDM}(z)$ of the groups found by the
algorithm. We averaged the $n_{fid}(z)$ distribution over 20
independent VVDS-like mocks to obtain its mean value, plotted as a
continuous line in Figure \ref{nz_variance}. In the Figure, the mean
$n_{VDM}(z)$ for the same 20 independent mocks is shown as black
points, with vertical bars being the {\it rms} among the 20 mocks. The
plot shows that the difference between $n_{VDM}(z)$ and $n_{fid}(z)$,
despite the presence of fake and/or missing groups in the VDM
catalogue, is within the errors. A $\chi^2$ test between the two mean
distributions gives $\chi^2 = 1.4$. We therefore conclude that the two
$n(z)$ distributions are statistically consistent with each other,
even if there is a tendency for having more VDM reconstructed groups
at low redshift. We repeated the same test using only groups with at
least 5 members and with $\sigma\geq350$ km/s, that is those groups
for which we are sure we can compute a reliable velocity dispersion,
and we found also in this case that $n_{VDM}(z)$ and $n_{fid}(z)$ are
consistent with each other.

{\it Velocity dispersion.} As discussed above, the comparison between
group properties in the fiducial and in the reconstructed catalogue is
an important check to verify that VDM algorithm is not only able to
recover real groups, but also to preserve their characteristics. This
means that when we compare the two catalogues on a group-by-group
basis, the fractions of interlopers and missing galaxies modify group
properties only below some tolerance level.  The same has to hold also
for the fiducial and reconstructed statistical distributions of these
properties, when considering that the reconstructed catalogue contains
fake groups and it misses some groups.

\begin{figure} 
\centering
\includegraphics[width=7cm]{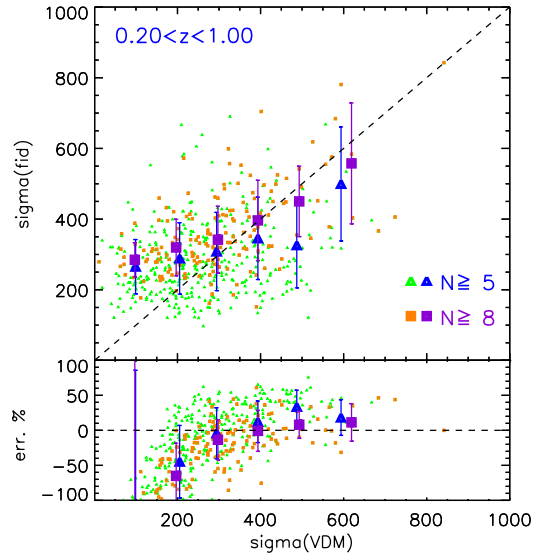}
\caption{Comparison between virial ($y$ axis) and VDM reconstructed
($x$ axis) group velocity dispersion. Only 2 ways matches are
considered in this plot. The upper panel shows the scatter plot, the
lower shows the percentage error. Green and blue triangles are groups
with at least 5 members, orange and purple squares groups with at
least 8 members; green and orange points are single groups, while blue
and purple symbols are the median (on $x$ axis) and mean (on $y$ axis)
values in bins of the property on the $x$ axis.  Vertical error bars
are {\it rms} of mean values. } 
\label{sigma_VIR_VDM_scatter}
\end{figure}

For each VDM group, we measured the velocity dispersion $\sigma_{VDM}$ of its galaxies
using Equation \ref{sigma_gapper}, correcting it as in Equation
\ref{sigma_gapper_err}.  Figure \ref{sigma_VIR_VDM_scatter} shows the
comparison between the velocity dispersion in the reconstructed groups
($\sigma_{VDM}$) and the {\it virial} velocity dispersion
($\sigma_{vir}$ quoted in the simulations) of the fiducial groups in
VVDS-like mocks on a group-by-group basis. Only {\it two-ways} matches
are considered. The Figure is divided in two panels as Figure
\ref{sigma_ratio_vsm_vir_33_275}: the upper part shows the scatter
plot, the lower shows the percentage error, computed as in Figure
\ref{sigma_ratio_vsm_vir_33_275}.  Green and blue triangles are groups
with at least 5 members, orange and purple squares groups with at
least 8 members; green and orange points are single groups, while blue
and purple symbols are the median (on $x$ axis) and mean (on $y$ axis)
values in bins of the property on the $x$ axis.  Vertical error bars
are {\it rms} of mean values.

This scatter plot shows the following: on a group-by-group basis, for
$\sigma_{VDM}\geq350$ km/s, close to the intrinsic limit set by the
flux-limited nature of the VVDS catalogue, the correlation between
$\sigma_{VDM}$ and $\sigma_{vir}$ is such that $\sigma_{VDM}$
overestimates $\sigma_{vir}$, but on average always by $\lesssim 30$\%
for groups with at least 5 members, while this overestimate is on
average $\lesssim 10$\% for groups with at least 8 members.  In fact,
we have shown in Section \ref{sigma_tex} that the velocity dispersion
$\sigma_{meas}$ that one can measure in groups within a VVDS-like data
sample is not a reliable estimator of $\sigma_{vir}$ for
$\sigma_{meas} \leq 350$ km/s.

\begin{figure} 
\centering
\includegraphics[width=7cm]{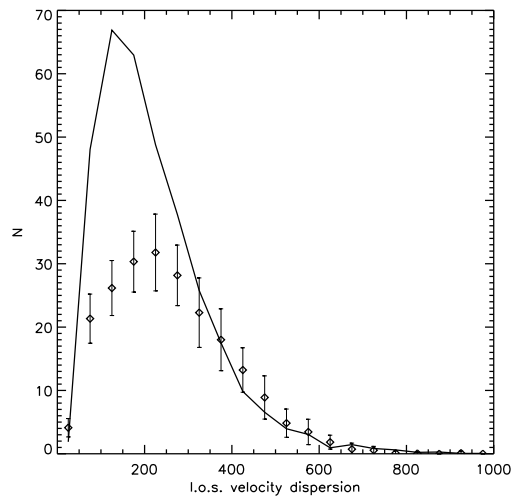} \caption{Mean
distribution of virial line of sight velocity dispersion (continuous
line), computed as the average over 20 VVDS-like mocks. The mean
distribution of $\sigma$ of groups reconstructed by the VDM, averaged
over the same 20 mocks, is over-plotted as black points, with vertical
bars corresponding to its {\it rms}.}  
\label{sigma_VIR_VDM_distrib}
\end{figure}

Besides the group-by-group comparison, it is also interesting the
analysis of the velocity dispersion distributions, thus including
unrecovered and fake groups in the fiducial and in the reconstructed
catalogues respectively.  Figure \ref{sigma_VIR_VDM_distrib} shows the
comparison between $n(\sigma_{vir})$ and $n(\sigma_{VDM})$
distributions (the solid line and the black diamonds
respectively). Values on the $y$ axis are averaged over 20 VVDS-like
mocks. Vertical bars associated to $\sigma_{VDM}$ points are their
{\it rms} over the 20 mocks. One can notice that the area below the
two distributions is different. This mainly because in $\sigma_{VDM}$
distribution we excluded groups for which we were not able to measure
$\sigma$, \ie groups for which we imposed $\sigma_{VDM}=0$. This
comparison shows that the two distribution agree for $\sigma \geq 350$
km/s, as confirmed through a $\chi^2$ test between the two mean
distributions for $\sigma \geq 350$ km/s.

\begin{figure} 
\centering
\includegraphics[width=7cm]{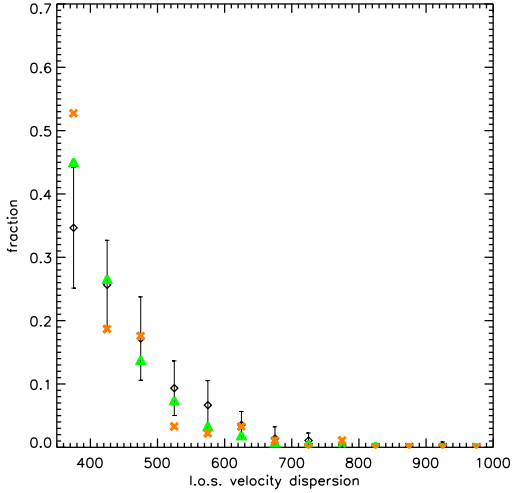}
\caption{Normalized mean distribution of
$\sigma_{VDM}$ (black diamonds) for $\sigma \geq 350$ km/s. It is the
same distribution as in Figure \ref{sigma_VIR_VDM_distrib}, but it is
normalized by the total number of groups with $\sigma \geq 350$
km/s. Overplotted green triangles represent the normalized mean
$\sigma_{vir}$ distribution of fiducial groups in MILLENNIUM mock
catalogue with flux limits at $I_{AB}=24$ and with 100\% sampling rate
($M(100,0)$ mock catalogues, see Subsection \ref{mocks_description}),
and the orange crosses are the normalized mean $\sigma_{vir}$
distribution of fiducial groups in complete light cones of the
MILLENNIUM Simulation, \ie catalogues with no flux limits.  For each
distribution, the redshift range considered is $0.2\leq z \leq 1.0$.}
\label{sigma_VDM_ALLnorm_distrib} 
\end{figure}

As a further test for the recovered $\sigma_{VDM}$ distribution, we
compared it with the $n(\sigma_{vir})$ in mock catalogues with the
same flux limits as VVDS-02h sample but with 100\% sampling rate (the
$M(100,0)$ catalogues presented in Subsection
\ref{mocks_description}), and with the $n(\sigma_{vir})$ of mock
catalogues with no flux limits (the complete light cones from which
$M(100,0)$ catalogues have been extracted).  In Figure
\ref{sigma_VDM_ALLnorm_distrib}, we show the normalized mean
$n(\sigma_{VDM})$ (black diamonds) for $\sigma \geq 350$ km/s. It is
the same distribution as in Figure \ref{sigma_VIR_VDM_distrib}, but it
is normalized by the total numbers of groups with $\sigma \geq 350$
km/s. Overplotted green triangles represent the normalized mean
$n(\sigma_{vir})$ of fiducial groups in $M(100,0)$ mock catalogues,
and the orange crosses are the normalized mean $n(\sigma_{vir})$
distribution of fiducial groups in complete light cones of the
MILLENNIUM Simulation.  For each distribution, the redshift range
considered is $0.2\leq z \leq 1.0$.  Considering these normalized
distributions for $\sigma \geq 350$ km/s, a $\chi^2$ test between the
$n(\sigma_{VDM})$ and the $n(\sigma_{vir})$ for $M(100,0)$ catalogues
leaves us with $\chi^2$ such that we can conclude that the two
distributions are statistically in agreement. We obtain the same
result when we apply the same test between $n(\sigma_{VDM})$ and
$n(\sigma_{vir})$ for the complete catalogues. This means that the 
$n(\sigma_{VDM})$ of the groups reconstructed by our algorithm is unbiased with respect 
to the $n(\sigma_{vir})$ of groups in the complete light cones.

We repeated the tests shown in Figures \ref{sigma_VIR_VDM_distrib} and
\ref{sigma_VDM_ALLnorm_distrib} also using only groups with at least 5
members, and we found similar results.

As discussed in Section \ref{the_algorithm}, one of the primary goals
of the VDM is to be able to recover the virial line of sight
velocity dispersion of group galaxies, at least above some minimum threshold. This is
not achieved, for example, by other commonly used group-finding
algorithms, such as the FOF method (see Section \ref{the_algorithm}).
The comparisons of $n(\sigma)$ distributions between reconstructed and
fiducial groups presented in this Section show that this aim has been
successfully obtained in a deep redshift survey such as VVDS, at least
up to $z=1$. Moreover, VVDS redshift measurement error and sampling
rate imposed an {\it a priori} lower limit for a reliable measurement
of the line of sight velocity dispersion of group galaxies ($\sigma \geq 350$
km/s, see Subsection \ref{sigma_tex}). We have shown in Figures
\ref{sigma_VIR_VDM_distrib} and \ref{sigma_VDM_ALLnorm_distrib} that
the finding group algorithm we used not only can recover a reliable
$n(\sigma)$ distribution above some minimum $\sigma$, but also it does
not worsen the minimum $\sigma$ threshold imposed by the survey
strategy itself. This result has been reached thanks to the
flexibility of the 6 VDM parameters. Each of them has a specific role
in determining the choice of the group members, through an intuitive
localization of group barycenters (Phase I), a reliable estimate of
the central richness (Phase II) and a correct exploitation of group
scaling laws (Phase III).

{\it Sampling rate.} As we applied the algorithm to the VVDS-like
mocks, we optimized it for the whole observed area ($\sim 0.5$ deg$^2$
each), irrespective of the varying sampling rate across the
field. Nevertheless, we tested also how completeness and purity change
if computed separately in areas with very different sampling rate,
that is covered by 1, 2 or 4 passes of the spectrograph (hereafter
called `1p', `2p' and `4p' areas). For this test, we assigned each
group to the 1p, 2p or 4p area according to its {\it R.A-Dec.} 
position (computed as the median value of {\it R.A.} and {\it Dec.} of
the member galaxies), even if it extends over an area with a sudden
drop/increase of the sampling rate. Considering the whole redshift
range $0.2\leq z \leq 1.0$, in the 4p area we find $C_1 = 0.67 \pm
0.04$, $C_2 = 0.54 \pm 0.04$, $P_1 = 0.55 \pm 0.03$ and $P_2 = 0.48
\pm 0.08$, while in the (1+2)p areas $C_1 =
0.58 \pm 0.04$, $C_2 = 0.52 \pm 0.04$, $P_1 = 0.57 \pm 0.03$ and $P_2
= 0.49 \pm 0.04$.

While $C_2$, $P_1$ and $P_2$ differences are inside error bars, one
can notice a larger worsening in $C_1$ when we decrease the number of
spectrograph passes, \ie the sampling rate.  Moreover, analyzing the
dependence of $C$ and $P$ on group richness, we can add that in the 4p
area completeness is higher even for $N \geq 5$. Moreover, we also
notice that in 4p area there is a higher overmerging, especially for
$N \leq 4$, while in the (1+2)p area fragmentation is increased for $N
\geq 5$.

\subsection{High purity parameters}

With the {\it best set} of parameters, we can obtain from VVDS-02h
data a group catalogue with high completeness, but it has been shown
that only $\sim50$\% of groups is pure.  This means that each group
identified by the algorithm has, on average, only 50\% probability of
being a real group.  It could be useful to identify the subsample of
groups that has an even higher probability of being real. Thus, we
optimized the group-finding algorithm a second time, in this case
maximizing purity (but paying attention not to reduce the new
recovered group catalogue to a few `super-secure' groups). The so
called {\it high-purity} parameter set is the following:

- $\mathcal{R}_{I}=0.10$ $h^{-1}$Mpc 

- $\mathcal{L}_{I}=5.0$ $h^{-1}$Mpc 

- $\mathcal{R}_{II}=0.6$ $h^{-1}$Mpc 

- $\mathcal{L}_{II}=5.0$ $h^{-1}$Mpc 

- $r=0.55$ $h^{-1}$Mpc 

- $l=14.0$ $h^{-1}$Mpc

Table \ref{table_stat_highpurity} shows $C$ and $P$ for the {\it
high-purity} parameter set. Necessarily, $C$ is very low, but now each
group identified by the algorithm has $\sim 70$\% of probability of
being real, and the interlopers fraction $f_I$ decreases from $\sim
40$\% to $\sim 25$\% with respect to the one obtained with the {\it
best set} of parameters (see Table \ref{table_stat}).

\begin{table} 
\caption{Quality statistics ($C_1$, $C_2$, $P_1$, $P_2$,
$S_{gal}$ and $f_{I}$, see text for details) of the group catalogue 
reconstructed by the algorithm with the {\it high-purity} parameter
set, for two different redshift bins and for the whole redshift
range, considering groups with N$\geq$2. Each parameter is computed 
as the mean over the 20 VVDS-like
mocks, and the associated error is its $rms$. The same parameters for 
groups with N$\geq$3 are consistent, within error bars,
with those presented here.}
\label{table_stat_highpurity} 
\centering 
\begin{tabular}{c c c c}
 \multicolumn{4}{c}{Quality statistics for N$\geq$2}\\  
\hline\hline 
Quality parameter & $0.2\leq z \leq 0.6$ & $0.6\leq z \leq 1.0$ &  $0.2\leq z \leq 1.0$\\    
\hline                        
   $C_1$          & $0.29 \pm 0.03$      & $0.20 \pm 0.03$      &  $0.24 \pm 0.02$     \\
   $C_2$          & $0.24 \pm 0.03$      & $0.17 \pm 0.02$      &  $0.20 \pm 0.02$     \\
   $P_1$          & $0.75 \pm 0.04$      & $0.73 \pm 0.06$      &  $0.74 \pm 0.04$     \\
   $P_2$          & $0.66 \pm 0.07$      & $0.69 \pm 0.06$      &  $0.67 \pm 0.05$      \\
   $S_{gal}$      & $0.32 \pm 0.03$      & $0.21 \pm 0.04$      &  $0.28 \pm 0.03$        \\
   $f_{I}$        & $0.24 \pm 0.04$      & $0.27 \pm 0.05$      &  $0.25 \pm 0.03$        \\
\hline                                   
\end{tabular}
\end{table}

%
%

\section{VVDS-02h field group catalogue}\label{real_VVDS_catalogue}

\begin{figure} 
\centering
\includegraphics[width=8cm]{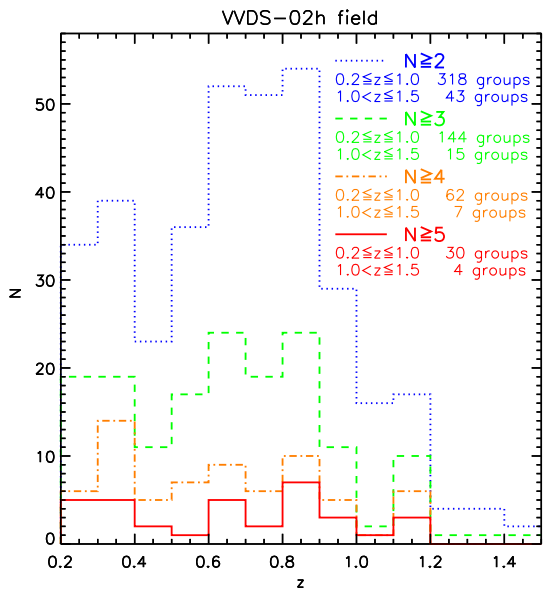} 
\caption{The
redshift distribution of groups in the VVDS-02h sample, found using
the {\it best set} of parameters. Different line styles are for
different cuts in group richness, as indicated. The total number of
groups with the corresponding richness is quoted in the labels, for
two different redshift ranges.}  
\label{Nz_VVDS_real} 
\end{figure}

We applied the group-finding algorithm to the VVDS-02h sample
described in Section \ref{real_data}, using the {\it best set} of
parameters. We defined the redshift and the position in the {\it
R.A.-Dec.} plane of each group as the median values of redshift, Right
Ascension and Declination of the group members. Figure
\ref{Nz_VVDS_real} shows the redshift distribution of the identified
groups, with different line styles for different cuts in group
richness, as indicated in the Figure. It is clear that beyond $z \sim
1$ there is a significant drop in the number of recovered groups,
irrespectively of their richness, as expected from Figure
\ref{slos_potentaility}.  This drop in the redshift distribution
may partly be related also to the choice of optimizing the algorithm
only up to $z=1$ (see Section \ref{algorithm_optimization}). We
applied the VDM to our galaxy sample also using the {\it high-purity}
set of parameters. With the {\it best set} of parameters, the
algorithm identified 318 groups with 2 or more members in the redshift
range $0.2 \leq z \leq 1.0$, one third of them  having also been detected with the
{\it high-purity} set. The identified groups comprise  $\sim 19$\% of the galaxies
in our sample. Comparing this percentage
with the fraction of galaxies that reside in groups in VVDS-like mock
catalogues, we found that it is consistent with both the fraction of
galaxies residing in {\it fiducial} groups ($\sim 20$ \%) and the
percentage of galaxies residing in {\it reconstructed} groups ($\sim
22$\%).

For each group we estimated the line of sight velocity dispersion
$\sigma$.  We used the {\it gapper } method, as described in Section
\ref{sigma_tex}, and we corrected it for the redshift measurement
error subtracting it in quadrature as in Equation
\ref{sigma_gapper_err}.  We set $\sigma=0$ km/s for those groups with
a measured $\sigma_G$ (from Equation \ref{sigma_gapper}) lower than
the redshift error. $\sim25$\% of
groups with $\sigma \geq 350$ km/s have been detected by the algorithm also with the
{\it high-purity} parameter set.

It is worth noticing that, given the small value of the parameter
$r$, driving the projected dimension of the search cylinder in Phase III (see
Section \ref{phaseIII}), the typical projected radius within which the
full set of group members is selected is always $<1$ $h^{-1}$Mpc.

Detailed group catalogue statistics are shown in Table
\ref{table_groups_VVDSreal}. The number of groups that has been found in VVDS-02h field is
quoted. Different rows are for different values of velocity
dispersion, different columns for different richnesses.  
Numbers in brackets indicate the number of groups that
have been identified by the algorithm also with the {\it high-purity}
set of parameters (even if with less members).

We tested the reliability of the reconstructed  catalogue 
by recomputing the groups
excluding galaxies with flag=2 and 9, \ie using only galaxies whose
redshift has a high likelihood ($>95$\%) of being  correct. 
We verified that, with respect
to our original group catalogue, 80\% (/77\%/75\%) of the groups with
at least 5 (/4/3) members are still recovered. These means that for
these recovered groups the galaxies with flag=2 and 9 were not in the
{\it seed} of the group, \ie in the first set of galaxies recovered in
the Phase I of the algorithm (see Section \ref{phaseI}).

Table \ref{table_listgroups_VVDSreal} lists all the groups identified 
in the redshift window
$0.2 \leq z \leq 1.0$. It is worth noticing that the quoted number of
members has to be considered as a lower limit for the real richness,
as the sampling rate of our survey is not 100\%. The groups
labeled with a star near their ID are those recovered also when using
only galaxies with flag=3 and 4. We apply this label only to groups
with at least 3 members.  The group members are presented  in  
Table \ref{table_listgalgroups_VVDSreal}.  Note that the galaxy
ID is the same used to identify galaxies in the public VVDS release at http://cencosw.oamp.fr. 

\begin{figure} 
\centering
\includegraphics[width=8cm]{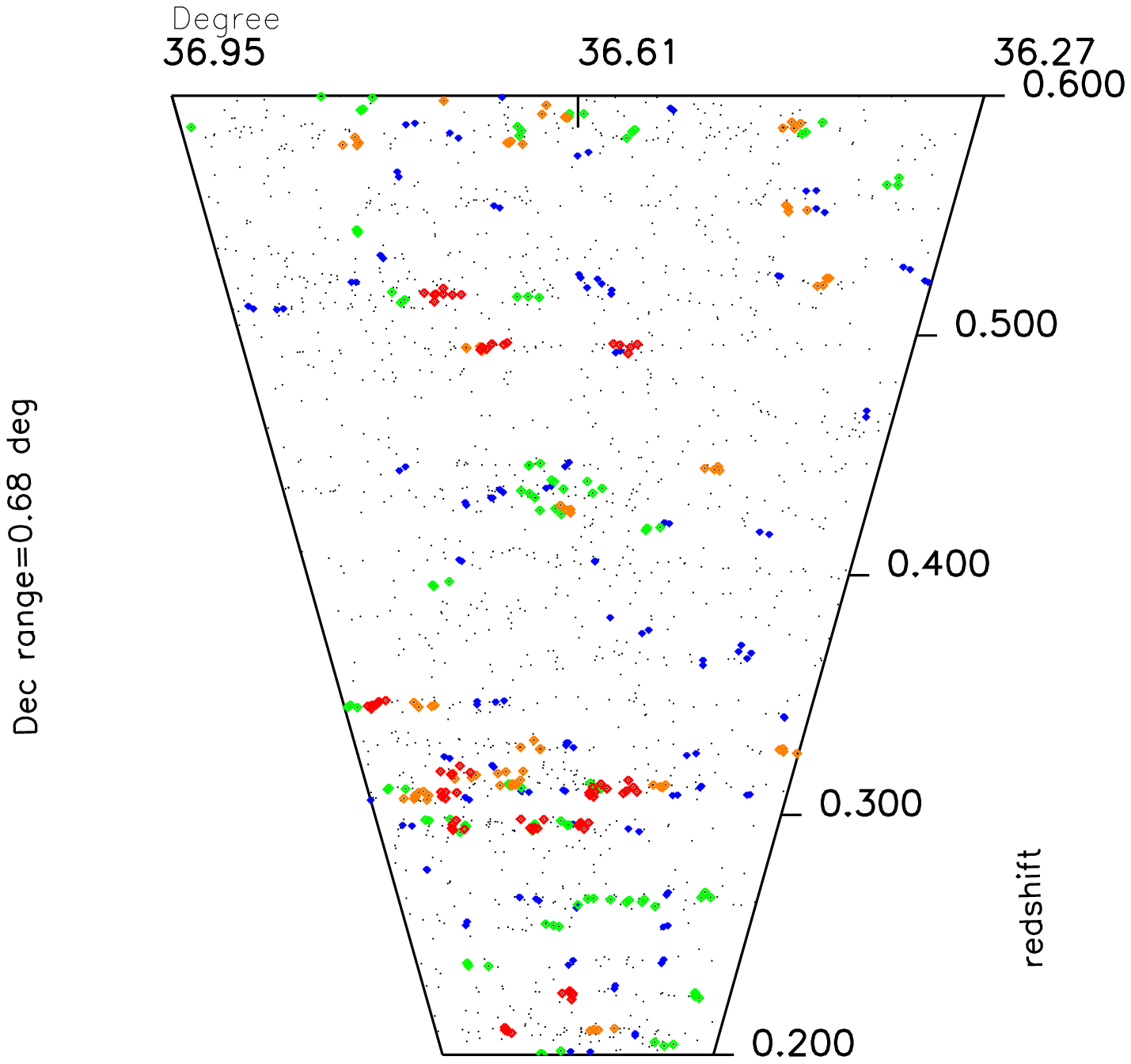}
\includegraphics[width=8cm]{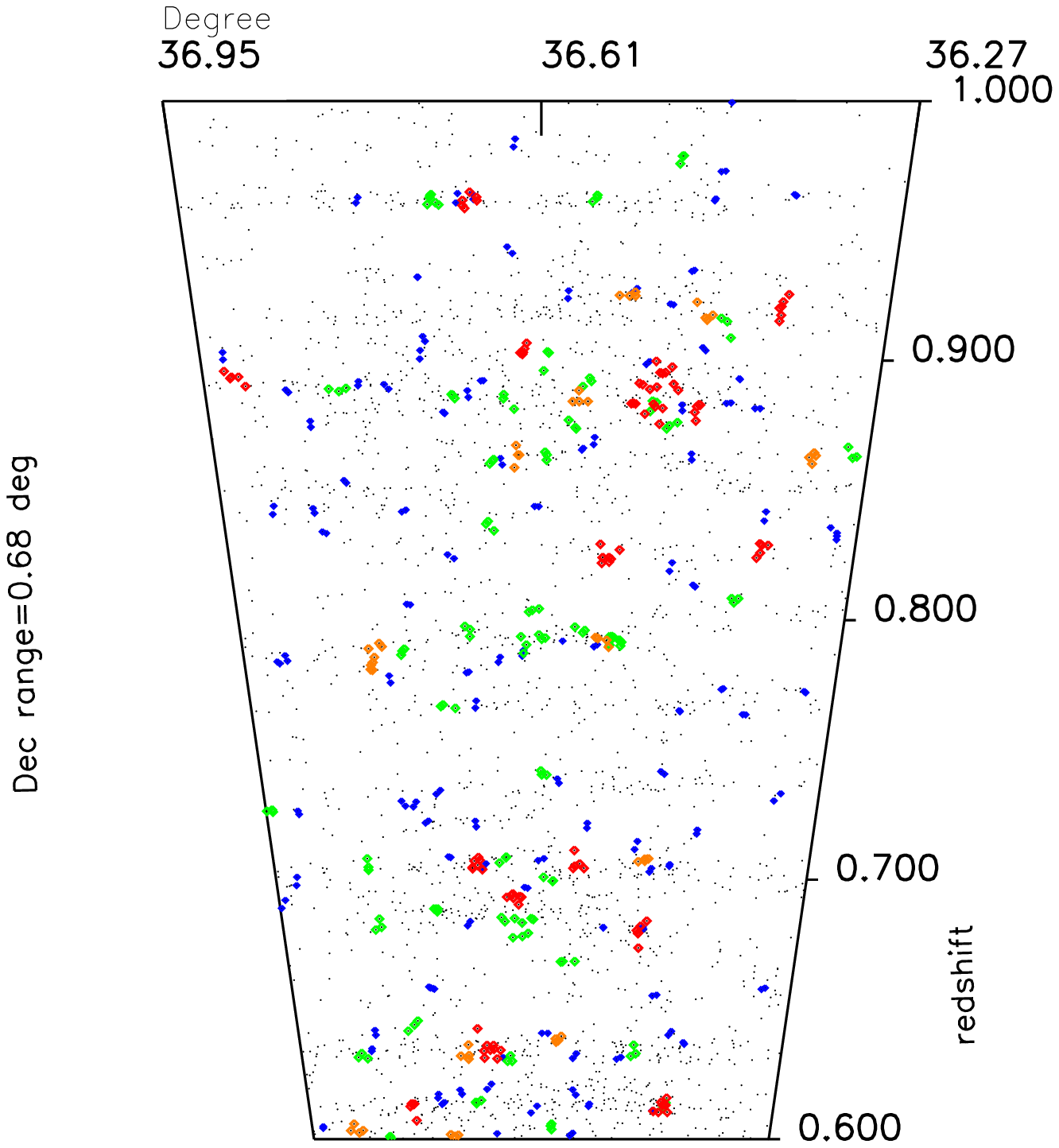}
\caption{Two-dimensional VVDS galaxy distribution as a function of
Right Ascension and redshift (points are compressed on the Declination
dimension). Each plot shows a different redshift bin ($0.2\leq z \leq
0.6$ and $0.6\leq z \leq 1.0$ in upper and lower panel
respectively). Black dots are field galaxies, coloured dots are group
members, according to the group catalogue obtained with the {\it best
set} of parameters. The colour code is the following: blue dots are
pair members, green are triplet members, orange are quartet members
and red dots are members of groups with 5 or more members.}
\label{cono_best} 
\end{figure}

In Figure \ref{cono_best}, the two-dimensional VVDS galaxy
distribution is shown, with galaxy positions projected on Right
Ascension and redshift.  Each plot shows a different redshift bin, as
quoted on the $y$ axis. Black dots are field galaxies, while coloured
dots are group members (blue dots are pair members, green are triplet
members, orange are quartet members and red dots are galaxies included in groups
with 5 or more members).

\begin{table*} 
\caption{Number of VVDS-02h groups reconstructed by the
algorithm using the {\it best set} of parameters in VVDS-02h field,
for $0.2\leq z \leq 1.0$. Statistics are quoted as a function of the
number of group members (columns) and of measured line of sight
velocity dispersion of group galaxies ($\sigma$, in km/s, rows). Numbers in brackets
indicate the number of groups found also applying the algorithm with
the {\it high-purity} set of parameters.}
\label{table_groups_VVDSreal} 
\centering 
\vspace{0.2cm}
\begin{tabular}{c |c c c c c c c c | c}
                 
\hline\hline 

$\sigma$ (km/s)        & \multicolumn{9}{c}{Group members}\\    
                              &   2               &   3              &    4        &   5         &    6       &   7        &  8         &  9  &  ALL  \\    

\hline                                   

  $\sigma = 0^a$        &  89(23)          &   24(10)     &     8(2)     & -      &  -       &  -      & -    &   -  &   121(35)  \\                     
  $0 < \sigma < 350$        &   61(25)          &   39(18)     &     18(6)     & 8(5)       &  3(2)       &  3(3)      &  2(1)   & -    &   134(60)    \\                 
  $\sigma \geq 350$      &    24(0)          &   19(6)     &     6(3)     &5(1)         &    4(2)      &  3(2)      &  1(0)    &  1(1)     &  63(15)    \\                   

\hline 
\multicolumn{8}{c}{} & Total:       & 318(110)\\    
\hline 
\hline 

\multicolumn{10}{l}{$^a$) We refer the reader to Section \ref{real_VVDS_catalogue} for the meaning of $\sigma = 0$}\\

\end{tabular}

\end{table*}

\begin{table} 
\caption{List of groups recovered in the VVDS-02h field in the range $0.2 \leq z \leq 1.0$. 
Columns are the
following: 1) group ID; 2) R.A.; 3) declination; 4) redshift; 5) number of detected members; 6)
l.o.s. velocity dispersion $\sigma$; 7) possible high purity. R.A. and
declination are in degrees and $\sigma$ in km/s. R.A., declination and
redshift are the median values of all the galaxies in each
group. The star near the group ID label those groups found by the algorithm 
also when using only galaxies with flag 3
and 4 (see text for further details). 
Column 7 labels with an `H' those groups detected by the
algorithm also with the {\it high-purity} set of parameters. See the electronic edition
for the complete list of VVDS-02h groups.}  
\label{table_listgroups_VVDSreal} 
\centering
\vspace{0.2cm} 
\begin{tabular}{|c c c c c c c |}
                 
\hline\hline 

grID & R.A. & Dec. & $z$ & N & $\sigma$ & Purity\\    
 &  $deg.$ & $deg.$ & &  & $km/s$ & \\    

\hline                                   
124$^{*}$  &        36.56310 &  -4.31748  &   0.5850 &  3  &   351   &	     H \\
144$^{*}$  &        36.79910 &  -4.59669  &   0.6135 &  5  &   428    &	     H \\
224$^{*}$  &        36.80323 &  -4.67940  &   0.7898 &  4  &   392   &	       \\

\hline 
\hline

\end{tabular}
\end{table}

\begin{table} 
\caption{List of group galaxies belonging to the groups listed in Table 
\ref{table_listgroups_VVDSreal}. Columns are the
following: 1) galaxy ID; 2) R.A.; 3) declination; 4) redshift; 5) redshift quality flag 
(see Section \ref{real_data}); 
6) ID of the group to which the galaxy belongs 7) total number of group members. 
R.A. and declination are in degrees. See the electronic edition
for the complete list of VVDS-02h group galaxies.}  
\label{table_listgalgroups_VVDSreal} 
\centering
\vspace{0.2cm} 
\begin{tabular}{|c c c c c c c |}
                 
\hline\hline 

galID$^{a}$ & R.A. & Dec. & $z$ & z flag  & grID & N \\    
 &  $deg.$ & $deg.$ & &  &  & \\    

\hline                                   

  20309041  &  36.56095  & -4.31812  &	 0.5859  &  4	&      124    &      3  \\
  20309502  &  36.56310  & -4.31748  &	 0.5850  &  4	&      124    &      3	\\
  20310401  &  36.56771  & -4.31544  &	 0.5824  &  2	&      124    &      3	\\
  20176187  &  36.79656  & -4.61242  &	 0.6072  &  4	&      144    &      5	\\
  20183000  &  36.79910  & -4.59744  &	 0.6135  &  4	&      144    &      5	\\
  20183332  &  36.79801  & -4.59669  &	 0.6136  &  4	&      144    &      5	\\
  20184297  &  36.80199  & -4.59303  &	 0.6137  &  2	&      144    &      5	\\
  20184706  &  36.80420  & -4.59215  &	 0.6126  &  3	&      144    &      5	\\
  20146543  &  36.80323  & -4.68008  &	 0.7857  &  3	&      224    &      4	\\
  20146933  &  36.80903  & -4.67979  &	 0.7890  &  3	&      224    &      4	\\
  20147204  &  36.79638  & -4.67940  &	 0.7911  &  4	&      224    &      4	\\
  20151406  &  36.79351  & -4.66935  &	 0.7898  &  3	&      224    &      4	\\

\hline 
\hline 

\multicolumn{7}{l}{$^a$) The galaxy ID refers to the public VVDS release at} \\
\multicolumn{7}{l}{~~~~ http://cencosw.oamp.fr} \\

\end{tabular}
\end{table}

\subsection{Line of sight velocity dispersion of group galaxies}\label{VVDS_group_prop}

It is interesting to check whether the real universe looks like the
simulated one. In this section we compare the VVDS catalogue with the
Millennium-based mock catalogues.

\begin{figure}
\centering
\includegraphics[width=7cm]{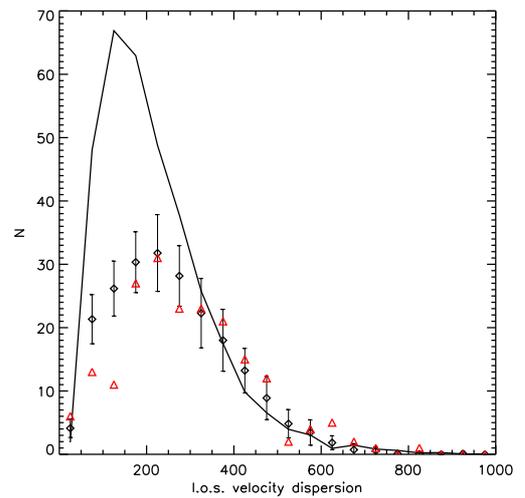} 
\caption{As
in Figure \ref{sigma_VIR_VDM_distrib}, but in this case the velocity
dispersion distribution of VVDS-02h field groups is also shown with
red triangles.}  
\label{ns_distrib_VVDS_real} 
\end{figure}

We compared the $n(\sigma)$ distributions of real and simulated
groups. Figure \ref{ns_distrib_VVDS_real} shows the $n(\sigma)$
distribution for all VVDS-02h groups in the redshift range $0.2 \leq z
\leq 1.0$ (red triangles) and the $n(\sigma)$ distribution for
VVDS-like mock catalogues. As in Figure \ref{sigma_VIR_VDM_distrib},
the continuous line is the distribution of $\sigma_{vir}$ of {\it
fiducial} groups, while black points represent the mean distribution
for {\it reconstructed} groups, with vertical bars being the {\it rms}
of the 20 mock catalogues. In this plot we consider the
$\sigma$ measured with the gapper method (for both mocks and real
data) and not the {\it virial} velocity dispersion. Moreover, we are excluding
groups with measured $\sigma$ equal to 0, because this value
indicates that we have not been able to measure it due to the redshift
measurement error (see Equation \ref{sigma_gapper_err}). This is the
reason why the area under $n(\sigma_{vir})$ in the plot is larger than
the area under the other distributions. In this Figure one can notice
a consistency in the $n(\sigma)$ distributions of real and mock group
catalogues, at least for $\sigma> 350$ km/s.

The relatively large number of groups for which the velocity
dispersion estimated through Equation \ref{sigma_gapper_err} is
formally negative is probably due to the fact that we did not take
into account possible dependences of the mean redshift error on the
properties of the galaxies (i.e. magnitude, presence of emission lines
etc.). It is likely that for many of these groups the redshift error
associated to their galaxy members is somewhat smaller than the
adopted average value (275 km/s). Nevertheless, we are reassured by
the fact that none of the groups with $N\geq5$ has $\sigma=0$.

\subsection{Comparison with other group catalogues in the same field}\label{comparison_other_cat}

Several group catalogues have already been compiled from different
kind of observations and with different methods in the sky area
covered by the VVDS-02h field. For example, X-ray clusters have been
identified from XMM-Newton images and then spectroscopically confirmed
\citep{andreon2004,valtchanov2004, andreon2005,
willis2005,willis2005err, bremer2006,pierre2006}. The
matched-filter technique has been used as well \citep{olsen2007},
together with a weak lensing search \citep{gavazzir2007} and structure
identification through photometric redshift \citep{mazure2007}. All
these latter methods were applied to photometric data from CFHTLS.

Among the X-ray clusters of the XMM-LSS, only 8 clusters fall in the
VVDS-02h field area in the redshift bin $0.2 \leq z \leq 1.0$: XLSSC
005, XLSSC 013 and XLSSC 025 from C1 catalogue, XLSSC 038 form C2
catalogue and then the clusters \emph{a, b, c, d} from C3 catalogue
(see Table 3 in \citealp{pierre2006}).  We find that both clusters
\emph{b} and \emph{c} have a counterpart in our VDM catalogue (with 6
and 8 detected members respectively), with an almost perfect match in
their barycenters. Clusters XLSSC 013 and XLSSC 025 have possible
counterparts at the same z (with 4 and 3 members respectively), but
their barycenters in ra-dec have a shift of $\sim200$ $h^{-1}$
kpc. Inspecting these two groups more in details, we find that the
possible XLSSC 025 counterpart is dominated by a massive galaxy that
is distant from XLSSC 025 barycenter $\sim100$ $h^{-1}$ kpc, showing
that in this case a better match would have been obtained if we had
computed a mass-weighted barycenter. On the contrary, for XLSSC 013
counterpart we do not identify any dominant galaxy. This shift of
$\sim200$ $h^{-1}$ kpc could also be explained by the following: we
studied the distances between the barycenters of VDM groups and their
corresponding fiducial groups, and we found that their distribution is
a Gaussian centered at $\sim0$ with a scatter of $\sim 200$ $h^{-1}$
kpc.  Finally, we do not find counterparts for XLSSC 005, XLSSC 038,
\emph{a} and \emph{d} in our catalogue. They fall inside our low
sampling rate areas (covered only by 1 or 2 passes of the
spectrograph), and a further inspection confirmed that the sampling
rate in those regions does not allow our algorithm to find at least
two galaxies inside the volume enclosed in the Phase I cylinder.

We concluded this comparison with XMM-LSS detections inspecting the
relation between optical and X-ray properties of the four groups for
which there exists a (possible) XMM counterpart. In particular, we
considered the relation between the X-ray luminosity $L_{X}$ presented
in Table 5 of \cite{pierre2006} and the velocity dispersions $\sigma$
we have measured. We verified that groups XLSSC 013, $b$ and $c$ have
a $\sigma$-$L_{X}$ relation well in agreement with the linear fit in
the plane $\sigma$-$L_{X}$ presented in Figure 13 of
\cite{popesso2005_III}. For group XLSSC 025 we measure a $\sigma$ that
would be too low for its quoted $L_{X}$, according to the shown
relation, but as its $\sigma$ is of the order of 200 km/s it does not
reside in the $\sigma$ range that we consider reliably measured.

It is worth noticing that our richest groups (10 groups with at least 7
members) do not match with XMM-LSS clusters, except one that is the
counterpart of the X-ray selected group \emph{c}. There are at least
three reasons why an optical group may have not been detected in
X-ray: a) it may falls on the boundaries of a XMM-LSS pointing, thus
in a region where the X-ray detector is affected by vignetting;
b) it may have a redshift much higher than the mean $z$ reachable by
the performed X-ray observations; c) it may have a low surface
brightness, that corresponds to a shallow potential well of the mass
distribution, thus making X-ray detection more difficult. We inspected
our richest groups, and we found that all of them fall in at least one
of these three categories. In particular, we verified that the $N(z)$
distribution of all X-ray clusters in the above-cited works is peaked
at $z\sim 0.4$, while the $N(z)$ distribution of our richest groups is
quite flat and reaches $z\sim0.9$, with 5 groups with
$z\geq0.7$. Moreover, at least half of our richest groups do not have
a dominant member, that is a galaxy with luminosity and/or stellar
mass much higher than the others. The VVDS-02h field sampling rate
could be enough to explain this lacking of dominant galaxies, but in
principle we can not reject the hypothesis that the dominant galaxy in
(some of) these groups may not exist, and in this latter case we are
allowed to think that these groups have a real low X-ray surface
brightness.

We compared our group catalogue also with the ones in
\cite{gavazzir2007}, \cite{olsen2007} and \cite{mazure2007}.

Among the about 20 clusters in \cite{olsen2007} that are inside the
sky area and redshift range that we have explored, roughly half fall
inside regions with too low sampling rate for our Phase I cylinder to
be able to detect at least a pair; two of them (ID 30 and 42) fall
very near in redshift to two wide structure at $z\sim0.32$ and
$z\sim0.45$ , within which our algorithm detects (possibly fragmenting
them) a few groups. Finally, considering the depth of the redshift bins in which
Olsen's groups can reside ($\Delta z \sim 0.1$) due to the use of
photometric redshifts, we find that for 5 groups in Olsen's catalogue
there exists a counterpart in our catalogue.

Among the about 30 structures detected by \cite{mazure2007} in the
redshift range $0.2<z<1.0$, we find that about 20 fall inside regions
with too low sampling rate for our finding group algorithm (13 of
which in the 1 pass area); a few of them reside in redshift slices
($z\sim0.3$, $z\sim0.7$ and $z\sim0.9$) where a wide (in ra-dec)
structure is also present, that has been possibly fragmented by our
algorithm, producing in our catalogue more than one
counterpart. Finally, three of the structures detected by
\cite{mazure2007} have a possible direct counterpart in our catalogue
(general ID 5, 19 and 21, see Table 3 in \citealp{mazure2007}).

Finally, the 3 structures detected by \cite{gavazzir2007} and that
fall inside VVDS-02h field are in very low sampling rate areas, thus
in regions where our algorithm did not detect any group.

In this comparison we also took into account that in the CFHTLS data
that have been used in the three above-mentioned works there are
masked sky regions that have not been used for group finding, as it is
shown for example in Figure 1 in \cite{mazure2007} and Figure 9 in
\cite{olsen2007}. We find that $\sim 5$\% of our groups in the range
$0.2 \leq z \leq 1.0$ fall in those regions.  Moreover, we observe
that roughly half of this masked area falls inside the region that in
VVDS-02h field has the highest sampling rate (the central area
highlighted in Figure \ref{VVDS02h_map_passes}), and that this
higher-sampling region covers only $\sim 25$\% of the VVDS-02h
field. Thus, the percentage of our groups falling in the masked areas
increases to $\sim 8$\% for groups with at least 3 members and to 20\%
for our 10 richest groups (those with at least 7 members).

%
%

\section{The U-B colour of group galaxies}\label{gal_group_properties}

Having reconstructed a catalogue of groups at high $z$, we want now to
use it to study the dependence of galaxy properties on environment,
and also its evolution with cosmic time.  More specifically, we aim at
investigating if physical properties of group galaxies are different
from the properties of the entire sample of galaxies, up to $z\sim 1$. Are
the relations that we see in groups at low redshift already present at
$z\sim1$? Is there any unambiguous signature of time evolution in
known scaling relations characterizing galaxies in cluster
environments?  In this paper, we will not carry on an exhaustive
analysis of this topic, that will be possibly the goal of a future
work. The main aim of this paper is to present the VVDS-02h field 
group catalogue and discuss its reliability. So, in this Section we 
simply want to show the potentiality of our group catalogue for 
studies related to environmental effects on galaxy properties on group 
scales.

As we want to investigate the redshift evolution of the properties of
group galaxies, we have to study a group sample homogeneous at all
$z$. We thus require that the groups we use for this analysis have at
least two members brighter than a luminosity limit that allows us to
be complete up to $z=1$. This luminosity limit evolves with redshift.
Following roughly the evolution of $M^{*}$, the characteristic magnitude 
of the luminosity function \citep{ilbert2005}, we set this limit as $M_B \leq -18.9-1.1 z$. Our
`group galaxy' sample is composed by those galaxies, brighter than
this limit, in groups with at least two members brighter than this
limit itself.  Moreover, we define a `total' galaxy sample considering
all galaxies brighter than this limit (including also those in groups).

Once defined the sample, as a first step we studied the fraction of
`blue' galaxies ($f_b$ from now on) in both the group and total
samples, in the range $0.2 \leq z \leq 1.0$. The general blueing of
cluster galaxy population for increasing redshift, first shown by
\cite{bo1978} and \cite{bo1984} and known as the Butcher-Oemler
effect, has been widely confirmed in following studies (see for
example \citealp{margoniner2001_fb, depropris2003,
gerke2007_groupsblue}). Nevertheless, nowadays there is no full
agreement about the origin of this blueing. It can be related to
environmental effects (\eg \citealp{dressler1997}), but it has also
been suggested that it is consistent with the overall ageing of all
galaxies, irrespective of their environment
\citep{andreon2004b,andreon2006}.

According to our criteria, a galaxy is `blue' if it has a colour 
$U-B \leq 1$. This threshold
has been chosen as it corresponds roughly to the minimum (\ie the
green valley) in the bimodal $U-B$ colour distribution. Moreover, this
colour cut has been kept constant at all redshifts as we found that
the green valley colour does not evolve much in the $z$ range
considered. For the computation of the $U$- and $B$-band absolute
magnitudes we refer the reader to \cite{ilbert2005}.

Since our goal is to study $f_b$ as a function of redshift, we 
first verified that the failure rate in redshift measurement does not depend on redshift for
specific galaxy colours. We assigned to each galaxy a ``photometric 
type'' according to the scheme proposed by \cite{zucca2006_VVDS_LF}. The classification 
is carried out by
fitting the Spectral Energy Distribution of galaxies to six templates (four observed spectra, 
\citealp{CWW1980}, and two starburts SEDs, \citealp{BC1993}).
We then proceeded as in \cite{franzetti2007}, by defining a broad bimodal classification.
We considered E/S0 and early spirals as `early type', and late-type
spirals, irregular and starburst types as a `late-type'.  The
relation between this classification scheme and the colour U-B that we use
to compute $f_b$ is monotonic, with bluer
colours being associated to `late type' templates. In particular, our `early type'
population constitutes $>90$\% of the galaxies with $U-B>1$.
We computed the `late type' galaxy
fraction in both our spectroscopic sample and in the photometric
parent catalogue, in three redshift bins in the range $0.1\leq
z\leq1.0$ (using photometric redshifts for the parent catalogue, see Section  \ref{real_data} for their
determination). As already found by \cite{franzetti2007}, who carried on a similar 
analysis on wider redshift
intervals up to $z\sim2$, the `late type' fraction is $~3$\% higher in the
spectroscopic sample and this increment does not depend on redshift. 
This result implies that any trend of $f_b$ with
redshift is not due to a measurement bias in our sample.

\begin{figure} 
\centering
\includegraphics[width=8cm]{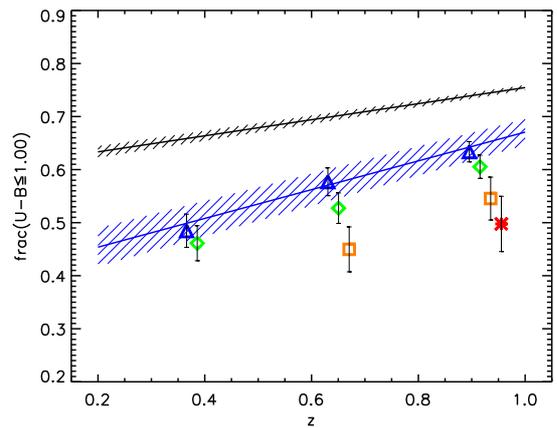} 
\caption{Fraction of
blue galaxies ($U-B \leq 1$)  for 
group galaxies (blue empty triangles) in three different redshift
bins: $0.2 \leq z \leq 0.5$, $0.5 \leq z \leq 0.7$ and $0.7 \leq z
\leq 1.0$. The linear fit of this three 
points is overplotted as a blue line, while the upper black line is the 
linear fit for
$f_b$ computed within the `total' sample. he dashed areas along the 
two linear fits show the locus
where the linear fits could lie considering their 1-$\sigma$ error on both 
intercept and slope. See text for more details. Other
symbols are for group galaxies in groups with
increasing corrected richness:  $Ncorr\geq 7,14,20$ for green diamonds,
orange squares and red stars 
respectively.  } 
\label{group_properties_plot} 
\end{figure}

Figure \ref{group_properties_plot} shows $f_b$ for the group galaxies
(blue triangles) in three different redshift bins: $0.2
\leq z \leq 0.5$, $0.5 \leq z \leq 0.7$ and $0.7 \leq z \leq 1.0$. The
vertical error bars are the $1\sigma$ confidence levels associated to
$f_b$, computed with the usual approximation of the formula for
binomial statistics given in \cite{gehrels86_binomial}: $\sigma^2=f_b
f_r/n$, where $f_r=1-f_b$ and $n$ is the total number of galaxies in
the redshift bin.

As reference, we plot the linear fit of the three points as a
blue line, while the upper black line is the linear fit for $f_b$
computed within the `total' sample. $f_b$ is clearly lower in groups
than in the total sample.  The slopes of the two fits, together with
their $1\sigma$ confidence levels, are $0.27 \pm 0.07$ and $0.15 \pm
0.02$ for the group and total sample respectively. Although they are
both significantly different from zero, and the group slope is
steeper, they are compatible with each other, the group sample slope
being steeper only at a $1.6 \sigma$ significance level. At this stage
of investigation we are only able to confirm the different overall
value of $f_b$ between group galaxies and the total sample, but not
their possible different evolution.  We verified that these
results are stable against the variation of the U-B
threshold adopted to define blue galaxies  (by $\pm 0.05$ mag).  
They did not change significantly also when we implemented a colour cut that
depends on luminosity, following the mild dependence on magnitudes of the green
valley locus. 
Although we do not
detect any redshift dependence of the green valley locus up to $z=1$, we
also allowed the colour cut to vary by 0.1 mag redward for any redshift decrease 
of  $dz=1.0$ (as
suggested by \citealp{blanton2006_blueseq} and as adopted by
\citealp{gerke2007_groupsblue}). As a matter of fact, even in this case
the two slopes are appreciably different from zero, and their
relative difference significant at the $\sim 1.5 \sigma$ level.

We compared our results with those presented by
\cite{gerke2007_groupsblue}, who studied the fraction of blue galaxies
in groups and in the field within different subsamples extracted from
the DEEP2 data set. In their Sample I, the one with a selection most
similar to ours, they find that $f_b$ is lower in groups than in the
field, but they do not detect any significant  evolution of $f_b$ with $z$
neither in groups nor in the field. Anyway, evolutionary effect are 
much more difficult to quantify in that sample since the redshift
range as well as the luminosity range covered is narrower 
with respect to that explored in this study. 

Our results are in agreement with those presented by
\cite{iovino2009_fblue}, who studied the evolution of $f_b$ in groups
and in the field within the zCOSMOS-10k sample (see also, for completeness, 
the analysis of \citealp{kovac2009_morph} concerning the fraction
of early type galaxies in groups). This agreement is based on the comparison  
with their Sample II, the one with a luminosity
cut most similar to ours, and it holds for both the group and the total sample.
Interestingly, \cite{iovino2009_fblue}
find that in their luminosity limited sample galaxy colour still
depends on environment at $z\sim1$ (with a trend similar to what we find in this work), but at
the same redshift they do not converge to the same conclusion when 
the blue fraction is recovered from a (stellar) mass limited sample 
($\log(M/M{\odot}) \geq 10.8$).
They explain this result suggesting that red galaxies of such stellar 
masses, already in place at $z\sim1$, may rise from internal mechanisms of
evolution, on which environment has no influence. We refer the reader to 
\cite{iovino2009_fblue} for more details.

\cite{cucciati2006} carried on a similar analysis using the same
VVDS-02h data set that we use in this work. They studied the
colour-density relation up to $z=1.5$, with the local density computed
within Gaussian filters with $\sigma=5$ h$^{-1}$Mpc. They found that
the colour-density relation becomes weaker for increasing redshift
(the evolution of $f_b$ being faster in high densities), and that at
$z\sim1$ no significant colour-density relation is detected, for
galaxies with $M_B \leq -20$ (that is equivalent to the threshold we
use in this work). Taken at face value, our results are not compatible with
these previous findings, as we find that at $z\sim1$ $f_b$ is
still different in groups and in the total sample. This difference can
be explained with the fact that we are exploring higher
densities/smaller scales ($<1$ $h^{-1}$Mpc, see Section
\ref{real_VVDS_catalogue}). For example, there are several studies in
literature suggesting that environmental effects on large scales are
only a weaker residual of the ones acting on smaller scales (\eg
\citealp{kauffmann2004, blanton2006}). The same hypothesis is
suggested by \cite{cooper2007_col_env}, when comparing the
colour-density relation found in the DEEP2 data set with the one
presented in \cite{cucciati2006}. They still find a colour-density
relation at $z\sim1$, but on smaller scales than those investigated by
\cite{cucciati2006}.

A direct comparison of results obtained on the basis of heterogeneous  
definitions of the local environment (as for example 
density field maxima as opposed to groups) is not straightforward.  \cite{cooper2007_col_env}
showed, for example,  that the evolution of the colour-density relation is
continuous in the range $0.4<z<1.3$, while 
\cite{gerke2007_groupsblue}, who used  the same DEEP2 data set 
but a different definition of environment based on groups,
found that the evolution of 
$f_b$ in groups is flat in the range $0.7<z<1.0$, and it steepens for
$1.0<z<1.3$. Nevertheless, the two works agree on the fact that at
$z\sim1.3$ the colour-density relation seems to disappear.  No need to 
emphasize that the physics associated with different environmental 
definition has still to be fully understood.

Also a direct comparison of our results with those presented in other 
works up to $z\sim1$ is not trivial. Interpretation is hampered
by the the non-homogeneity of group catalogues selected  
according to different selection criteria. As a consequence, the picture 
emerging from these studies is complex and sometimes even contradictory. As 
this Section is meant to give a general idea of the kind of studies that
can be potentially carried out with our group sample, we will not
enter in details. We rather refer the reader to \cite{poggianti2006} and
\cite{andreon2006}  for a more in-depth
discussion about the status of the art and the problems related to
uncontrolled selection effects.

As a second step, we examined the behavior of $f_b$
in groups characterized by different degrees of richness. There is still no
agreement in literature about the dependence of $f_b$ on cluster
properties. For example, $f_b$ is found both to depend on cluster
richness \citep{margoniner2001_fb, goto2003_fb}, or to be independent 
of it, as well as of line of sight velocity dispersion and
mass \citep{Depropris2004,goto2005_fb,popesso2007_VI}. To address this issue, 
we associated to each group the number of members found by the
algorithm. Nevertheless, due to the survey characteristic (sampling
rate, spectra signal to noise ratio, etc.), the observed number has to be
corrected to recover the real number of members within the flux limit 
of the survey ($I_{AB}\leq 24$). We did this by weighting each galaxies  
with the `target sampling rate' and the `spectroscopic success rate' 
of the survey (see \citealp{ilbert2005}). We then
modulated this mean weight with a finer correction which takes into account that
the sampling rate is not uniform in the field. This is done by using the 
$\Psi(\alpha,\delta)$ selection function  described in
\cite{cucciati2006}. This way, for each group we computed a corrected
richness ($N_{c}$), which is the sum of the weights of those galaxies
brighter than the evolving luminosity limit described above.  In
Figure \ref{group_properties_plot},  green diamonds, orange
squares and red stars show $f_b$ within groups with $N_{c} \geq $
7, 14 and 20. We can see in the plot a general trend of
decreasing $f_b$ for increasing $N_{c}$, at any redshift explored.
Given the error bars, this decrement is not significant when
considering single steps in $N_{c}$, but the overall tendency is
clear. Nevertheless, it does not seem that the $f_b$ redshift
evolution is different for different values of $N_{c}$.  These
results are in agreement with those found  by 
\cite{iovino2009_fblue} in their analysis of the zCOSMOS-10k group 
sample, but we extended them to fainter magnitudes.  
It would be indeed interesting to study the dependence on $N_{c}$ also for other
galaxy properties, as it has been done for example for star formation
rate (SFR) and specific SFR (see for example
\citealp{popesso2007_VI}). This could give hints on how different
galaxies properties may be affected by different environments. We
defer this study to a future work.

%
%
\section{Summary and conclusions}\label{Conclusions}

We have compiled a homogeneous catalogue of optical groups
identified in the VVDS-02 field by means of the VDM algorithm, in the range $0.2 \leq z \leq 1.0$.
 
We used mock catalogues simulating the VVDS survey  to 
optimize the performances of  the group-finding algorithm (maximizing 
the completeness and the purity of the resulting group catalogue) as well as to minimize possible selection effects. Our main results are here summarized.

\begin{itemize}

\item[-] Using the mock catalogues, we verified that the
VVDS-02h survey sampling rate allows us to recover at least $50$\% of
the groups (with a virial line of sight velocity
dispersion $\sigma_{vir} \geq 350$ km/s) that are
potentially present in the parent photometric catalogue up to $z=1$.
 
\item[-] We tested how well $\sigma_{vir}$ of the halo mass particles
can be estimated using sparsely sampled galaxy velocities. We verified that 
with this method, given the characteristics of our survey (flux limit, 
sampling rate, redshift measurement error)  we are able to recover a sensible value of 
$\sigma_{vir}$ for $\sigma_{vir} \geq 350$ km/s.

\item[-] Applying the optimized algorithm to the VVDS real data set, we
obtained a catalogue of 318 groups of galaxies (with at least two
members) in the range $0.2 \leq z \leq 1.0$. Among these groups, 63 have a measured 
line of sight velocity dispersion greater than 350 km/s. The group catalogue 
is characterized by an overall
completeness of $\sim60$\% and a purity of $\sim50$\%. 
Nearly $19$\% of the total population of galaxies live in these systems.

\item[-] the number density distribution as a function of both redshift ($n(z)$) and 
velocity dispersion ($n(\sigma)$) of the VVDS groups with $\sigma>350$ km/s
scales in qualitative agreement with the analogous statistics recovered from the
mock catalogues. 

\item[-] We studied the fraction $f_b$ of blue galaxies ($U-B \leq
1$) in the range $0.2\leq z \leq 1$. We used a luminosity-limited subsample 
of galaxies extracted from our data ($M_B \leq -18.9-1.1 z$), complete up to $z=1$. 
We found that $f_b$  is significantly
lower in groups than in the global galaxy population. Moreover, $f_b$
increases as a function of  redshift irrespectively of the environment, with 
marginal evidence for a faster growth rate in groups.  We also
analysed how $f_b$ varies as a function of group richness, finding that, 
at any  redshift explored, $f_b$ decreases in systems with increasing richness.
\end{itemize}

Further explorations of the properties of VVDS groups is left to future
works. We only anticipate that the high degree of completeness of the catalogue
can be potentially exploited for extracting cosmological information via, 
for example, cluster counts techniques. The high level of purity  makes the
VVDS group sample  ideal also for astrophysical studies which aim at tracing 
various physical properties of galaxies as a function of local density and environment. 
We also mention that the cross-correlation studies of 
our optically-selected catalogue with samples inferred in the same field  
with independent techniques will help to gain insights 
not only on cluster selection biases but also on the physics at work 
within these extreme environments.


\begin{acknowledgements}

OC thanks Stefano Andreon for stimulating discussions. We thank the
referee for helpful comments. This research has been developed within
the framework of the VVDS consortium and it has been partially
supported by the CNRS-INSU and its Programme National de Cosmologie
(France), by the Italian Ministry (MIUR) grants COFIN2000 (MM02037133)
and COFIN2003 (num.2003020150) and by PRIN-INAF 2005 (CRA
1.06.08.10). The VLT-VIMOS observations have been carried out on
guaranteed time (GTO) allocated by the European Southern Observatory
(ESO) to the VIRMOS consortium, under a contractual agreement between
the Centre National de la Recherche Scientifique of France, heading a
consortium of French and Italian institutes, and ESO, to design,
manufacture and test the VIMOS instrument. A. Pollo  also acknowledges 
financial support from the Polish Ministry of Science,  
grant PBZ/MNiSW/07/2006/34A.
  
Based on observations obtained with MegaPrime/MegaCam, a joint project
of CFHT and CEA/DAPNIA, at the Canada-France-Hawaii Telescope (CFHT)
which is operated by the National Research Council (NRC) of Canada,
the Institut National des Science de l'Univers of the Centre National
de la Recherche Scientifique (CNRS) of France, and the University of
Hawaii. This work is based in part on data products produced at
TERAPIX and the Canadian Astronomy Data Centre as part of the
Canada-France-Hawaii Telescope Legacy Survey, a collaborative project
of NRC and CNRS
  
The Millennium Simulation databases used in this paper and the web
application providing online access to them were constructed as part
of the activities of the German Astrophysical Virtual Observatory.
  
\end{acknowledgements}




\bibliographystyle{aa}
\bibliography{biblio}

\begin{thebibliography}{111}
\expandafter\ifx\csname natexlab\endcsname\relax\def\natexlab#1{#1}\fi

\bibitem[{{Abell}(1958)}]{abell1958}
{Abell}, G.~O. 1958, \apjs, 3, 211

\bibitem[{{Adami} {et~al.}(2005){Adami}, {Mazure}, {Ilbert}, {Cappi},
  {Bottini}, {Garilli}, {Le Brun}, {Le F{\`e}vre}, {Maccagni}, {Picat},
  {Scaramella}, {Scodeggio}, {Tresse}, {Vettolani}, {Zanichelli}, {Arnaboldi},
  {Arnouts}, {Bardelli}, {Bolzonella}, {Charlot}, {Ciliegi}, {Contini},
  {Covone}, {Foucaud}, {Franzetti}, {Gavignaud}, {Guzzo}, {Iovino}, {Lauger},
  {McCracken}, {Marano}, {Marinoni}, {Meneux}, {Merighi}, {Paltani},
  {Pell{\`o}}, {Pollo}, {Pozzetti}, {Radovich}, {Zamorani}, {Zucca}, {Bondi},
  {Bongiorno}, {Busarello}, {Gregorini}, {Mathez}, {Mellier}, {Merluzzi},
  {Ripepi}, \& {Rizzo}}]{adami2005_CDFS}
{Adami}, C., {Mazure}, A., {Ilbert}, O., {et~al.} 2005, \aap, 443, 805

\bibitem[{{Allen} {et~al.}(2002){Allen}, {Schmidt}, \& {Fabian}}]{allen2002}
{Allen}, S.~W., {Schmidt}, R.~W., \& {Fabian}, A.~C. 2002, \mnras, 334, L11

\bibitem[{{Andreon} {et~al.}(2004{\natexlab{a}}){Andreon}, {Lobo}, \&
  {Iovino}}]{andreon2004b}
{Andreon}, S., {Lobo}, C., \& {Iovino}, A. 2004{\natexlab{a}}, \mnras, 349, 889

\bibitem[{{Andreon} {et~al.}(2009){Andreon}, {Maughan}, {Trinchieri}, \&
  {Kurk}}]{andreon2008_z19}
{Andreon}, S., {Maughan}, B., {Trinchieri}, G., \& {Kurk}, J. 2009, \aap, 507,
  147

\bibitem[{{Andreon} {et~al.}(2006){Andreon}, {Quintana}, {Tajer}, {Galaz}, \&
  {Surdej}}]{andreon2006}
{Andreon}, S., {Quintana}, H., {Tajer}, M., {Galaz}, G., \& {Surdej}, J. 2006,
  \mnras, 365, 915

\bibitem[{{Andreon} {et~al.}(2005){Andreon}, {Valtchanov}, {Jones}, {Altieri},
  {Bremer}, {Willis}, {Pierre}, \& {Quintana}}]{andreon2005}
{Andreon}, S., {Valtchanov}, I., {Jones}, L.~R., {et~al.} 2005, \mnras, 359,
  1250

\bibitem[{{Andreon} {et~al.}(2004{\natexlab{b}}){Andreon}, {Willis},
  {Quintana}, {Valtchanov}, {Pierre}, \& {Pacaud}}]{andreon2004}
{Andreon}, S., {Willis}, J., {Quintana}, H., {et~al.} 2004{\natexlab{b}},
  \mnras, 353, 353

\bibitem[{{Bahcall}(1981)}]{bahcall1981_Msigma}
{Bahcall}, N.~A. 1981, \apj, 247, 787

\bibitem[{Barber {et~al.}(1996)Barber, Dobkin, \&
  Huhdanpaa}]{barber1996quickhull}
Barber, C.~B., Dobkin, D.~P., \& Huhdanpaa, H. 1996, ACM Transactions on
  Mathematical Software, 22, 469

\bibitem[{{Beers} {et~al.}(1990){Beers}, {Flynn}, \& {Gebhardt}}]{beers1990}
{Beers}, T.~C., {Flynn}, K., \& {Gebhardt}, K. 1990, \aj, 100, 32

\bibitem[{{Berlind} {et~al.}(2006){Berlind}, {Frieman}, {Weinberg}, {Blanton},
  {Warren}, {Abazajian}, {Scranton}, {Hogg}, {Scoccimarro}, {Bahcall},
  {Brinkmann}, {Gott}, {Kleinman}, {Krzesinski}, {Lee}, {Miller}, {Nitta},
  {Schneider}, {Tucker}, \& {Zehavi}}]{berlind2006}
{Berlind}, A.~A., {Frieman}, J., {Weinberg}, D.~H., {et~al.} 2006, \apjs, 167,
  1

\bibitem[{{Binney} \& {Tremaine}(1988)}]{binney_tremaine1988}
{Binney}, J. \& {Tremaine}, S. 1988, {Galactic Dynamics (Princeton Univiversity
  Press)}

\bibitem[{{Blaizot} {et~al.}(2005){Blaizot}, {Wadadekar}, {Guiderdoni},
  {Colombi}, {Bertin}, {Bouchet}, {Devriendt}, \& {Hatton}}]{Blaizot2005}
{Blaizot}, J., {Wadadekar}, Y., {Guiderdoni}, B., {et~al.} 2005, \mnras, 360,
  159

\bibitem[{{Blanton}(2006)}]{blanton2006_blueseq}
{Blanton}, M.~R. 2006, \apj, 648, 268

\bibitem[{{Blanton} {et~al.}(2006){Blanton}, {Eisenstein}, {Hogg}, \&
  {Zehavi}}]{blanton2006}
{Blanton}, M.~R., {Eisenstein}, D., {Hogg}, D.~W., \& {Zehavi}, I. 2006, \apj,
  645, 977

\bibitem[{{Borgani} {et~al.}(1997){Borgani}, {Gardini}, {Girardi}, \&
  {Gottlober}}]{borgani1997}
{Borgani}, S., {Gardini}, A., {Girardi}, M., \& {Gottlober}, S. 1997, New
  Astronomy, 2, 119

\bibitem[{{Borgani} {et~al.}(1999){Borgani}, {Girardi}, {Carlberg}, {Yee}, \&
  {Ellingson}}]{borgani1999}
{Borgani}, S., {Girardi}, M., {Carlberg}, R.~G., {Yee}, H.~K.~C., \&
  {Ellingson}, E. 1999, \apj, 527, 561

\bibitem[{{Bremer} {et~al.}(2006){Bremer}, {Valtchanov}, {Willis}, {Altieri},
  {Andreon}, {Duc}, {Fang}, {Jean}, {Lonsdale}, {Pacaud}, {Pierre}, {Shupe},
  {Surace}, \& {Waddington}}]{bremer2006}
{Bremer}, M.~N., {Valtchanov}, I., {Willis}, J., {et~al.} 2006, \mnras, 371,
  1427

\bibitem[{{Bruzual A.} \& {Charlot}(1993)}]{BC1993}
{Bruzual A.}, G. \& {Charlot}, S. 1993, \apj, 405, 538

\bibitem[{{Butcher} \& {Oemler}(1984)}]{bo1984}
{Butcher}, H. \& {Oemler}, A. 1984, \apj, 285, 426

\bibitem[{{Butcher} \& {Oemler}(1978)}]{bo1978}
{Butcher}, H. \& {Oemler}, Jr., A. 1978, \apj, 226, 559

\bibitem[{{Carlberg} {et~al.}(2001){Carlberg}, {Yee}, {Morris}, {Lin}, {Hall},
  {Patton}, {Sawicki}, \& {Shepherd}}]{carlberg2001}
{Carlberg}, R.~G., {Yee}, H.~K.~C., {Morris}, S.~L., {et~al.} 2001, \apj, 552,
  427

\bibitem[{{Coleman} {et~al.}(1980){Coleman}, {Wu}, \& {Weedman}}]{CWW1980}
{Coleman}, G.~D., {Wu}, C., \& {Weedman}, D.~W. 1980, \apjs, 43, 393

\bibitem[{{Cooper} {et~al.}(2007){Cooper}, {Newman}, {Coil}, {Croton}, {Gerke},
  {Yan}, {Davis}, {Faber}, {Guhathakurta}, {Koo}, {Weiner}, \&
  {Willmer}}]{cooper2007_col_env}
{Cooper}, M.~C., {Newman}, J.~A., {Coil}, A.~L., {et~al.} 2007, \mnras, 376,
  1445

\bibitem[{{Cucciati} {et~al.}(2006){Cucciati}, {Iovino}, {Marinoni}, {Ilbert},
  {Bardelli}, {Franzetti}, {Le F{\`e}vre}, {Pollo}, {Zamorani}, {Cappi},
  {Guzzo}, {McCracken}, {Meneux}, {Scaramella}, {Scodeggio}, {Tresse}, {Zucca},
  {Bottini}, {Garilli}, {Le Brun}, {Maccagni}, {Picat}, {Vettolani},
  {Zanichelli}, {Adami}, {Arnaboldi}, {Arnouts}, {Bolzonella}, {Charlot},
  {Ciliegi}, {Contini}, {Foucaud}, {Gavignaud}, {Marano}, {Mazure}, {Merighi},
  {Paltani}, {Pell{\`o}}, {Pozzetti}, {Radovich}, {Bondi}, {Bongiorno},
  {Busarello}, {de La Torre}, {Gregorini}, {Lamareille}, {Mathez}, {Mellier},
  {Merluzzi}, {Ripepi}, {Rizzo}, {Temporin}, \& {Vergani}}]{cucciati2006}
{Cucciati}, O., {Iovino}, A., {Marinoni}, C., {et~al.} 2006, \aap, 458, 39

\bibitem[{{Davis} {et~al.}(2003){Davis}, {Faber}, {Newman}, {Phillips},
  {Ellis}, {Steidel}, {Conselice}, {Coil}, {Finkbeiner}, {Koo}, {Guhathakurta},
  {Weiner}, {Schiavon}, {Willmer}, {Kaiser}, {Luppino}, {Wirth}, {Connolly},
  {Eisenhardt}, {Cooper}, \& {Gerke}}]{Davis2003}
{Davis}, M., {Faber}, S.~M., {Newman}, J., {et~al.} 2003, in Discoveries and
  Research Prospects from 6- to 10-Meter-Class Telescopes II. Edited by
  Guhathakurta, Puragra. Proceedings of the SPIE, Volume 4834, pp. 161-172
  (2003)., ed. P.~{Guhathakurta}, 161--172

\bibitem[{{De Lucia} \& {Blaizot}(2007)}]{delucia_blaizot2007}
{De Lucia}, G. \& {Blaizot}, J. 2007, \mnras, 375, 2

\bibitem[{{De Propris} {et~al.}(2004){De Propris}, {Colless}, {Peacock},
  {Couch}, {Driver}, {Balogh}, {Baldry}, {Baugh}, {Bland-Hawthorn}, {Bridges},
  {Cannon}, {Cole}, {Collins}, {Cross}, {Dalton}, {Efstathiou}, {Ellis},
  {Frenk}, {Glazebrook}, {Hawkins}, {Jackson}, {Lahav}, {Lewis}, {Lumsden},
  {Maddox}, {Madgwick}, {Norberg}, {Percival}, {Peterson}, {Sutherland}, \&
  {Taylor}}]{Depropris2004}
{De Propris}, R., {Colless}, M., {Peacock}, J.~A., {et~al.} 2004, \mnras, 351,
  125

\bibitem[{{De Propris} {et~al.}(2003){De Propris}, {Stanford}, {Eisenhardt}, \&
  {Dickinson}}]{depropris2003}
{De Propris}, R., {Stanford}, S.~A., {Eisenhardt}, P.~R., \& {Dickinson}, M.
  2003, \apj, 598, 20

\bibitem[{{Delaunay}(1934)}]{delaunay1934}
{Delaunay}, B. 1934, Bull. Acad. Sci. USSR, 7, 793

\bibitem[{{Donahue} {et~al.}(2002){Donahue}, {Scharf}, {Mack}, {Lee},
  {Postman}, {Rosati}, {Dickinson}, {Voit}, \& {Stocke}}]{donahue2002_Xrayopt}
{Donahue}, M., {Scharf}, C.~A., {Mack}, J., {et~al.} 2002, \apj, 569, 689

\bibitem[{{Dressler}(1980)}]{dressler1980}
{Dressler}, A. 1980, \apj, 236, 351

\bibitem[{{Dressler} {et~al.}(1997){Dressler}, {Oemler}, {Couch}, {Smail},
  {Ellis}, {Barger}, {Butcher}, {Poggianti}, \& {Sharples}}]{dressler1997}
{Dressler}, A., {Oemler}, A.~J., {Couch}, W.~J., {et~al.} 1997, \apj, 490, 577

\bibitem[{{Eke} {et~al.}(2004){Eke}, {Baugh}, {Cole}, {Frenk}, {Norberg},
  {Peacock}, {Baldry}, {Bland-Hawthorn}, {Bridges}, {Cannon}, {Colless},
  {Collins}, {Couch}, {Dalton}, {de Propris}, {Driver}, {Efstathiou}, {Ellis},
  {Glazebrook}, {Jackson}, {Lahav}, {Lewis}, {Lumsden}, {Maddox}, {Madgwick},
  {Peterson}, {Sutherland}, \& {Taylor}}]{eke2004}
{Eke}, V.~R., {Baugh}, C.~M., {Cole}, S., {et~al.} 2004, \mnras, 348, 866

\bibitem[{{Ettori} {et~al.}(2009){Ettori}, {Morandi}, {Tozzi}, {Balestra},
  {Borgani}, {Rosati}, {Lovisari}, \& {Terenziani}}]{ettori2009}
{Ettori}, S., {Morandi}, A., {Tozzi}, P., {et~al.} 2009, \aap, 501, 61

\bibitem[{{Ettori} {et~al.}(2003){Ettori}, {Tozzi}, \& {Rosati}}]{ettori2003}
{Ettori}, S., {Tozzi}, P., \& {Rosati}, P. 2003, \aap, 398, 879

\bibitem[{{Finoguenov} {et~al.}(2010){Finoguenov}, {Watson}, {Tanaka},
  {Simpson}, {Cirasuolo}, {Dunlop}, {Peacock}, {Farrah}, {Akiyama}, {Ueda},
  {Smol{\v c}i{\'c}}, {Stewart}, {Rawlings}, {van Breukelen}, {Almaini},
  {Clewley}, {Bonfield}, {Jarvis}, {Barr}, {Foucaud}, {McLure}, {Sekiguchi}, \&
  {Egami}}]{finoguenov2010}
{Finoguenov}, A., {Watson}, M.~G., {Tanaka}, M., {et~al.} 2010, \mnras, 403,
  2063

\bibitem[{{Franzetti} {et~al.}(2007){Franzetti}, {Scodeggio}, {Garilli},
  {Vergani}, {Maccagni}, {Guzzo}, {Tresse}, {Ilbert}, {Lamareille}, {Contini},
  {Le F{\`e}vre}, {Zamorani}, {Brinchmann}, {Charlot}, {Bottini}, {Le Brun},
  {Picat}, {Scaramella}, {Vettolani}, {Zanichelli}, {Adami}, {Arnouts},
  {Bardelli}, {Bolzonella}, {Cappi}, {Ciliegi}, {Foucaud}, {Gavignaud},
  {Iovino}, {McCracken}, {Marano}, {Marinoni}, {Mazure}, {Meneux}, {Merighi},
  {Paltani}, {Pell{\`o}}, {Pollo}, {Pozzetti}, {Radovich}, {Zucca}, {Cucciati},
  \& {Walcher}}]{franzetti2007}
{Franzetti}, P., {Scodeggio}, M., {Garilli}, B., {et~al.} 2007, \aap, 465, 711

\bibitem[{{Garilli} {et~al.}(1999){Garilli}, {Maccagni}, \&
  {Andreon}}]{garilli1999}
{Garilli}, B., {Maccagni}, D., \& {Andreon}, S. 1999, \aap, 342, 408

\bibitem[{{Gavazzi} {et~al.}(2009){Gavazzi}, {Adami}, {Durret}, {Cuillandre},
  {Ilbert}, {Mazure}, {Pell{\'o}}, \& {Ulmer}}]{gavazzi2009}
{Gavazzi}, R., {Adami}, C., {Durret}, F., {et~al.} 2009, \aap, 498, L33

\bibitem[{{Gavazzi} \& {Soucail}(2007)}]{gavazzir2007}
{Gavazzi}, R. \& {Soucail}, G. 2007, \aap, 462, 459

\bibitem[{{Gehrels}(1986)}]{gehrels86_binomial}
{Gehrels}, N. 1986, \apj, 303, 336

\bibitem[{{Gerke} {et~al.}(2005){Gerke}, {Newman}, {Davis}, {Marinoni}, {Yan},
  {Coil}, {Conroy}, {Cooper}, {Faber}, {Finkbeiner}, {Guhathakurta}, {Kaiser},
  {Koo}, {Phillips}, {Weiner}, \& {Willmer}}]{gerke2005_groups}
{Gerke}, B.~F., {Newman}, J.~A., {Davis}, M., {et~al.} 2005, \apj, 625, 6

\bibitem[{{Gerke} {et~al.}(2007){Gerke}, {Newman}, {Faber}, {Cooper}, {Croton},
  {Davis}, {Willmer}, {Yan}, {Coil}, {Guhathakurta}, {Koo}, \&
  {Weiner}}]{gerke2007_groupsblue}
{Gerke}, B.~F., {Newman}, J.~A., {Faber}, S.~M., {et~al.} 2007, \mnras, 376,
  1425

\bibitem[{{Gilbank} {et~al.}(2004){Gilbank}, {Bower}, {Castander}, \&
  {Ziegler}}]{gilbank2004_Xrayopt}
{Gilbank}, D.~G., {Bower}, R.~G., {Castander}, F.~J., \& {Ziegler}, B.~L. 2004,
  \mnras, 348, 551

\bibitem[{{Girardi} {et~al.}(1993){Girardi}, {Biviano}, {Giuricin},
  {Mardirossian}, \& {Mezzetti}}]{girardi1993_gapper}
{Girardi}, M., {Biviano}, A., {Giuricin}, G., {Mardirossian}, F., \&
  {Mezzetti}, M. 1993, \apj, 404, 38

\bibitem[{{Girardi} {et~al.}(2000){Girardi}, {Borgani}, {Giuricin},
  {Mardirossian}, \& {Mezzetti}}]{girardi2000}
{Girardi}, M., {Borgani}, S., {Giuricin}, G., {Mardirossian}, F., \&
  {Mezzetti}, M. 2000, \apj, 530, 62

\bibitem[{{Giuricin} {et~al.}(2001){Giuricin}, {Samurovi{\'c}}, {Girardi},
  {Mezzetti}, \& {Marinoni}}]{giuricin2001}
{Giuricin}, G., {Samurovi{\'c}}, S., {Girardi}, M., {Mezzetti}, M., \&
  {Marinoni}, C. 2001, \apj, 554, 857

\bibitem[{{Gladders} \& {Yee}(2000)}]{gladders2000}
{Gladders}, M.~D. \& {Yee}, H.~K.~C. 2000, \aj, 120, 2148

\bibitem[{{Goto}(2005)}]{goto2005_fb}
{Goto}, T. 2005, \mnras, 356, L6

\bibitem[{{Goto} {et~al.}(2003){Goto}, {Okamura}, {Yagi}, {Sheth}, {Bahcall},
  {Zabel}, {Crouch}, {Sekiguchi}, {Annis}, {Bernardi}, {Chong}, {G{\'o}mez},
  {Hansen}, {Kim}, {Knudson}, {McKay}, \& {Miller}}]{goto2003_fb}
{Goto}, T., {Okamura}, S., {Yagi}, M., {et~al.} 2003, \pasj, 55, 739

\bibitem[{{Hansen} {et~al.}(2005){Hansen}, {McKay}, {Wechsler}, {Annis},
  {Sheldon}, \& {Kimball}}]{hansen2005}
{Hansen}, S.~M., {McKay}, T.~A., {Wechsler}, R.~H., {et~al.} 2005, \apj, 633,
  122

\bibitem[{{Huchra} \& {Geller}(1982)}]{huchra1982h}
{Huchra}, J.~P. \& {Geller}, M.~J. 1982, \apj, 257, 423

\bibitem[{{Ilbert} {et~al.}(2006){Ilbert}, {Arnouts}, {McCracken},
  {Bolzonella}, {Bertin}, {Le F{\`e}vre}, {Mellier}, {Zamorani}, {Pell{\`o}},
  {Iovino}, {Tresse}, {Le Brun}, {Bottini}, {Garilli}, {Maccagni}, {Picat},
  {Scaramella}, {Scodeggio}, {Vettolani}, {Zanichelli}, {Adami}, {Bardelli},
  {Cappi}, {Charlot}, {Ciliegi}, {Contini}, {Cucciati}, {Foucaud}, {Franzetti},
  {Gavignaud}, {Guzzo}, {Marano}, {Marinoni}, {Mazure}, {Meneux}, {Merighi},
  {Paltani}, {Pollo}, {Pozzetti}, {Radovich}, {Zucca}, {Bondi}, {Bongiorno},
  {Busarello}, {de La Torre}, {Gregorini}, {Lamareille}, {Mathez}, {Merluzzi},
  {Ripepi}, {Rizzo}, \& {Vergani}}]{ilbert2006zphot}
{Ilbert}, O., {Arnouts}, S., {McCracken}, H.~J., {et~al.} 2006, \aap, 457, 841

\bibitem[{{Ilbert} {et~al.}(2005){Ilbert}, {Tresse}, {Zucca}, {Bardelli},
  {Arnouts}, {Zamorani}, {Pozzetti}, {Bottini}, {Garilli}, {Le Brun}, {Le
  F{\`e}vre}, {Maccagni}, {Picat}, {Scaramella}, {Scodeggio}, {Vettolani},
  {Zanichelli}, {Adami}, {Arnaboldi}, {Bolzonella}, {Cappi}, {Charlot},
  {Contini}, {Foucaud}, {Franzetti}, {Gavignaud}, {Guzzo}, {Iovino},
  {McCracken}, {Marano}, {Marinoni}, {Mathez}, {Mazure}, {Meneux}, {Merighi},
  {Paltani}, {Pello}, {Pollo}, {Radovich}, {Bondi}, {Bongiorno}, {Busarello},
  {Ciliegi}, {Lamareille}, {Mellier}, {Merluzzi}, {Ripepi}, \&
  {Rizzo}}]{ilbert2005}
{Ilbert}, O., {Tresse}, L., {Zucca}, E., {et~al.} 2005, \aap, 439, 863

\bibitem[{{Iovino} {et~al.}(2010){Iovino}, {Cucciati}, {Scodeggio}, {Knobel},
  {Kova{\v c}}, {Lilly}, {Bolzonella}, {Tasca}, {Zamorani}, {Zucca}, {Caputi},
  {Pozzetti}, {Oesch}, {Lamareille}, {Halliday}, {Bardelli}, {Finoguenov},
  {Guzzo}, {Kampczyk}, {Maier}, {Tanaka}, {Vergani}, {Carollo}, {Contini},
  {Kneib}, {Le F{\`e}vre}, {Mainieri}, {Renzini}, {Bongiorno}, {Coppa}, {de La
  Torre}, {de Ravel}, {Franzetti}, {Garilli}, {Le Borgne}, {Le Brun},
  {Mignoli}, {Pell{\`o}}, {Peng}, {Perez-Montero}, {Ricciardelli}, {Silverman},
  {Tresse}, {Abbas}, {Bottini}, {Cappi}, {Cassata}, {Cimatti}, {Koekemoer},
  {Leauthaud}, {Maccagni}, {Marinoni}, {McCracken}, {Memeo}, {Meneux},
  {Porciani}, {Scaramella}, {Schiminovich}, \& {Scoville}}]{iovino2009_fblue}
{Iovino}, A., {Cucciati}, O., {Scodeggio}, M., {et~al.} 2010, \aap, 509, A40+

\bibitem[{{Iovino} {et~al.}(2005){Iovino}, {McCracken}, {Garilli}, {Foucaud},
  {Le F{\`e}vre}, {Maccagni}, {Saracco}, {Bardelli}, {Busarello}, {Scodeggio},
  {Zanichelli}, {Paioro}, {Bottini}, {Le Brun}, {Picat}, {Scaramella},
  {Tresse}, {Vettolani}, {Adami}, {Arnaboldi}, {Arnouts}, {Bolzonella},
  {Cappi}, {Charlot}, {Ciliegi}, {Contini}, {Franzetti}, {Gavignaud}, {Guzzo},
  {Ilbert}, {Marano}, {Marinoni}, {Mazure}, {Meneux}, {Merighi}, {Paltani},
  {Pell{\`o}}, {Pollo}, {Pozzetti}, {Radovich}, {Zamorani}, {Zucca}, {Bertin},
  {Bondi}, {Bongiorno}, {Cucciati}, {Gregorini}, {Mathez}, {Mellier},
  {Merluzzi}, {Ripepi}, \& {Rizzo}}]{iovino2005}
{Iovino}, A., {McCracken}, H.~J., {Garilli}, B., {et~al.} 2005, \aap, 442, 423

\bibitem[{{Kauffmann} {et~al.}(2004){Kauffmann}, {White}, {Heckman},
  {M{\'e}nard}, {Brinchmann}, {Charlot}, {Tremonti}, \&
  {Brinkmann}}]{kauffmann2004}
{Kauffmann}, G., {White}, S.~D.~M., {Heckman}, T.~M., {et~al.} 2004, \mnras,
  353, 713

\bibitem[{{Kepner} {et~al.}(1999){Kepner}, {Fan}, {Bahcall}, {Gunn}, {Lupton},
  \& {Xu}}]{kepner1999}
{Kepner}, J., {Fan}, X., {Bahcall}, N., {et~al.} 1999, \apj, 517, 78

\bibitem[{{Kneib} {et~al.}(2003){Kneib}, {Hudelot}, {Ellis}, {Treu}, {Smith},
  {Marshall}, {Czoske}, {Smail}, \& {Natarajan}}]{kneib2003}
{Kneib}, J., {Hudelot}, P., {Ellis}, R.~S., {et~al.} 2003, \apj, 598, 804

\bibitem[{{Knobel} {et~al.}(2009){Knobel}, {Lilly}, {Iovino}, {Porciani},
  {Kova{\v c}}, {Cucciati}, {Finoguenov}, {Kitzbichler}, {Carollo}, {Contini},
  {Kneib}, {LeF{\`e}vre}, {Mainieri}, {Renzini}, {Scodeggio}, {Zamorani},
  {Bardelli}, {Bolzonella}, {Bongiorno}, {Caputi}, {Coppa}, {de la Torre}, {de
  Ravel}, {Franzetti}, {Garilli}, {Kampczyk}, {Lamareille}, {LeBorgne},
  {LeBrun}, {Maier}, {Mignoli}, {Pello}, {Peng}, {Montero}, {Ricciardelli},
  {Silverman}, {Tanaka}, {Tasca}, {Tresse}, {Vergani}, {Zucca}, {Abbas},
  {Bottini}, {Cappi}, {Cassata}, {Cimatti}, {Fumana}, {Guzzo}, {Koekemoer},
  {Leauthaud}, {Maccagni}, {Marinoni}, {McCracken}, {Memeo}, {Meneux}, {Oesch},
  {Pozzetti}, \& {Scaramella}}]{knobel2009_groups}
{Knobel}, C., {Lilly}, S.~J., {Iovino}, A., {et~al.} 2009, \apj, 697, 1842

\bibitem[{{Koester} {et~al.}(2007){Koester}, {McKay}, {Annis}, {Wechsler},
  {Evrard}, {Rozo}, {Bleem}, {Sheldon}, \& {Johnston}}]{koester2007}
{Koester}, B.~P., {McKay}, T.~A., {Annis}, J., {et~al.} 2007, \apj, 660, 221

\bibitem[{{Kovac} {et~al.}(2009){Kovac}, {Lilly}, {Knobel}, {Bolzonella},
  {Iovino}, {Carollo}, {Scarlata}, {Sargent}, {Cucciati}, {Zamorani},
  {Pozzetti}, {Tasca}, {Scodeggio}, {Kampczyk}, {Peng}, {Oesch}, {Zucca},
  {Finoguenov}, {Contini}, {Kneib}, {Le Fevre}, {Mainieri}, {Renzini},
  {Bardelli}, {Bongiorno}, {Caputi}, {Coppa}, {de la Torre}, {de Ravel},
  {Franzetti}, {Garilli}, {Lamareille}, {Le Borgne}, {Le Brun}, {Maier},
  {Mignoli}, {Pello}, {Perez Montero}, {Ricciardelli}, {Silverman}, {Tanaka},
  {Tresse}, {Vergani}, {Abbas}, {Bottini}, {Cappi}, {Cassata}, {Cimatti},
  {Fumana}, {Guzzo}, {Koekemoer}, {Leauthaud}, {Maccagni}, {Marinoni},
  {McCracken}, {Memeo}, {Meneux}, {Porciani}, {Scaramella}, \&
  {Scoville}}]{kovac2009_morph}
{Kovac}, K., {Lilly}, S.~J., {Knobel}, C., {et~al.} 2009, ArXiv e-prints
  0909.2032

\bibitem[{{Le F{\`e}vre} {et~al.}(2004){Le F{\`e}vre}, {Mellier}, {McCracken},
  {Foucaud}, {Gwyn}, {Radovich}, {Dantel-Fort}, {Bertin}, {Moreau},
  {Cuillandre}, {Pierre}, {Le Brun}, {Mazure}, \& {Tresse}}]{lefevre2004b}
{Le F{\`e}vre}, O., {Mellier}, Y., {McCracken}, H.~J., {et~al.} 2004, \aap,
  417, 839

\bibitem[{{Le F{\`e}vre} {et~al.}(2005){Le F{\`e}vre}, {Vettolani}, {Garilli},
  {Tresse}, {Bottini}, {Le Brun}, {Maccagni}, {Picat}, {Scaramella},
  {Scodeggio}, {Zanichelli}, {Adami}, {Arnaboldi}, {Arnouts}, {Bardelli},
  {Bolzonella}, {Cappi}, {Charlot}, {Ciliegi}, {Contini}, {Foucaud},
  {Franzetti}, {Gavignaud}, {Guzzo}, {Ilbert}, {Iovino}, {McCracken}, {Marano},
  {Marinoni}, {Mathez}, {Mazure}, {Meneux}, {Merighi}, {Paltani}, {Pell{\`o}},
  {Pollo}, {Pozzetti}, {Radovich}, {Zamorani}, {Zucca}, {Bondi}, {Bongiorno},
  {Busarello}, {Lamareille}, {Mellier}, {Merluzzi}, {Ripepi}, \&
  {Rizzo}}]{lefevre2005a}
{Le F{\`e}vre}, O., {Vettolani}, G., {Garilli}, B., {et~al.} 2005, \aap, 439,
  845

\bibitem[{{Ledlow} {et~al.}(2003){Ledlow}, {Voges}, {Owen}, \&
  {Burns}}]{ledlow2003_Xrayopt}
{Ledlow}, M.~J., {Voges}, W., {Owen}, F.~N., \& {Burns}, J.~O. 2003, \aj, 126,
  2740

\bibitem[{{Lemson} \& {Virgo Consortium}(2006)}]{lemson2006_database}
{Lemson}, G. \& {Virgo Consortium}, t. 2006, ArXiv Astrophysics e-prints

\bibitem[{{Lilly} {et~al.}(2007){Lilly}, {F{\`e}vre}, {Renzini}, {Zamorani},
  {Scodeggio}, {Contini}, {Carollo}, {Hasinger}, {Kneib}, {Iovino}, {Le Brun},
  {Maier}, {Mainieri}, {Mignoli}, {Silverman}, {Tasca}, {Bolzonella},
  {Bongiorno}, {Bottini}, {Capak}, {Caputi}, {Cimatti}, {Cucciati}, {Daddi},
  {Feldmann}, {Franzetti}, {Garilli}, {Guzzo}, {Ilbert}, {Kampczyk}, {Kovac},
  {Lamareille}, {Leauthaud}, {Borgne}, {McCracken}, {Marinoni}, {Pello},
  {Ricciardelli}, {Scarlata}, {Vergani}, {Sanders}, {Schinnerer}, {Scoville},
  {Taniguchi}, {Arnouts}, {Aussel}, {Bardelli}, {Brusa}, {Cappi}, {Ciliegi},
  {Finoguenov}, {Foucaud}, {Franceschini}, {Halliday}, {Impey}, {Knobel},
  {Koekemoer}, {Kurk}, {Maccagni}, {Maddox}, {Marano}, {Marconi}, {Meneux},
  {Mobasher}, {Moreau}, {Peacock}, {Porciani}, {Pozzetti}, {Scaramella},
  {Schiminovich}, {Shopbell}, {Smail}, {Thompson}, {Tresse}, {Vettolani},
  {Zanichelli}, \& {Zucca}}]{lilly2007_zcosmos}
{Lilly}, S.~J., {F{\`e}vre}, O.~L., {Renzini}, A., {et~al.} 2007, \apjs, 172,
  70

\bibitem[{{Lilly} {et~al.}(2009){Lilly}, {LeBrun}, {Maier}, {Mainieri},
  {Mignoli}, {Scodeggio}, {Zamorani}, {Carollo}, {Contini}, {Kneib},
  {LeF{\`e}vre}, {Renzini}, {Bardelli}, {Bolzonella}, {Bongiorno}, {Caputi},
  {Coppa}, {Cucciati}, {de la Torre}, {de Ravel}, {Franzetti}, {Garilli},
  {Iovino}, {Kampczyk}, {Kovac}, {Knobel}, {Lamareille}, {LeBorgne}, {Pello},
  {Peng}, {P{\'e}rez-Montero}, {Ricciardelli}, {Silverman}, {Tanaka}, {Tasca},
  {Tresse}, {Vergani}, {Zucca}, {Ilbert}, {Salvato}, {Oesch}, {Abbas},
  {Bottini}, {Capak}, {Cappi}, {Cassata}, {Cimatti}, {Elvis}, {Fumana},
  {Guzzo}, {Hasinger}, {Koekemoer}, {Leauthaud}, {Maccagni}, {Marinoni},
  {McCracken}, {Memeo}, {Meneux}, {Porciani}, {Pozzetti}, {Sanders},
  {Scaramella}, {Scarlata}, {Scoville}, {Shopbell}, \&
  {Taniguchi}}]{lilly2009_zcosmos}
{Lilly}, S.~J., {LeBrun}, V., {Maier}, C., {et~al.} 2009, \apjs, 184, 218

\bibitem[{{Limousin} {et~al.}(2009){Limousin}, {Cabanac}, {Gavazzi}, {Kneib},
  {Motta}, {Richard}, {Thanjavur}, {Foex}, {Pello}, {Crampton}, {Faure},
  {Fort}, {Jullo}, {Marshall}, {Mellier}, {More}, {Soucail}, {Suyu},
  {Swinbank}, {Sygnet}, {Tu}, {Valls-Gabaud}, {Verdugo}, \&
  {Willis}}]{limousin2009}
{Limousin}, M., {Cabanac}, R., {Gavazzi}, R., {et~al.} 2009, \aap, 502, 445

\bibitem[{{Limousin} {et~al.}(2010){Limousin}, {Ebeling}, {Ma}, {Swinbank},
  {Smith}, {Richard}, {Edge}, {Jauzac}, {Kneib}, {Marshall}, \&
  {Schrabback}}]{limousin2010}
{Limousin}, M., {Ebeling}, H., {Ma}, C., {et~al.} 2010, \mnras, 474

\bibitem[{{Margoniner} {et~al.}(2001){Margoniner}, {de Carvalho}, {Gal}, \&
  {Djorgovski}}]{margoniner2001_fb}
{Margoniner}, V.~E., {de Carvalho}, R.~R., {Gal}, R.~R., \& {Djorgovski}, S.~G.
  2001, \apjl, 548, L143

\bibitem[{{Marinoni} {et~al.}(2002){Marinoni}, {Davis}, {Newman}, \&
  {Coil}}]{marinoni2002}
{Marinoni}, C., {Davis}, M., {Newman}, J.~A., \& {Coil}, A.~L. 2002, \apj, 580,
  122

\bibitem[{{Marinoni} \& {Hudson}(2002)}]{marinoni2002_ML}
{Marinoni}, C. \& {Hudson}, M.~J. 2002, \apj, 569, 101

\bibitem[{{Materne}(1978)}]{materne1978}
{Materne}, J. 1978, \aap, 63, 401

\bibitem[{{Mazure} {et~al.}(2007){Mazure}, {Adami}, {Pierre}, {Le F{\`e}vre},
  {Arnouts}, {Duc}, {Ilbert}, {Lebrun}, {Meneux}, {Pacaud}, {Surdej}, \&
  {Valtchanov}}]{mazure2007}
{Mazure}, A., {Adami}, C., {Pierre}, M., {et~al.} 2007, \aap, 467, 49

\bibitem[{{McCracken} {et~al.}(2003){McCracken}, {Radovich}, {Bertin},
  {Mellier}, {Dantel-Fort}, {Le F{\`e}vre}, {Cuillandre}, {Gwyn}, {Foucaud}, \&
  {Zamorani}}]{mccracken2003}
{McCracken}, H.~J., {Radovich}, M., {Bertin}, E., {et~al.} 2003, \aap, 410, 17

\bibitem[{{Meneux} {et~al.}(2008){Meneux}, {Guzzo}, {Garilli}, {Le F{\`e}vre},
  {Pollo}, {Blaizot}, {De Lucia}, {Bolzonella}, {Lamareille}, {Pozzetti},
  {Cappi}, {Iovino}, {Marinoni}, {McCracken}, {de la Torre}, {Bottini}, {Le
  Brun}, {Maccagni}, {Picat}, {Scaramella}, {Scodeggio}, {Tresse}, {Vettolani},
  {Zanichelli}, {Abbas}, {Adami}, {Arnouts}, {Bardelli}, {Bongiorno},
  {Charlot}, {Ciliegi}, {Contini}, {Cucciati}, {Foucaud}, {Franzetti},
  {Gavignaud}, {Ilbert}, {Marano}, {Mazure}, {Merighi}, {Paltani}, {Pell{\`o}},
  {Radovich}, {Vergani}, {Zamorani}, \& {Zucca}}]{meneux2008_sm}
{Meneux}, B., {Guzzo}, L., {Garilli}, B., {et~al.} 2008, \aap, 478, 299

\bibitem[{{Miller} {et~al.}(2005){Miller}, {Nichol}, {Reichart}, {Wechsler},
  {Evrard}, {Annis}, {McKay}, {Bahcall}, {Bernardi}, {Boehringer}, {Connolly},
  {Goto}, {Kniazev}, {Lamb}, {Postman}, {Schneider}, {Sheth}, \&
  {Voges}}]{miller2005}
{Miller}, C.~J., {Nichol}, R.~C., {Reichart}, D., {et~al.} 2005, \aj, 130, 968

\bibitem[{Mirtich(1996)}]{mirtich1996}
Mirtich, B. 1996, journal of graphics tools, 1, 31

\bibitem[{{Morandi} {et~al.}(2010){Morandi}, {Pedersen}, \&
  {Limousin}}]{morandi2010}
{Morandi}, A., {Pedersen}, K., \& {Limousin}, M. 2010, \apj, 713, 491

\bibitem[{{Newman} \& {Davis}(2002)}]{newman2002}
{Newman}, J.~A. \& {Davis}, M. 2002, \apj, 564, 567

\bibitem[{{Oemler}(1974)}]{oemler1974}
{Oemler}, A.~J. 1974, \apj, 194, 1

\bibitem[{{Olsen} {et~al.}(2007){Olsen}, {Benoist}, {Cappi}, {Maurogordato},
  {Mazure}, {Slezak}, {Adami}, {Ferrari}, \& {Martel}}]{olsen2007}
{Olsen}, L.~F., {Benoist}, C., {Cappi}, A., {et~al.} 2007, \aap, 461, 81

\bibitem[{{Pierre} {et~al.}(2006){Pierre}, {Pacaud}, {Duc}, {Willis},
  {Andreon}, {Valtchanov}, {Altieri}, {Galaz}, {Gueguen}, {Le F{\`e}vre},
  {F{\`e}vre}, {Ponman}, {Sprimont}, {Surdej}, {Adami}, {Alshino}, {Bremer},
  {Chiappetti}, {Detal}, {Garcet}, {Gosset}, {Jean}, {Maccagni}, {Marinoni},
  {Mazure}, {Quintana}, \& {Read}}]{pierre2006}
{Pierre}, M., {Pacaud}, F., {Duc}, P.-A., {et~al.} 2006, \mnras, 372, 591

\bibitem[{{Poggianti} {et~al.}(2006){Poggianti}, {von der Linden}, {De Lucia},
  {Desai}, {Simard}, {Halliday}, {Arag{\'o}n-Salamanca}, {Bower}, {Varela},
  {Best}, {Clowe}, {Dalcanton}, {Jablonka}, {Milvang-Jensen}, {Pello},
  {Rudnick}, {Saglia}, {White}, \& {Zaritsky}}]{poggianti2006}
{Poggianti}, B.~M., {von der Linden}, A., {De Lucia}, G., {et~al.} 2006, \apj,
  642, 188

\bibitem[{{Pollo} {et~al.}(2005){Pollo}, {Meneux}, {Guzzo}, {Le F{\`e}vre},
  {Blaizot}, {Cappi}, {Iovino}, {Marinoni}, {McCracken}, {Bottini}, {Garilli},
  {Le Brun}, {Maccagni}, {Picat}, {Scaramella}, {Scodeggio}, {Tresse},
  {Vettolani}, {Zanichelli}, {Adami}, {Arnaboldi}, {Arnouts}, {Bardelli},
  {Bolzonella}, {Charlot}, {Ciliegi}, {Contini}, {Foucaud}, {Franzetti},
  {Gavignaud}, {Ilbert}, {Marano}, {Mathez}, {Mazure}, {Merighi}, {Paltani},
  {Pell{\`o}}, {Pozzetti}, {Radovich}, {Zamorani}, {Zucca}, {Bondi},
  {Bongiorno}, {Busarello}, {Gregorini}, {Lamareille}, {Mellier}, {Merluzzi},
  {Ripepi}, \& {Rizzo}}]{pollo2005}
{Pollo}, A., {Meneux}, B., {Guzzo}, L., {et~al.} 2005, \aap, 439, 887

\bibitem[{{Popesso} {et~al.}(2007{\natexlab{a}}){Popesso}, {Biviano},
  {B{\"o}hringer}, \& {Romaniello}}]{popesso2007_V}
{Popesso}, P., {Biviano}, A., {B{\"o}hringer}, H., \& {Romaniello}, M.
  2007{\natexlab{a}}, \aap, 461, 397

\bibitem[{{Popesso} {et~al.}(2005){Popesso}, {Biviano}, {B{\"o}hringer},
  {Romaniello}, \& {Voges}}]{popesso2005_III}
{Popesso}, P., {Biviano}, A., {B{\"o}hringer}, H., {Romaniello}, M., \&
  {Voges}, W. 2005, \aap, 433, 431

\bibitem[{{Popesso} {et~al.}(2007{\natexlab{b}}){Popesso}, {Biviano},
  {Romaniello}, \& {B{\"o}hringer}}]{popesso2007_VI}
{Popesso}, P., {Biviano}, A., {Romaniello}, M., \& {B{\"o}hringer}, H.
  2007{\natexlab{b}}, \aap, 461, 411

\bibitem[{{Popesso} {et~al.}(2004){Popesso}, {B{\"o}hringer}, {Brinkmann},
  {Voges}, \& {York}}]{popesso2004_I}
{Popesso}, P., {B{\"o}hringer}, H., {Brinkmann}, J., {Voges}, W., \& {York},
  D.~G. 2004, \aap, 423, 449

\bibitem[{{Postman} \& {Geller}(1984)}]{postman1984}
{Postman}, M. \& {Geller}, M.~J. 1984, \apj, 281, 95

\bibitem[{{Radovich} {et~al.}(2004){Radovich}, {Arnaboldi}, {Ripepi},
  {Massarotti}, {McCracken}, {Mellier}, {Bertin}, {Zamorani}, {Adami},
  {Bardelli}, {Le F{\`e}vre}, {Foucaud}, {Garilli}, {Scaramella}, {Vettolani},
  {Zanichelli}, \& {Zucca}}]{radovich2004}
{Radovich}, M., {Arnaboldi}, M., {Ripepi}, V., {et~al.} 2004, \aap, 417, 51

\bibitem[{{Richard} {et~al.}(2010){Richard}, {Smith}, {Kneib}, {Ellis},
  {Sanderson}, {Pei}, {Targett}, {Sand}, {Swinbank}, {Dannerbauer}, {Mazzotta},
  {Limousin}, {Egami}, {Jullo}, {Hamilton-Morris}, \& {Moran}}]{richard2010}
{Richard}, J., {Smith}, G.~P., {Kneib}, J., {et~al.} 2010, \mnras, 313

\bibitem[{{Rosati} {et~al.}(2002){Rosati}, {Borgani}, \& {Norman}}]{rosati2002}
{Rosati}, P., {Borgani}, S., \& {Norman}, C. 2002, \araa, 40, 539

\bibitem[{{Sheldon} {et~al.}(2009){Sheldon}, {Johnston}, {Masjedi}, {McKay},
  {Blanton}, {Scranton}, {Wechsler}, {Koester}, {Hansen}, {Frieman}, \&
  {Annis}}]{sheldon2009_ML}
{Sheldon}, E.~S., {Johnston}, D.~E., {Masjedi}, M., {et~al.} 2009, \apj, 703,
  2232

\bibitem[{{Springel} {et~al.}(2005){Springel}, {White}, {Jenkins}, {Frenk},
  {Yoshida}, {Gao}, {Navarro}, {Thacker}, {Croton}, {Helly}, {Peacock}, {Cole},
  {Thomas}, {Couchman}, {Evrard}, {Colberg}, \& {Pearce}}]{springel2005_MILL}
{Springel}, V., {White}, S.~D.~M., {Jenkins}, A., {et~al.} 2005, \nat, 435, 629

\bibitem[{{Sunyaev} \& {Zeldovich}(1980)}]{sunyaev_zeldovich1980}
{Sunyaev}, R.~A. \& {Zeldovich}, I.~B. 1980, \araa, 18, 537

\bibitem[{{Sunyaev} \& {Zeldovich}(1972)}]{sunyaev_zeldovich1972}
{Sunyaev}, R.~A. \& {Zeldovich}, Y.~B. 1972, Comments on Astrophysics and Space
  Physics, 4, 173

\bibitem[{{Temporin} {et~al.}(2008){Temporin}, {Iovino}, {Bolzonella},
  {McCracken}, {Scodeggio}, {Garilli}, {Bottini}, {Le Brun}, {Le F{\`e}vre},
  {Maccagni}, {Picat}, {Scaramella}, {Tresse}, {Vettolani}, {Zanichelli},
  {Adami}, {Arnouts}, {Bardelli}, {Cappi}, {Charlot}, {Ciliegi}, {Contini},
  {Cucciati}, {Foucaud}, {Franzetti}, {Gavignaud}, {Guzzo}, {Ilbert}, {Marano},
  {Marinoni}, {Mazure}, {Meneux}, {Merighi}, {Paltani}, {Pell{\`o}}, {Pollo},
  {Pozzetti}, {Radovich}, {Vergani}, {Zamorani}, {Zucca}, {Bondi}, {Bongiorno},
  {Brinchmann}, {de la Torre}, {Lamareille}, {Mellier}, \&
  {Walcher}}]{temporin2008}
{Temporin}, S., {Iovino}, A., {Bolzonella}, M., {et~al.} 2008, \aap, 482, 81

\bibitem[{{Treu} {et~al.}(2003){Treu}, {Ellis}, {Kneib}, {Dressler}, {Smail},
  {Czoske}, {Oemler}, \& {Natarajan}}]{treu2003}
{Treu}, T., {Ellis}, R.~S., {Kneib}, J.-P., {et~al.} 2003, \apj, 591, 53

\bibitem[{{Tully}(1980)}]{tully1980h}
{Tully}, R.~B. 1980, \apj, 237, 390

\bibitem[{{Valtchanov} {et~al.}(2004){Valtchanov}, {Pierre}, {Willis}, {Dos
  Santos}, {Jones}, {Andreon}, {Adami}, {Altieri}, {Bolzonella}, {Bremer},
  {Duc}, {Gosset}, {Jean}, \& {Surdej}}]{valtchanov2004}
{Valtchanov}, I., {Pierre}, M., {Willis}, J., {et~al.} 2004, \aap, 423, 75

\bibitem[{{Voronoi}(1908)}]{voronoi1908}
{Voronoi}, G.~F. 1908, J. Reine Angew. Math., 134, 198

\bibitem[{{Weinmann} {et~al.}(2006){Weinmann}, {van den Bosch}, {Yang}, \&
  {Mo}}]{weinmann06}
{Weinmann}, S.~M., {van den Bosch}, F.~C., {Yang}, X., \& {Mo}, H.~J. 2006,
  \mnras, 366, 2

\bibitem[{{Willis} {et~al.}(2005{\natexlab{a}}){Willis}, {Pacaud},
  {Valtchanov}, {Pierre}, {Ponman}, {Read}, {Andreon}, {Altieri}, {Quintana},
  {Dos Santos}, {Birkinshaw}, {Bremer}, {Duc}, {Galaz}, {Gosset}, {Jones}, \&
  {Surdej}}]{willis2005}
{Willis}, J.~P., {Pacaud}, F., {Valtchanov}, I., {et~al.} 2005{\natexlab{a}},
  \mnras, 363, 675

\bibitem[{{Willis} {et~al.}(2005{\natexlab{b}}){Willis}, {Pacaud},
  {Valtchanov}, {Pierre}, {Ponman}, {Read}, {Andreon}, {Altieri}, {Quintana},
  {Santos}, {Birkinshaw}, {Bremer}, {Duc}, {Galaz}, {Gosset}, {Jones}, \&
  {Surdej}}]{willis2005err}
{Willis}, J.~P., {Pacaud}, F., {Valtchanov}, I., {et~al.} 2005{\natexlab{b}},
  \mnras, 364, 751

\bibitem[{{Zhang} {et~al.}(2006){Zhang}, {B{\"o}hringer}, {Finoguenov},
  {Ikebe}, {Matsushita}, {Schuecker}, {Guzzo}, \& {Collins}}]{zhang2006_baryon}
{Zhang}, Y., {B{\"o}hringer}, H., {Finoguenov}, A., {et~al.} 2006, \aap, 456,
  55

\bibitem[{{Zucca} {et~al.}(2006){Zucca}, {Ilbert}, {Bardelli}, {Tresse},
  {Zamorani}, {Arnouts}, {Pozzetti}, {Bolzonella}, {McCracken}, {Bottini},
  {Garilli}, {Le Brun}, {Le F{\`e}vre}, {Maccagni}, {Picat}, {Scaramella},
  {Scodeggio}, {Vettolani}, {Zanichelli}, {Adami}, {Arnaboldi}, {Cappi},
  {Charlot}, {Ciliegi}, {Contini}, {Foucaud}, {Franzetti}, {Gavignaud},
  {Guzzo}, {Iovino}, {Marano}, {Marinoni}, {Mazure}, {Meneux}, {Merighi},
  {Paltani}, {Pell{\`o}}, {Pollo}, {Radovich}, {Bondi}, {Bongiorno},
  {Busarello}, {Cucciati}, {Gregorini}, {Lamareille}, {Mathez}, {Mellier},
  {Merluzzi}, {Ripepi}, \& {Rizzo}}]{zucca2006_VVDS_LF}
{Zucca}, E., {Ilbert}, O., {Bardelli}, S., {et~al.} 2006, \aap, 455, 879

\bibitem[{{Zwicky} {et~al.}(1968){Zwicky}, {Herzog}, \&
  {Wild}}]{zwicky1968_clusters}
{Zwicky}, F., {Herzog}, E., \& {Wild}, P. 1968, {Catalogue of galaxies and of
  clusters of galaxies} (Pasadena: California Institute of Technology (CIT),
  1961-1968)

\end{thebibliography}

\end{document}